\documentclass[aps,prb,superscriptaddress,twocolumn]{revtex4-2}
\usepackage[cp1251]{inputenc}
\usepackage{amsmath}
\usepackage{amssymb}
\usepackage{amsfonts}
\usepackage{color}
\usepackage{graphicx}
\usepackage{lipsum,graphicx}
\usepackage{amsfonts}
\usepackage{textcomp}
\usepackage[normalem]{ulem}
\newcommand{\beq}{\begin{equation}}
	\newcommand{\eeq}{\end{equation}}
\newcommand{\bea}{\begin{eqnarray}}
	\newcommand{\eea}{\end{eqnarray}}

\usepackage{hyperref}  
\usepackage{hyperref}

\begin{document}
\title{Intrinsic Negative Magnetoresistance in Layered  AFM Semimetals: the Case of EuSn$_2$As$_2$}
\author{K. S. Pervakov}
\author{A.V. Sadakov}
\author{O. A. Sobolevskiy}
\author{V. A. Vlasenko}
\author{V. P. Martovitsky}
	\affiliation{V.~L.~Ginzburg Research Center at P.~N.~Lebedev Physical Institute, RAS, Moscow 119991, Russia.}
	\author{E. A. Sedov}
 	\affiliation{V.~L.~Ginzburg Research Center at P.~N.~Lebedev Physical Institute, RAS,  Moscow 119991, Russia.}
\affiliation{HSE University, Moscow 101000, Russia}
\author{E. I.  Maltsev}
	\author{N. P\'{e}rez}
	\affiliation{Leibniz Institute for Solid State and Materials Research, IFW,  
		01069 Dresden, Germany}
	\affiliation{Dresden-W\"{u}rzburg Cluster of Excellence ct.qmat, Dresden, Germany}
	\author{L. Veyrat}
	\affiliation{Leibniz Institute for Solid State and Materials Research, IFW,  
		01069 Dresden, Germany}
	\affiliation{Dresden-W\"{u}rzburg Cluster of Excellence ct.qmat, Dresden, Germany}
	\affiliation{Laboratoire National des Champs Magnetiques Intenses, CNRS-INSA-UJF-UPS,  F-31400 Toulouse, France}
\author{P. D. Grigoriev}
	\affiliation{L.D. Landau Institute of Theoretical Physics, RAS}
\author{N. S. Pavlov}
	\affiliation{Institute for Electrophysics,  RAS, Ekaterinburg, 620016, Russia}
	 	\affiliation{V.~L.~Ginzburg Research Center at P.~N.~Lebedev Physical Institute, RAS,  Moscow 119991, Russia.}
\author{I. A. Nekrasov}
	\affiliation{Institute for Electrophysics, RAS, Ekaterinburg, 620016, Russia}
\author{O. E. Tereshchenko}
	\affiliation{Institute for Semiconductor Physics, RAS, Novosibirsk, Russia}
\author{V. A. Golyashov}
	\affiliation{Institute for Semiconductor Physics, RAS, Novosibirsk, Russia}
\author{V. M. Pudalov}
 	\affiliation{V.~L.~Ginzburg Research Center at P.~N.~Lebedev Physical Institute, RAS,  Moscow 119991, Russia.}
	\affiliation{HSE University, Moscow 101000, Russia}

\begin{abstract}
Here, by applying a comprehensive approach including magnetic, transport measurements, ARPES band structure measurements,  DFT calculations,  and analytical theory consideration, we unveil the puzzling origin of the negative isotropic magnetoresistance in  the highly anisotropic semimetals, particularly, Eu$_2$Sn$_2$As$_2$ with AFM ordering of Eu atoms. The isotropic magnetoresistance developing along with the magnetization changes up to the complete spin polarization field was reported previously in several experimental studies, though its theoretical explanation was missing up to date.  Recently, we proposed a 
novel theoretical mechanism to describe the observed magnetoresistance in layered AFM compounds
by exchange splitting of the electron energy levels and by confining the electron wave functions with different spin projection in the vicinity of the respective magnetic layer. In this paper, we present more detailed experimental studies of the negative magnetoresistance with several samples of EuSn$_2$As$_2$ in order to identify its sample-independent features  including temperature dependence.
We also  substantiate the proposed theory by comparing it with magnetotransport data, with ARPES measurements of the energy band structure, and DFT energy spectrum calculations.
\end{abstract}

\maketitle

\section{Introduction}
The van der Waals  materials with magnetically  ordered atomic layers 
have inspired great scientific interests in different fields due to their tunable ground states
\cite{chang_Science_2013, zhang_PRL_2019, jo_PRB_2020, sanchez_PRB_2021}. These layered 
materials are attracting from fundamental point of view  because can host 
interesting  topological states, such as quantum anomalous Hall state, chiral anomaly, 
Chern insulator, higher-order topological M\"{o}bius insulator, etc. Among them, of a special interest are intrinsic  (stoichiometric) materials  with significant spin-orbit and exchange coupling,  since  disorder effects can be suppressed to a large extent, which is expected to facilitate the virgin quantum phenomena.

These compounds host sublattice of large spin atoms  (Eu, Mn, Co, etc), which order antiferromagnetically upon lowering temperature, and layers 
of Se, As, Te supplying electrons at the Fermi level. The most famous and most studied compound  is  MnBi$_2$Te$_4$ 
an intrinsic  topological insulator (TI)  with antiferromagnetic (AFM) ordering of Mn-ions, that manifests fascinating physics. 
Less explored (though not less interesting)  are  topologically trivial stoichiometric  
Eu-based semimetals, such as EuSn$_2$As$_2$  \cite{arguilla_InChemFront_2017, pakhira_PRB_2021, golov_JMMM_2022, lv_ASCAplElextronMat_2022, li_PRX_2019, chen_ChPhysLet_2020, li_PRB_2021}, 
EuSn$_2$P$_2$ \cite{pierantozzi_PNAS_2022, gui_ASCCentrSci_2019}, EuFe$_2$As$_2$ \cite{jiang_NJP_2009, sanchez_PRB_2021, golov_PRB_2022, talanov_JETPL_2023}, and Co-based semimetals - CaCo$_2$As$_2$ \cite{ying_PRB_2012}, 
and CsCo$_2$Se$_2$ \cite{yang_JMMM_2019}. These ``122-compounds'' are also  building blocks of more complex
remarkable materials, such as e.g. EuRbFe$_4$As$_4$ \cite{Eu-1144_ARPES, Eu-1144_STM}, the high-$T_c$-superconductor with AFM ordering of Eu atoms.

All the layered non-Weyl  AFM semimetals exhibit a  negative magnetoresistance (NMR). 
The most puzzling is the fact that the magnetoresistance in the highly anisotropic layered compounds is fully isotropic.
Despite the fact that this isotropic  negative magnetoresistance (NIMR) is well documented experimentally, its theoretical interpretation  was missing, until recently. In topologically trivial semimetals (EuSn$_2$As$_2$,
EuSn$_2$P$_2$, CaCo$_2$As$_2$, and CsCo$_2$Se$_2$) 
NIMR is approximately similar: it has a magnitude about 3-6\%, nearly parabolic shape $\delta R(H)/R(0) \propto -\alpha H^2$,  and is isotropic with respect to the field direction and current direction \cite{chen_ChPhysLet_2020, li_PRB_2021, ying_PRB_2012}. As field increases, this parabolic  negative magnetoresistance sharply terminates, exactly at the 
field of complete spin polarization $H_{\rm sf}$, and further changes to a conventional positive  magnetoresistance. Thus, the NMR  correlates closely with the  magnetization $M(H)$ characteristics \cite{pakhira_PRB_2021, chen_ChPhysLet_2020, li_PRB_2021}. 
Earlier, the negative magnetoresistance in layered AFM semimetals was suggested to associate  with either nontrivial topological properties,  or with scattering by magnons, domains,  magnetic impurities.
We showed in Ref.~\cite{PRL_tbp} and explain in more detail in this paper that none of the previously known mechanisms is capable to explain the observed negative isotropic magnetoresistance. 

In view of the remarkable magnetoresistance isotropy, we introduced in Ref.~\cite{PRL_tbp} a novel mechanism of NIMR that implies only point-like scattering of electrons and provides the isotropy of the NIMR. Within the framework of this mechanism, the 
driving force of the NIMR is the exchange splitting  of the energy bands in the AFM polarized sublattices. It  modifies spatial distribution of the wave function of spin-up and spin-down electrons and leads to the magnetization-dependent enhancement of scattering rate.
Though the close relationship between magnetization and magnetoresistance is well known also for
giant magnetoresistance (GMR) in layered Fe/Cr superlattices and for colossal magnetoresistance (CMR) in granular samples and in manganites \cite{dagotto_PhysRep_2001, dagotto_book,  yin_PRB_2000, baibich_PRL(1988)}, 
the novel mechanism \cite{PRL_tbp} is completely different and provides isotropic negative magnetoresistance.

In the current paper we provide more details of the proposed theory of NIMR \cite{PRL_tbp}, present extended experimental data on transport, magnetotransport, magnetization and band structure of EuSn$_2$As$_2$ single crystals. 
Firstly, we show that the observed  NIMR is an {\it intrinsic} (i.e. irrelevant to structural defects) property of the layered AFM materials with strong spin-orbit and exchange splitting.  This conclusion is based on transport measurements  with  samples of various thicknesses (from 140nm to 500nm),  magnetization and  ARPES measurements,  DMFT band structure calculations,  and  high precision  XRD and TEM examination of the crystal  perfection. 

Secondly,  we  test the novel theoretical  mechanism and show that it  qualitatively explains the observed NIMR. 
In the proposed  mechanism,   NIMR originates from the enhancement of the usual short-range scattering of conducting electrons by the alternating sublattice magnetization in the layered AFM crystal.  Therefore, this NIMR mechanism may be 
applicable not only to EuSn$_2$As$_2$, but also to a  wide class of the layered AFM semimetals, independent of their topology.

The paper is organized as follows.
In Section \ref{sec:samples} we describe  crystal lattice, samples and their characterization. In section \ref{sec:BS}
we present our results of band structure ARPES measurements and DMFT calculations, 
  in section \ref{sec:magnetization} we present the main experimental data 
 of magnetization, and in section \ref{sec:magnetotransport} - of magnetotransport measurements,  particularly, we show full isotropy of the NIMR with respect to the field direction and current direction. In section \ref{sec:theory} we introduce the basic idea of the model. 
In section \ref{sec:data comparison} we compare  our calculations with experimental data. Finally, we provide a brief summary and discussion of the results in section \ref{sec:conclusion}.  In Appendices I, II and III
we analyse difference in the NMR shape for various samples, discuss the relevance of various scattering mechanisms to NIMR, and present a detailed theoretical derivation of NIMR. 
Supplemental materials \cite{SM} contain additional information on the crystal synthesis,  XRD examination of the lattice  structure, energy spectrum measurements (ARPES) and calculations.

Although our experimental studies are performed with EuSn$_2$As$_2$, the magnetoresistance effect we explore  is   common for a number of the above listed layered compounds.
\section{Lattice structure. Samples and their characterization}
\label{sec:samples}
In EuSn$_2$As$_2$,  at $T<24$\,K the magnetic Eu-sublattice orders in the A-type AFM structure (AFMa) in which   Eu magnetic moments lie in the $ab$-plane and rotate by $\pi$ from layer to layer (Fig.~1). 
By combining first principles calculations and time-resolved tr-ARPES measurements, Li et al. \cite{li_PRX_2019} suggested that EuSn$_2$As$_2$ is an intrinsic AFM topological insulator.
Our low temperature magnetotransport as well as  ARPES studies and DFT calculations (Sec.~\ref{sec:BS}) however,  show that EuSn$_2$As$_2$ is the non-Weyl semimetal, rather than TI. 

\begin{figure*}
\includegraphics[width=230pt]{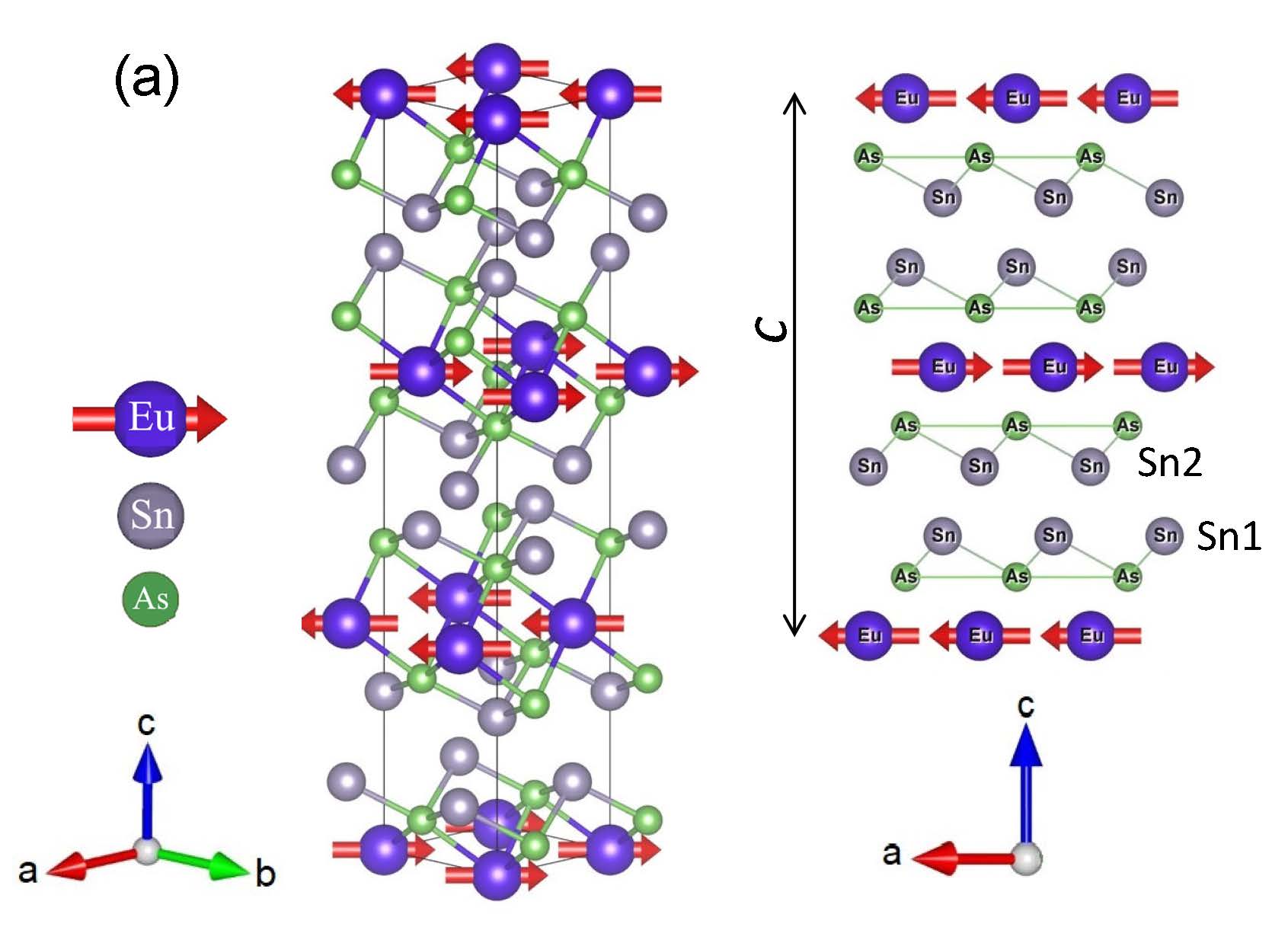}
	\includegraphics[width=85pt]{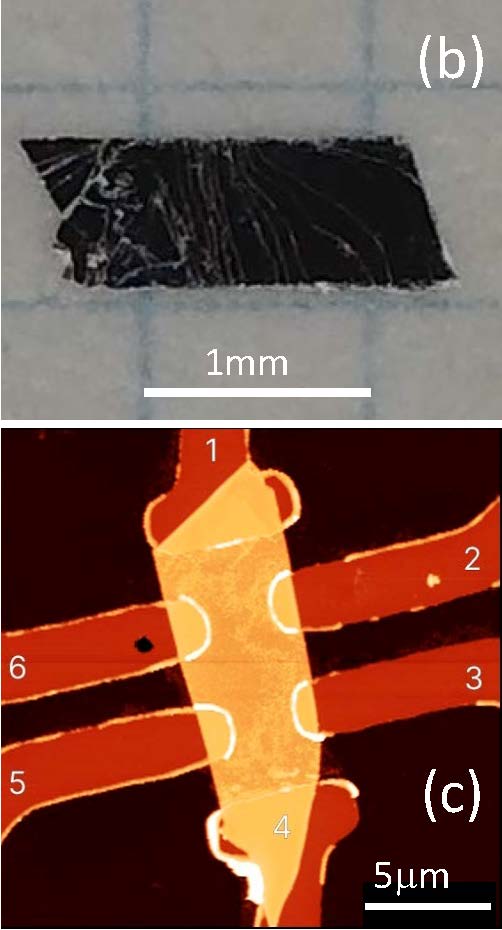}
	\caption{(a) Lattice structure of the EuSn$_2$As$_2$,  3D-view and 2D projection on the $a-c$ plane. Magenta arrows show 
		Eu-atoms magnetization direction in the A-type AFM ordered state. 
		(b) Typical bulk crystal. (c) Top view of the 140-nm-thick exfoliated flake with 
		Au contact stripes \cite{maltsev_tbp}. Horizontal white bar shows 5\,$\mu$m scale. 
	}
	\label{fig:crystal_structure}
\end{figure*}

The layered EuSn$_2$As$_2$ compound has a space group R-3m, where each trigonal Eu layer is sandwiched between two buckled honeycomb SnAs layers (Fig.~\ref{fig:crystal_structure}), located at a distance $c_0 \approx 2.4$\AA \ apart \cite{lv_ASCAplElextronMat_2022}. The overall height of the elementary cell is $c\approx 26.4$\AA 
(for more detail, see  SM \cite{SM}). The charge carriers are supplied to the band  from SnAs layers confined between the layers of Eu (see Section \ref{sec:BS}, and  SM \cite{SM}).  At low temperatures, below the Neel temperature $T_N \approx 22-24$K, the carriers experience the static  alternating magnetization of ferromagnetically ordered individual Eu layers.

\begin{figure}
		\includegraphics[width=230pt]{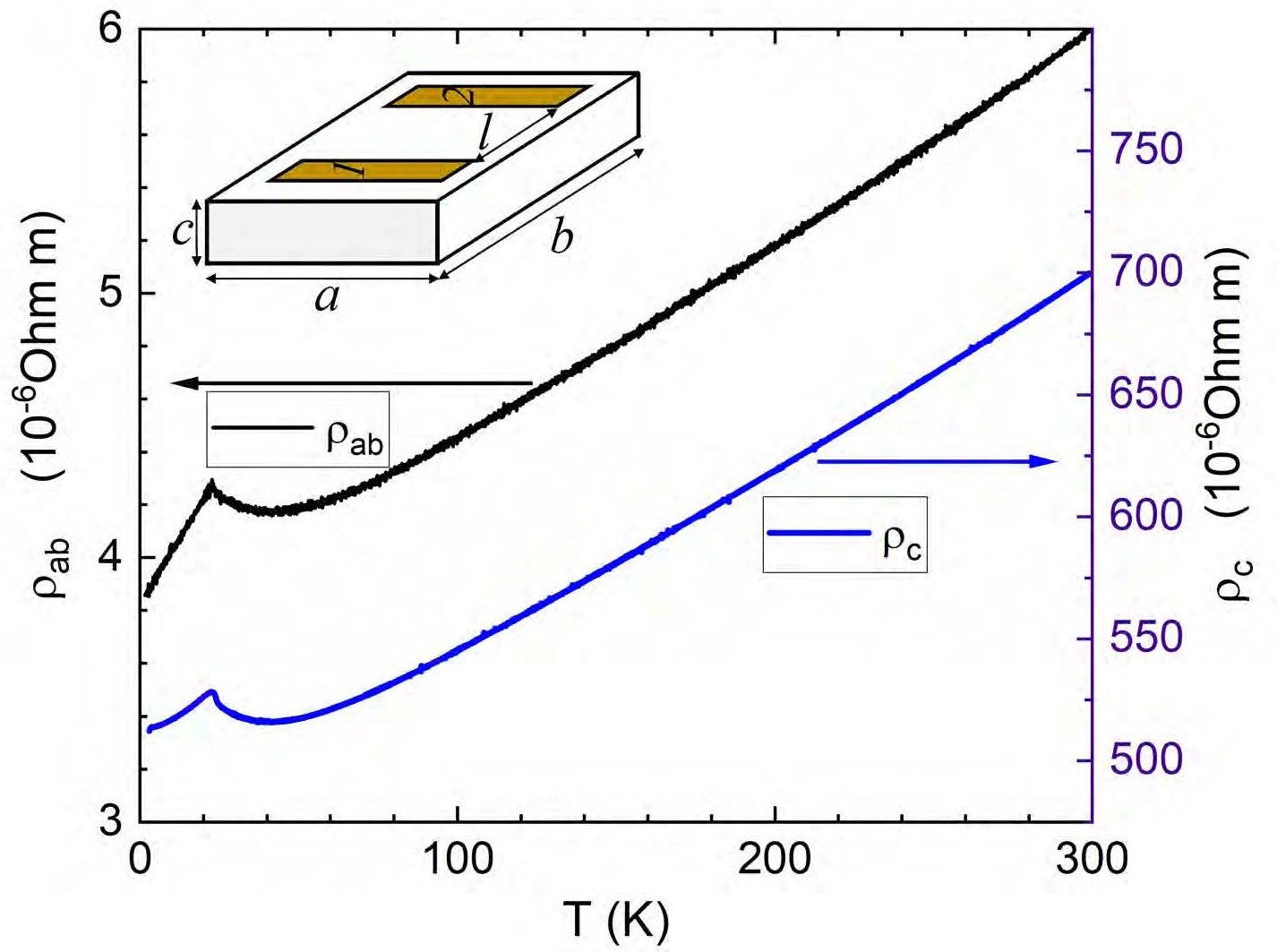}
	\caption{Temperature dependences of the resistivity components at zero field for  bulk sample \#2: in-plane  $\rho_{ab}$,  and  normal to the basal plane  $\rho_c$.  Insert shows  schematics of the sample and contact arrangement. $a=1.1$mm, $b=2.6$mm, $c\approx 0.1 \pm 0.03$mm, $l= 2$mm. {\em 1, 2} - current contact pads, and  {\em 3, 4} (underneath, invisible) - potential contacts. 
			}
	\label{fig:R(T)}
\end{figure}

Transport measurements were done with bulk crystals, approximately 
$ (1 -2.5)$\,mm in the $ab$-plane and $(0.1-0.5)$mm thick (Fig.~\ref{fig:R(T)}, inset),   and with exfoliated thin flakes (Fig.~\ref{fig:crystal_structure}b,c). Two bulk samples 
have been studied most intensively: 
 sample \#1-1  with a thickness of 0.5\,mm  after growth already had one flat mirror surface and was not cleaved.  Sample \#2  was cleaved down to thickness of  0.1mm and  shaped rectangular. Au/Ti contact pads were deposited through a contact mask on the face and bottom sides of the bulk samples. 30-$\mu$m Au-wires were attached to the contact pads using  conducting silver paint. 

We also studied the flake exfoliated from the bulk crystal with lateral dimensions $\sim 5\times 10\mu$m in the $ab$-plane and thickness 140 nm in the $c$ direction (Fig.~\ref{fig:crystal_structure}c).
Further studies on the flakes are presented elsewhere [Ref.~\cite{maltsev_tbp}].
Electrical leads and contact pads  to the flake were made of  Au/Ti film  evaporated by e-beam gun and patterned  using laser lithography.  Underneath the contacts the flake was thinned down to 40\,nm in order to provide more stable 
Au/Ti contacts.

The  resistance components were measured using four probe AC-technique: the diagonal $R_{xx}$ component was measured with bulk crystals and flakes, whereas  Hall resistance $R_{xy}$ was measured only with flakes, using lithographically defined four or more contacts. 
Two setups were used for bulk samples. For measuring in-plane resistance, $R_{ab}$, the bias current was applied through the contacts at  one $ab$-surface and voltage drop was  measured between two potential contacts deposited at one of the   $ab$-surfaces.  For measuring $R_c$ component, bias current was applied along  $c$ axis and  voltage drop was measured between potential contacts deposited at two opposite $ab$-sides. An example of  contacts configuration for flakes is shown in Fig.~\ref{fig:crystal_structure}c, and for bulk sample -- in the inset to Fig.~\ref{fig:R(T)}.
 Two resistance components $R_{xx}$ and $R_{xy}$ were measured in different magnetic sweeps due to limitation of the setup.

Temperature dependences of the diagonal resistance components, in-plane  $R_{ab}$, and normal to the plane, $R_c$,  in zero field are shown in Fig.~\ref{fig:R(T)}. Both components $R_{ab}$ and $R_c$ show a kink, 1-2\% high, at temperature of the AFM ordering $T_N\approx 24$K.
The increase in resistivity near $T_N$  is likely due to scattering of delocalized conduction 
electrons by the fluctuating localized Eu$^{2+}$ magnetic moments beginning to order in the vicinity of 
$T_N$  \cite{rojas_JAlComp_2010}. 
Such behavior in the vicinity of magnetic ordering temperature was previously observed for EuSn$_2$As$_2$  ~\cite{arguilla_InChemFront_2017, chen_ChPhysLet_2020, pakhira_PRB_2021}, for EuSn$_2$P$_2$ \cite{gui_ASCCentrSci_2019} and for EuIn$_2$As$_2$ \cite{yu_PRB_2020}.

Both  $\rho_{ab}$ and $\rho_c$ components  exhibit a typical semimetallic temperature dependence consistent with that reported earlier \cite{chen_ChPhysLet_2020, pakhira_PRB_2021, arguilla_InChemFront_2017,  li_PRB_2021}. The resistance gradually decreases with lowering temperature from 300K to 30K by a factor of $\approx 1.4$ for the bulk and for the 140nm-thick flake, displaying no contribution from the predicted bulk insulator and Dirac surface states \cite{li_PRX_2019}. 
Below the kink,  both resistivity components continue  decreasing down to our lowest $T=2$\,K, with a difference that  $R_{ab}$ decreases steeper than $R_c$. This is consistent with the gapless band structure revealed in our ARPES measurements and  band structure calculations (Section \ref{sec:BS}), 
and also consistent with previous studies in Ref.~\cite{li_PRX_2019}.  
The interlayer resistivity $\rho_c$ was measured only with bulk crystal; it is  a factor of 130 larger than the in-plane component, whereas the temperature dependences of both components are similar when scaled down by the anisotropy factor.
The huge resistivity anisotropy $\rho_c/\rho_{ab}\sim 130$ is qualitatively consistent with highly anisotropic crystal structure,
 where  $c\approx 26.4$\AA,  $a, b \approx 4.22$\AA  (see Fig.~\ref{fig:crystal_structure}, SM \cite{SM}, and also  Ref.~\cite{arguilla_InChemFront_2017}).

Our data for all studied samples  doesn't confirm a superconducting transition below 4.8K contrary to that reported in Ref.~\cite{sakuragi_arxiv}.
In measurements  with bulk crystals and   flakes we didn't observe also a nonmonotonic $\rho(H)$ and $M(H)$ behavior at $H>5$\,T as that reported in Ref.~\cite{li_PRB_2021}.

\section{Band structure ARPES measurements and DFT calculations}
\label{sec:BS}
\begin{figure}
	\includegraphics[width=240pt]{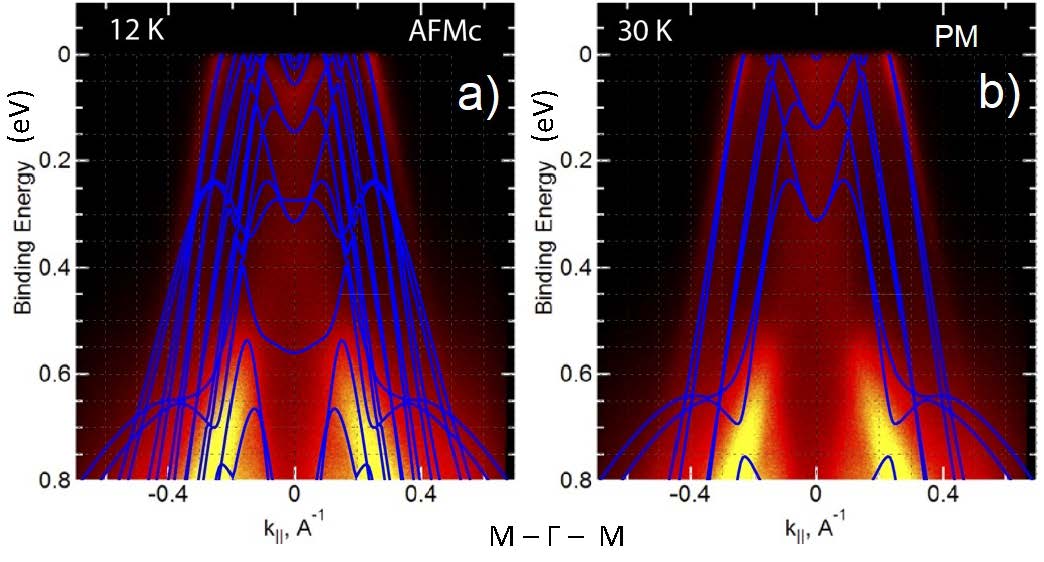}
	\caption{Electronic structure of EuSn$_2$As$_2$ bulk crystal below $E_F$ along 
		M-$\Gamma$-M direction. (a) in the AFM state at $T=12$K, and (b) in the PM state at $T=30$K.
		Color  intensity map - ARPES measurements  at  photon energy $h\nu = 21.2$\,eV. Blue lines  show the calculated dispersion in the  AFM and PM state. 
	}
	\label{fig:ARPES}
\end{figure}

Band structure of the crystal surface was investigated by angle resolved photoemission spectroscopy (ARPES). 
Measurements were done using He-Ia radiation ($h\nu = 21.22$\,eV) of a nonmonochromated He-discharge lamp (UVS-300, SPECS GMBH) and an hemispherical electron energy analyzer equipped with electrostatic deflector (ASTRAIOS-190, SPECS GMBH). The samples were cleaved in situ in a vacuum of $5\times 10^{-10}$\,mbar at room temperature. Low energy electron diffraction (LEED) measurements showed 
clear $1\times 1$ diffraction patterns and confirmed the cleaved crystal (0001) surfaces to be free from any reconstruction. The 
results are presented in Fig.~\ref{fig:ARPES} for the  $M-\Gamma-M$ direction for low binding energies 0 - 0.8eV. 

\begin{figure}[ht]
	\includegraphics[width=1.0\linewidth]{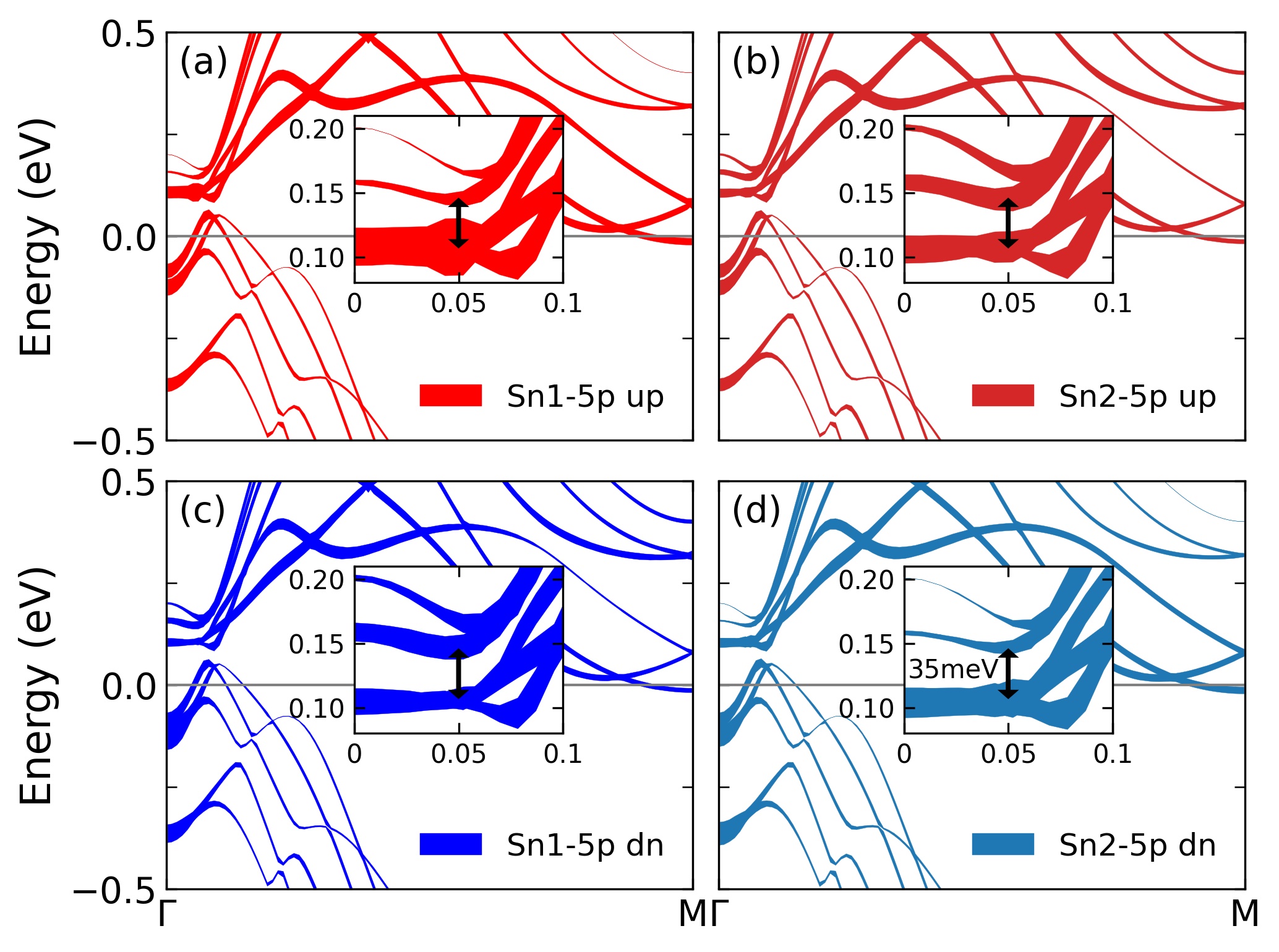}
	\includegraphics[width=1.0\linewidth]{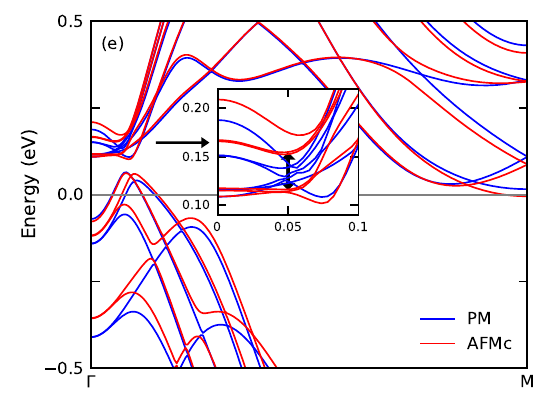}
	\caption{Band structure of EuSn$_2$As$_2$ with orbital and spin contributions for Sn-5p states from the 1st layer Sn1 (see  panels a,c) and from the 2nd layer Sn2 (panels  b,d).  The band width corresponds to orbital contribution to the band.
	(e) Comparison of the band structure in AFMc (red lines) and PM (blue lines) state  with the same double-size along  $c$ axes unit cell.  The insert shows an enlarged region of splitting. Verical arrows show the exchange splitting.
	}
	\label{fig:DFT_fbands}
\end{figure}

Data for a wider  energy range and for the $k_x-k_y$ cut at the Fermi level is given in SM \cite{SM}. 
When our ARPES data  are considered in the wide energy range from 0 to -4.5eV  \cite{SM}, the overall energy spectrum  is very similar to that  measured earlier  \cite{li_PRX_2019}, calculated in Ref.~\cite{lv_ASCAplElextronMat_2022}, and also shown in Fig.~\ref{fig:DFT_fbands}.
We  focus here on the low-energy  region in the vicinity of the Fermi level, since namely these quasiparticles participate in the multi-band semimetallic transport at low temperatures.  
 In the vicinity of the  $\Gamma$ point there is an electron-like pocket, surrounded by several enclosed hole bands (which  presumably have a corrugate cylinder shape). Other electron bands theoretically predicted to be around ${\rm M}$ point, and a surface topological states at $\Gamma$ point are not seen at $E\leq E_F$, in agreement with Ref.~\cite{li_PRX_2019}.
Both the hole- and electron levels  in the AFM state are significantly broadened, the feature that has not been discussed earlier.  In Figure \ref{fig:ARPES}, at the  $\Gamma-\rm{M}$ cross-section,  one can clearly see that broadening of  the hole  and  electron pocket contours 
persists down to low temperatures. 

The DFT band structure calculations were performed within the DFT+U approximation in the VASP software package~\cite{vasp}. The generalized gradient approximation (GGA) in the form of the Perdew-Burke-Ernzerhof (PBE) exchange-correlation functional~\cite{DFT_PBE} was employed. The onsite Coulomb interaction of Eu-$4f$ electrons was described with the DFT+U scheme with the Dudarev approach~\cite{DFT_U} (U=5.0\,eV same as in Ref.~\cite{li_PRX_2019}).	
A non-collinear magnetic DFT calculations  were performed with spin-orbit coupling.
The calculated electron energy dispersion in AFM magnetic state is shown in Figs.~\ref{fig:ARPES} by the blue lines. 
The main contribution to the  bands near the Fermi level comes from the  Sn-$p$ and As-$p$ orbitals. 
	
The most important result of these calculations for clarifying 
the origin of NIMR (and, particularly, for substantiating our theory of NIMR \cite{PRL_tbp}) is the band splitting around the Fermi level shown in more detail in Fig.~\ref{fig:DFT_fbands}.   In a noncollinear calculation, the spin components are not easily distinguished. 
For distinguishing the up and down spin components, we performed calculation in the collinear spin configuration that corresponds to the AFM-$c$ order.

The elementary cell of EuSn$_2$As$_2$ contains a pair of neighboring layers of Sn atoms (we call them Sn1 and Sn2 - see Fig.~\ref{fig:crystal_structure}a),
sandwiched  between Eu layers with opposite spin projection ($\uparrow$ and $\downarrow$, or ``left'' and ''right'' in Fig.~1a). 
The ferromagnetic magnetization of the nearest Eu-neighbor leads to magnetization of the Sn-5p state and to the spatial separation of the oppositely magnetized  $\downarrow$,$\uparrow$ Sn-5p states \cite{PRL_tbp}.  Contribution of As-4p
states to the bands at the Fermi level is not large, therefore we consider only Sn-5p orbital resolved bands.

Some of the energy levels  for Sn-5p states with $\downarrow$,$\uparrow$ spin coincide with each other (see Fig.~\ref{fig:DFT_fbands}).  
However, near the $\Gamma$-point and slightly below the Fermi level, the Sn-5p bands with $\downarrow$,$\uparrow$ spin projection are split in energy by about 30-40 meV.  The spliting is well seen above the Fermi level and the given above value
 corresponds to the Sn-5p state. The presence of bands splitting indicates that these states are affected by exchange interaction. Moreover, the Sn1 contribution with spin up is almost the same as Sn2 with spin down.  As a result, the 
Sn1 and Sn2 layers are predominantly magnetized in  opposite directions depending on which Eu atom is closer. This spatial separation of the oppositely magnetized Sn layeres is confirmed by our DFT calculations in Ref.~\cite{PRL_tbp}.

In order to illustrate that the level splitting originates mostly from  exchange, we compare in Fig.~\ref{fig:DFT_fbands}e 
the calculated 	band structure  in the AFM-c and paramagnetic (PM) states. 
One can see in the same region near the $\Gamma$-point and slightly below Fermi level (see the insert to Fig.~\ref{fig:DFT_fbands}e) that the PM bands (blue lines) almost coincide at $k\sim 0.05$\AA$^{-1}$. There are several PM bands due to the presence of 6 such pairs of Sn layers in the unit cell.  In contrast, in the AFM state, the bands at the same region  are split. The splitting may be due to hopping along $z$ axis (crystallographic $c$-direction) between layers and due to exchange interaction. Since the bands in this region collapse in the paramagnetic
solution, it suggests that the splitting has the exchange 	nature. In the earlier investigation Ref.~\cite{AFM_splitting}, 	the band splitting was also found in calculation without spin-orbit coupling for AFM solution (see the SST4 case~\cite{AFM_splitting}). 

The agreement of our calculations with Ref.~\cite{AFM_splitting} supports the exchange origin of the bands splitting in the  AFM state. The value of exchange interaction is estimated namely from splitting in this region, because the bands with opposite spin always coincide in the conventional AFM state. Therefore, the splitting of Sn-bands with opposite spin is a ``smoking gun evidence'' for the exchange interaction origin  of the splitting. 
		
To summarize this section:	Figure~\ref{fig:DFT_fbands}e clearly shows that 
in the	PM state (blue lines) the Sn- bands almost coincide, and a minor difference between them 
is due to the presence of 6 such pairs of Sn layers in the elementary cell.
In the AFM state, the bands  with opposite spins for one Sn atom are essentially  split (see  Fig.~\ref{fig:DFT_fbands}e and the inset).	We also note, that earlier in Ref.~\cite{AFM_splitting}  the band splitting without spin-orbit contribution  was also found in the AFM state   (SST4 case),  which supports our result.

\section{Magnetization measurements}
\label{sec:magnetization}

\begin{figure}[ht]
	\includegraphics[width=200pt]{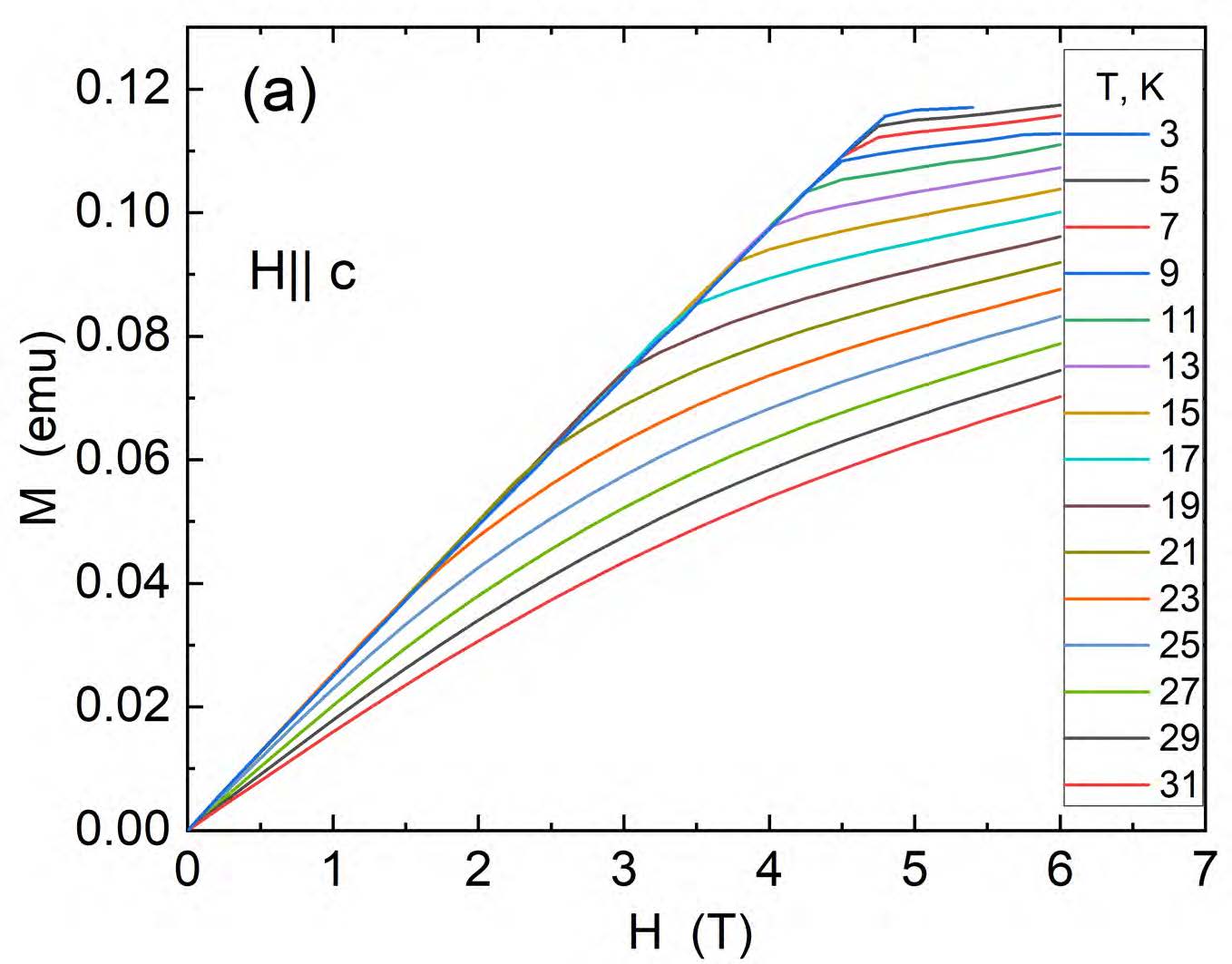}
	\includegraphics[width=200pt]{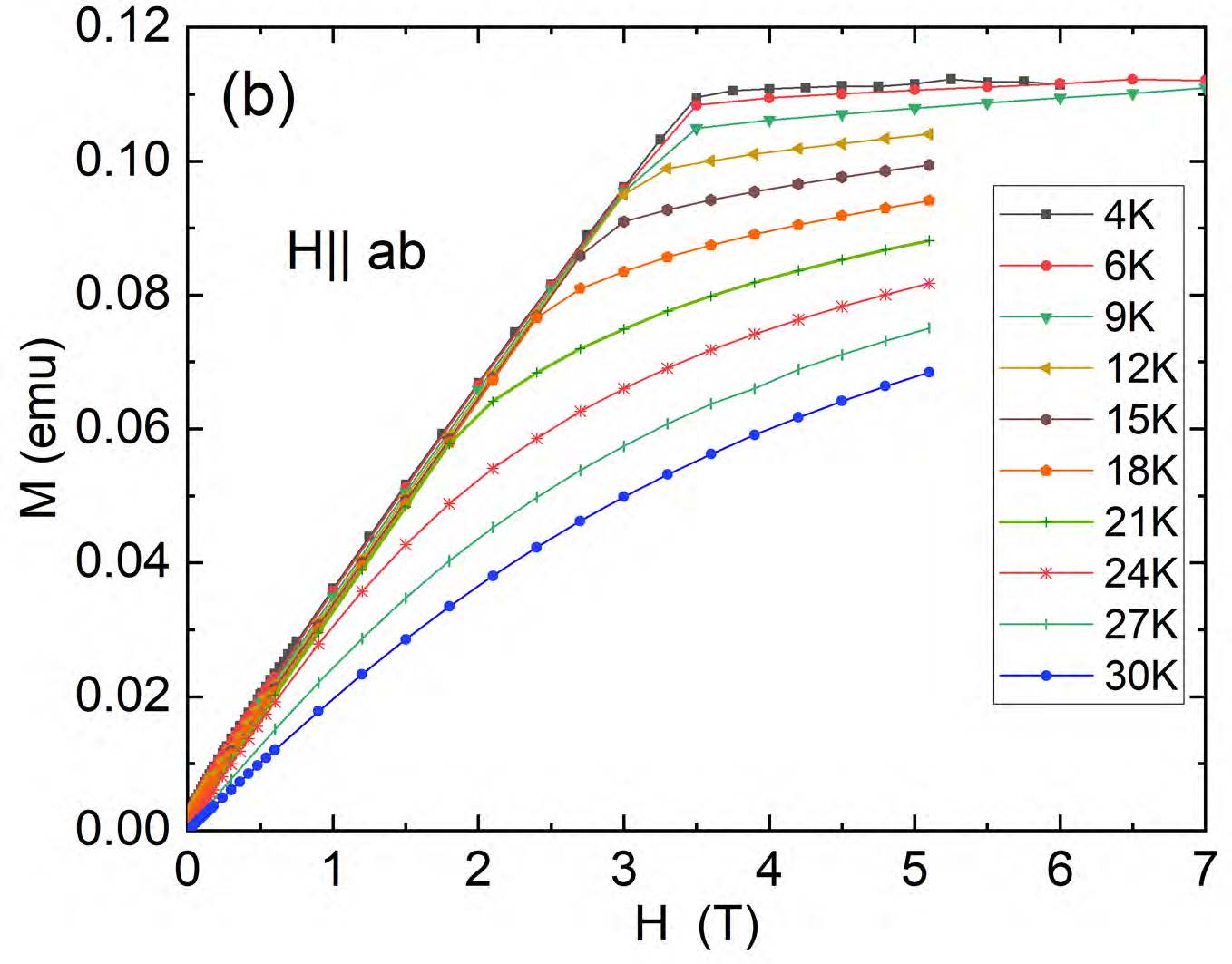}
	\caption{(a,b) DC-magnetization $M(H)$ for bulk crystal \#2  (with a mass of $1.8$\,mg)
		at various temperatures	for two field orientations.  Temperature values are indicated on the panels.} 
	\label{fig:M(HT)-main}
\end{figure}

Magnetization measurements were done using  MPMS-7 SQUID magnetometer,  with bulk samples, solely. 
Figure~\ref{fig:M(HT)-main} shows evolution of the DC-magnetization curves  $M(H_{ab},T)$ and   $M(H_s,T)$, measured for field orientations, in-plane $H\|ab$ and perpendicular to the plane $H\perp ab$ (or, equivalently, $H\|c$), respectively,  at various 
temperatures, from the antiferromagnetic state $T<T_N$  to paramagnetic state $T>T_N$.
The $M(H, T)$ curves  follow the conventional field dependence for the A-type easy-plane antiferromagnet 
\cite{macNeill_PRL_2021, gurevich-melkov_book}, fully consistent with our previous measurements on similar samples \cite{golov_JMMM_2022, PRL_tbp}, and also similar to those   reported  in other works for EuSn$_2$As$_2$ \cite{arguilla_InChemFront_2017, pakhira_PRB_2021, li_PRX_2019, chen_ChPhysLet_2020}.
The  magnetization  saturation corresponds to  the  full spin polarization; for our lowest temperatures this occurs at $H_s \approx 4.7$T when $H$ is parallel to the $c$-axis,  and at $H_s\approx 3.4$T when $H$ lies in the easy magnetization $ab$-plane.  
Extrapolated to $T=0$ the spin polarization fields equal  $H_s(0)=4.8$ and  $3.5$\,T, respectively

When  field is perpendicular to the $ab$-plane, the  magnetization curves $M(H\|c)$ at the lowest temperatures are precisely linear in the field range from zero and  up to  $H_s$  (Fig.~\ref{fig:M(HT)-main}a). In the field interval of 5-10\,T we didn't observe  a two-stage $M(H)$ saturation as that  reported in \cite{li_PRB_2021} above $\approx 4$\,T. For  field in the $ab$ plane, there is  a weak  nonlinearity of $M(H)$ in low fields less than $\approx 0.03$T, 
visible at the lowest temperatures in Fig.~\ref{fig:M(HT)-main}b. 
Similar nonlinearity  of $M(H)$  may be noticed also in the preceding  papers \cite{pakhira_PRB_2021, li_PRB_2021, chen_ChPhysLet_2020, li_PRX_2019}. 
We investigated the $M(H)$  nonlinearity   and   the $\chi(T)$ divergence in detail in Ref.~\cite{degtyarenko_tbp} and have shown that these effects are caused by planar defects in the crystal lattice which behave below $T\approx 24$K as weak ferromagnets embedded into the AFM matrix. 

\section{Magnetotransport results}
	\label{sec:magnetotransport}
Magnetotransport data were collected with PPMS-16 and CFMS-16 measuring systems in fields up to 16\,T, using 4-probe measuring set-up.

	\subsection{Hall resistance measurements}
	\label{sec:Hall}
\begin{figure}[h]
\includegraphics[width=\linewidth]{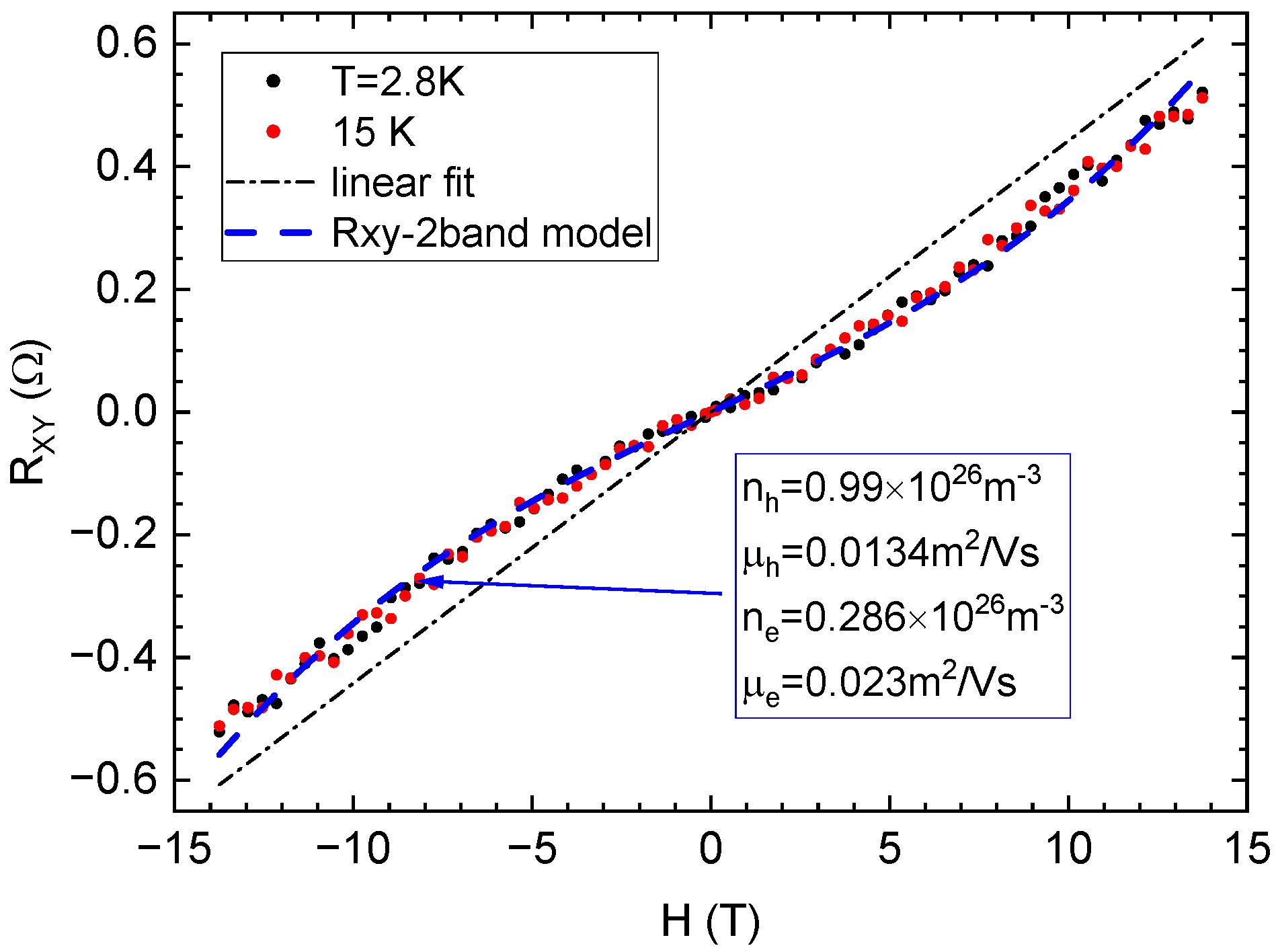}
\caption{ Hall resistance versus perpendicular magnetic field measured with 140nm-flake: data at two temperatures,  2.8 and 15K, 
and the best fit with Eq.~\ref{eq:(2)} (solid blue line). The fitting parameters ($n_h$, $n_e$, $\mu_h$, $\mu_e$) are given in Table 1. Dash-dotted  line  shows a single-band  fitting  at high fields.
		}
		\label{fig:hall}
	\end{figure}

Hall  voltage was measured only for the flakes, in fields directed perpendicular to the $ab$-plane. For Hall voltage measurements,  we used one of the two pairs of lithographically defined potential probes (see Fig.~\ref{fig:crystal_structure}c.
 The results shown in Fig.~\ref{fig:hall} demonstrate that the Hall resistance and, hence, carrier density,  
  almost don't vary with  temperature in the range from 2.8 to 15\,K. The  absence of $T$-dependence of the Hall effect is in line with the  diffusive (``metallic'') transport character and signals the negligible contribution of the  topological surface states (if any).  
	
	In   fields $H \sim \pm 5$T,   the Hall resistance $R_{xy}(H)$ changes  slope. 
	In higher fields $R_{xy}(H)$ tends to  become  linear, and from its  slope, $dR_{xy}/dH \approx 
4.4\times 10^{-2}$Ohm/T,  assuming the single band model
one could roughly estimate  the density value  $n \approx 
10.15\times 10^{26}$/m$^3$ (using the measured flake thickness $d=140$nm).
This density value is by a factor of 2  higher than that reported in Refs. ~\cite{chen_ChPhysLet_2020, pakhira_PRB_2021, li_PRB_2021}.  
However, as we show below, this estimate has little to do with reality, since the energy  
spectrum in the vicinity of $E_F$ (see Figs.~\ref{fig:ARPES}, \ref{fig:DFT_fbands}, and also Fig.~\ref{fig:band structure} of SM) 
 consists of several bands  (holes and electrons), whose  contributions to $R_{xy}$ significantly compensate each other. 
For simplicity, we consider only the two band case, where the charge transport is determined by four  parameters: $n_h, n_e$ and $\mu_h, \mu_e$ - the 
hole and electron densities and mobilities, respectively.

In order to solve  the four parameter problem, we turn to the  non-linear field dependence of the Hall voltage shown in Fig.~\ref{fig:hall}. The nonlinear $R_{xy}(H)$ dependence was noticed earlier and 	was attributed to the anomalous Hall effect in Refs.~\cite{yan_PRR_2022, li_PRB_2021}. 
We argue that for EuSn$_2$As$_2$  {\em the $R_{xy}(H)$ non-linearity, in the lowest approximation,  is a simple consequence of the two-band charge transport}.
 
Indeed,  the  energy spectrum (Figs.~\ref{fig:ARPES} and \ref{fig:DFT_fbands}) clearly shows the presence of the two major 	bands	at the Fermi level - the larger hole and the smaller electron pockets. Within the two-band concept, and considering, 
for simplicity, only the in-plane current direction $\mathbf{j}\|(ab)$, i.e. ignoring the case of $\mathbf{j}\|c$,
we arrive at the 2D  geometry of measurements. Thus, for $H\perp (ab)$,  $\mathbf{j} \in (ab)$ 	in the moderate field range  $\omega_{\rm c}\tau \equiv \mu H \leq 1$, by neglecting terms of the order  $(\omega_c\tau)^4$, 
 neglecting  electron-electron interaction corrections \cite{kuntsevich_PRB_2013}, and assuming isotropic scattering, 
the conductivity tensor  may be written as
	\beq
	\sigma_{jk} =\sum\limits_{i=1}^2 \frac{e_i n_i\mu_i}{1+(\mu_i H)^2}\begin{pmatrix}
		1& \pm\mu_i H\\
		\mp\mu_i H& 1
	\end{pmatrix}.
	\label{eq:(1)}
	\eeq 
Here,   $n_1\equiv n_h$, $n_2\equiv n_e$, $\mu_1\equiv \mu_h$, $\mu_2\equiv \mu_e$, and $e_i= \pm e$  - 
the elementary charge of  holes and electrons, respectively.
By inverting the conductivity tensor and keeping only terms up to  $(\omega_{\rm c}\tau)^3$ we obtain for the off-diagonal resistivity component: 
	\beq
\rho_{xy} = \frac{H}{e}\frac{(n_{h}\mu_{h}^{2} - n_{e}\mu_{e}^{2}) + (n_{h} - n_{e})(\mu_{h}\mu_{e} H)^{2}}{(n_{h}\mu_{h} + 
	n_{e}\mu_{e})^{2} + (n_{h} - n_{e})^{2} (\mu_{h}\mu_{e} H)^{2}}
	\label{eq:(2)}
	\eeq

In Fig.~\ref{fig:hall} the blue dashed curve shows fitting of the experimental   $R_{xy}(H)$ data with Eq.~\ref{eq:(2)} using the set of parameters given in  Table 1. As expected, the hole density is a factor of 3.5 larger than the electron density. 
Such relationship $n_h/n_e$ is approximately consistent with the partial sizes of the hole and electron bands in
the  energy spectrum (Fig.~\ref{fig:DFT_fbands}).   This set of parameters  provides also  the  zero field diagonal resistivity value 
  $\rho_{xx}^{\rm calc}=n_h\mu_h+n_e\mu_e=3.15\mu\Omega$m  that may be compared  with 
$\rho_{xx} = 3.85\pm 1\mu\Omega$m  measured for the bulk crystal  (Fig.~\ref{fig:R(T)}), and  $3.01\mu\Omega$m) for the flake.
Taking into account a 25\% uncertainty in the bulk crystal thickness, and  large contact pads area on the flake 
surface (Fig.~\ref{fig:crystal_structure}c), the agreement is satisfactory.

Given a good quantitative agreement between  Eq.~\ref{eq:(2)} and the Hall resistance data, we see no need to involve  additional assumptions (e.g. such as   ``coupling between the localized magnetic ordering and the conduction electronic states'' 	\cite{li_PRB_2021})  as a mechanism causing Hall resistance nonlinearity. 
	
	\begin{table}
		\caption{Numerical parameters obtained from fitting experimental data with the two-band model. 
			$n_{e,h}$ is in units of $[10^{20}/$cm$^3]$, $\mu_{e,h}$ in [cm$^2/$Vs].	}
\begin{tabular}{|c|c|c|c|c|}
			\hline \hline
			$n_h$             & $n_e$              & $\mu_h$  & $\mu_e$  & Data fitting: \\ \hline
            0.99 &    0.286 & 134 &  230 &   Hall: $R_{ab}(H\|c)$, flake 140nm\\ \hline
		\end{tabular}
	\end{table}

 For the above density value, and assuming 3D spin-degenerate system, we estimate $k_F^h=3.0\times 10^7$ cm$^{-1}$, $\lambda_F^h \approx 42$\AA, respectively. Thus, the  wave function for the holes extends 	at a distance much larger than the  spacing between two neighbouring Eu-planes, $\approx 6$\AA.  The electron wavelength $\lambda_F^e$ is even larger by a factor of $\sim 1.5$. The large extension of the carrier wave functions is an important requisite of our model of the negative magnetoresistance \cite{PRL_tbp}.
	
\subsection{Diagonal magnetoresistance}
\label{sec:magnetoresistance}

\begin{figure*}[ht]
	\includegraphics[width=180pt]{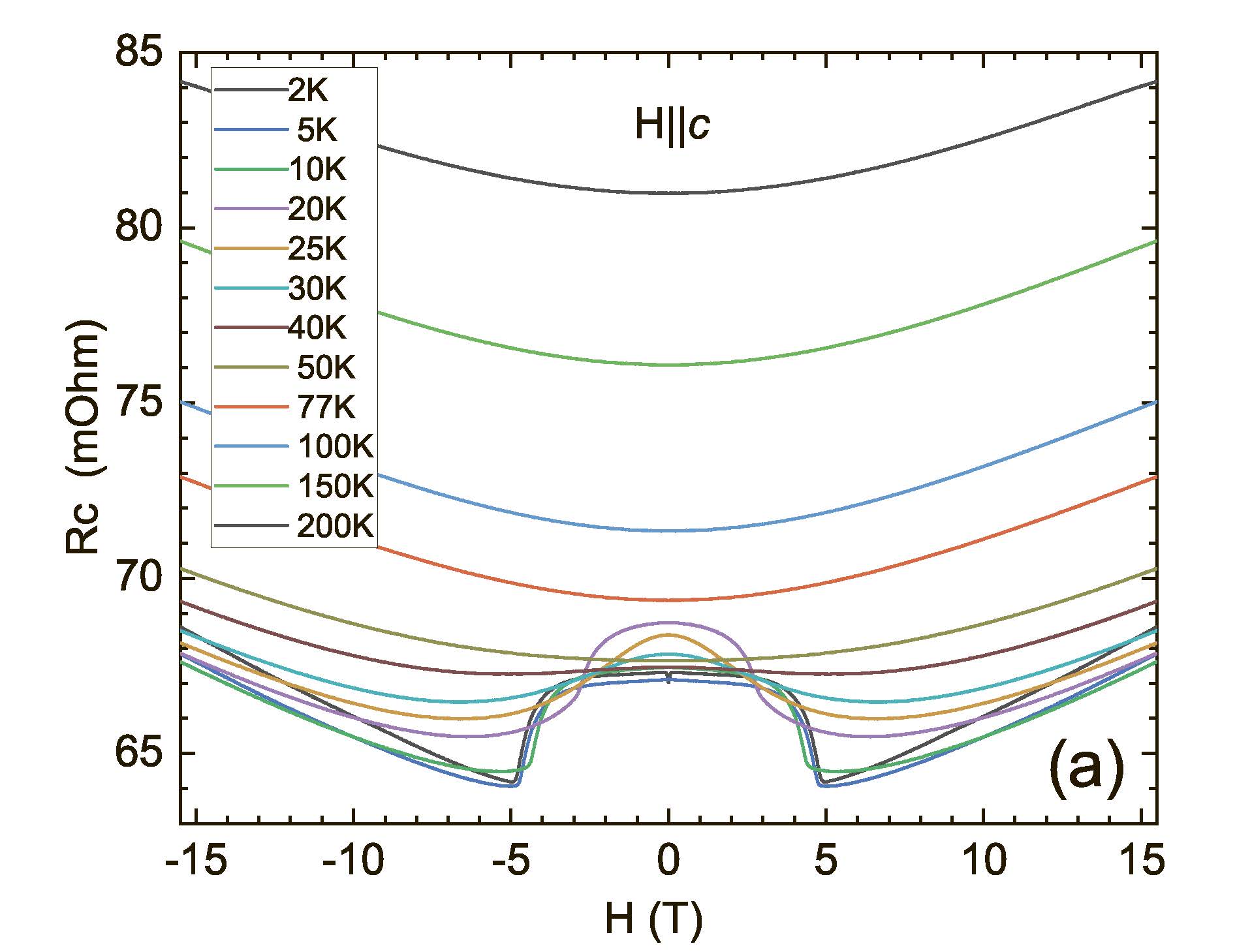}
	\includegraphics[width=180pt]{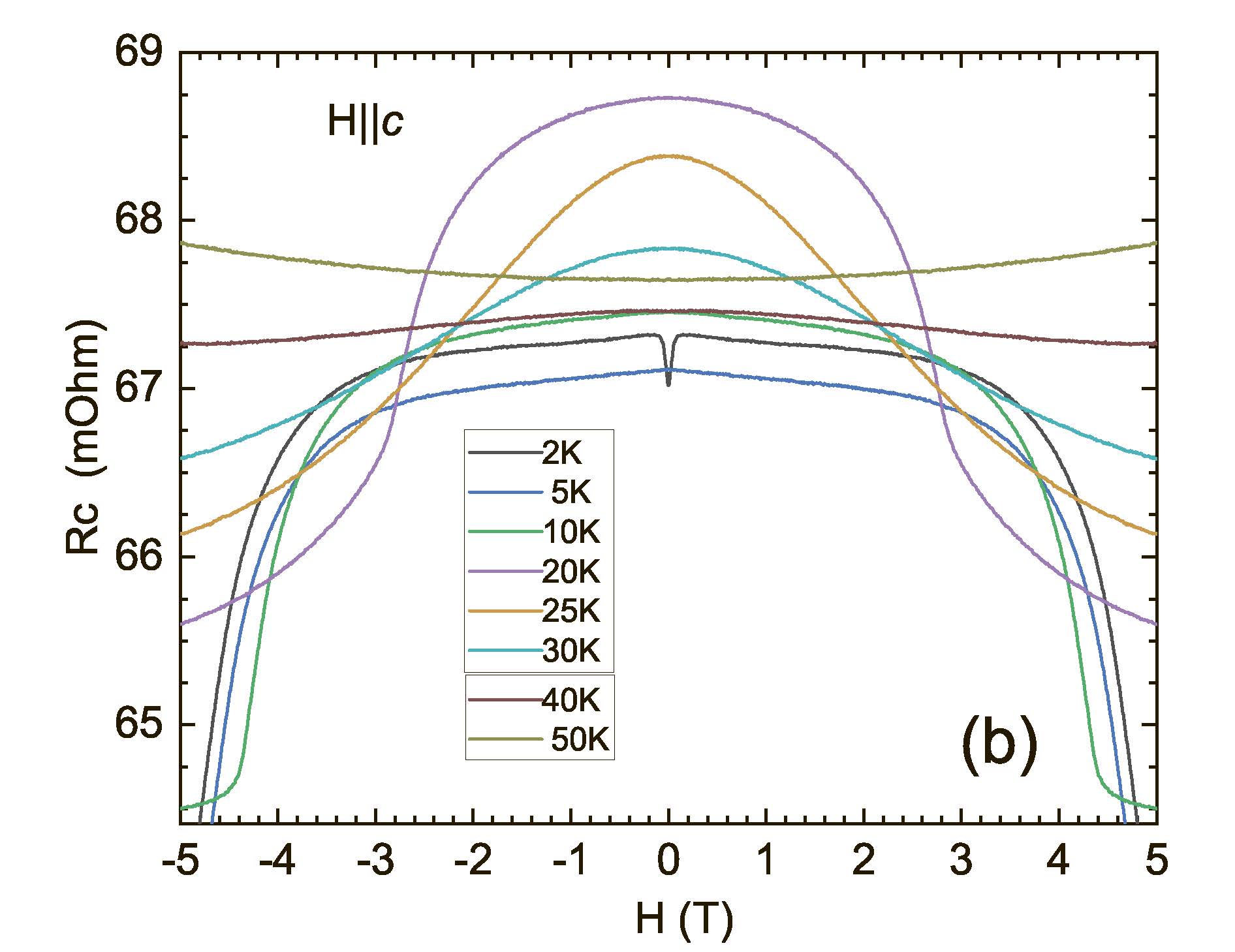}
	\includegraphics[width=180pt]{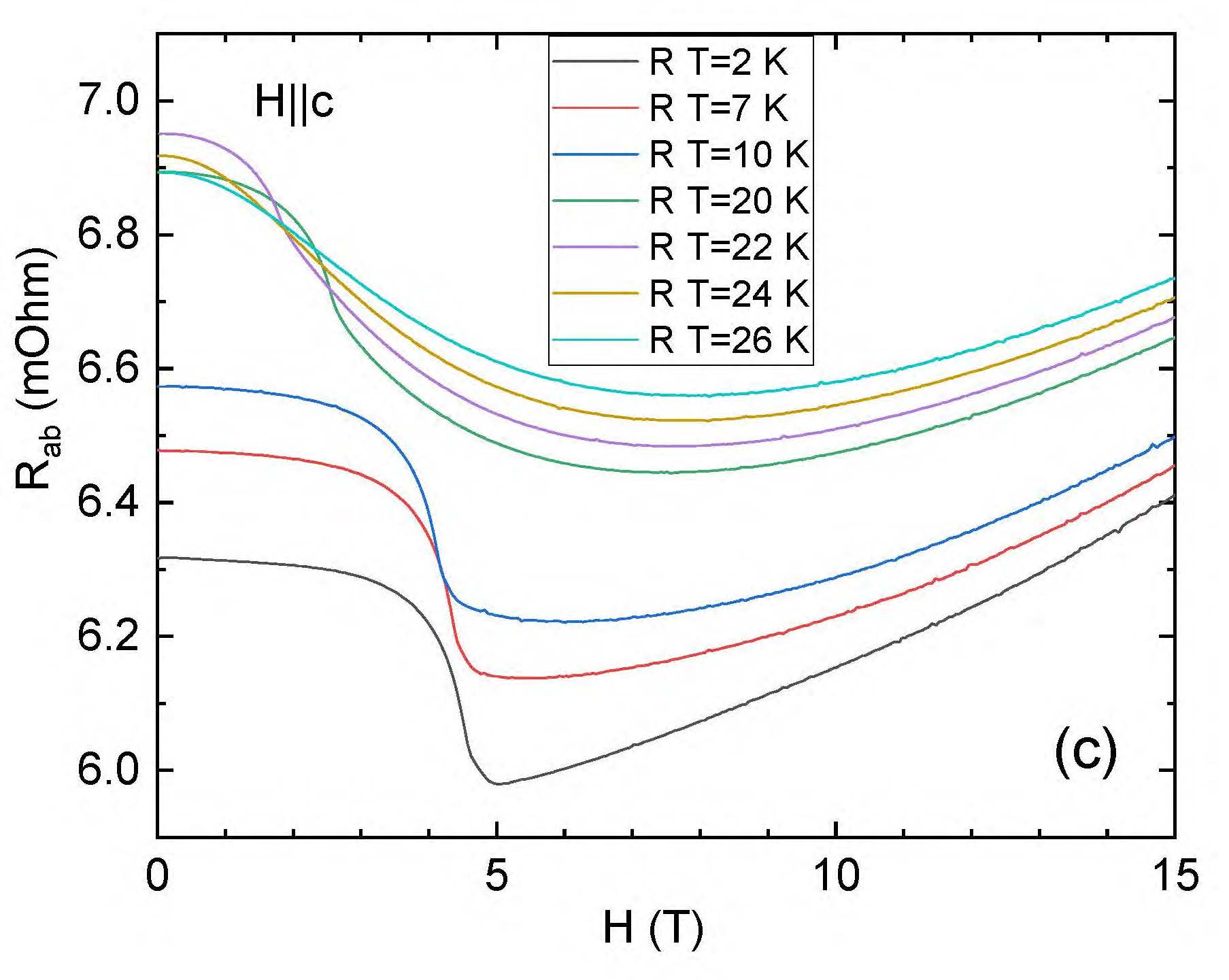}
	\includegraphics[width=180pt,height=142pt]{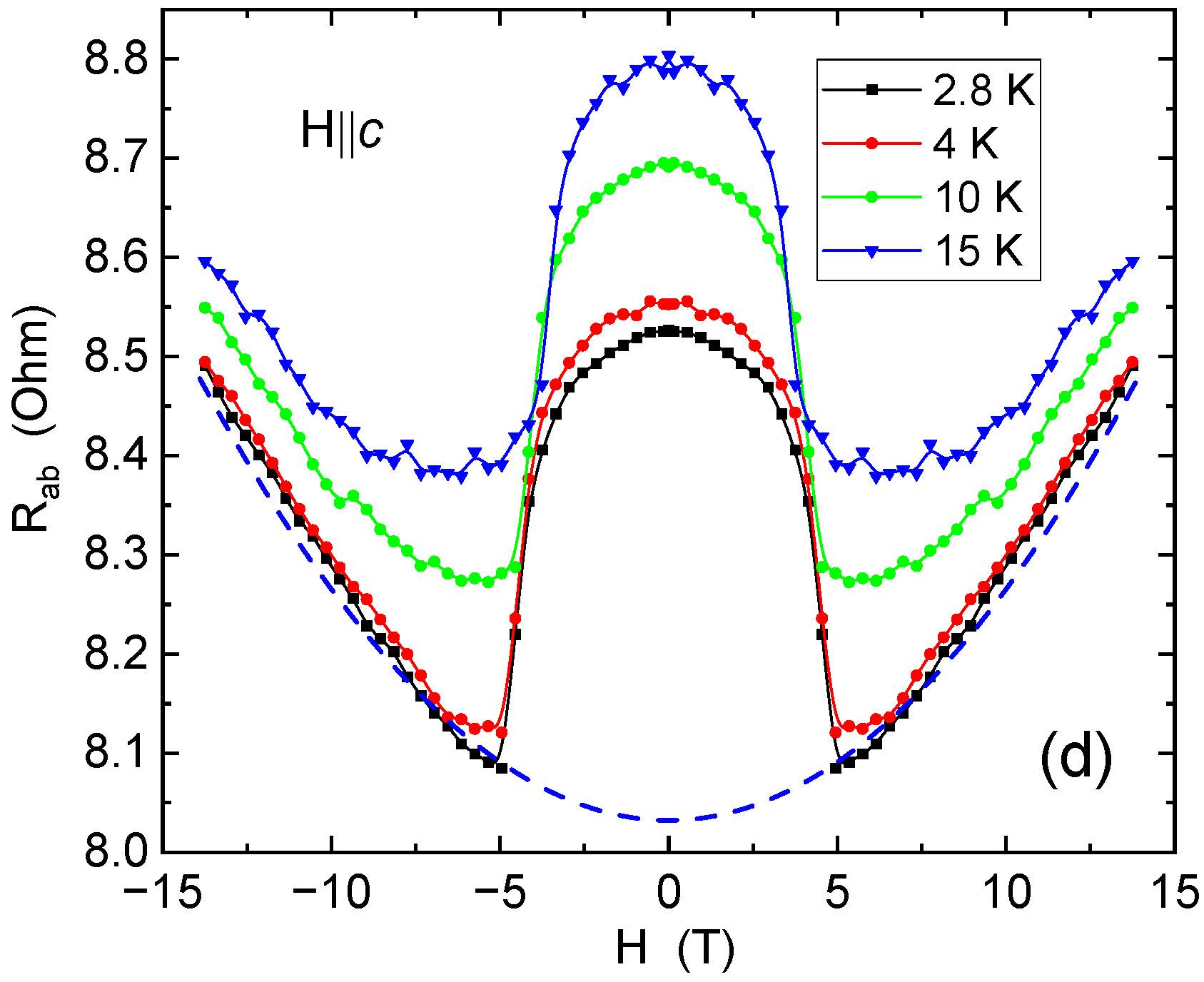}
	\caption{(a)-(c) Examples of the ``flattened''-type $R(H)$  dependences at various  temperatures for $H\|c$: 
		(a) $R_c(H\|c)$ for  bulk crystal \#2 at temperatures  from 2 to 200K, and in the wide field range; 
		(b) same, in the narrower field range for 7 temperature values;
		(c) $R_{ab}(H\|c)$ for the bulk crystal \#2 in the range 2  - 26K; 
		(d) $R_{ab}(H\|c)$ for the 140-nm thick flake at temperatures  
		from 2.8 to 15\,K. Blue dashed curve shows CPMR calculated using Eq.~\ref{eq:CPMR} with parameters from Table 1.
	}
	\label{fig:flattened_MR}
\end{figure*}

The overall evolution of the normal to the plane  $R_c(H,T)$ and  in-plane $R_{ab}(H,T)$ magnetoresistance with temperature and magnetic field    is shown in  Fig.~\ref{fig:flattened_MR}. In view of the data complexity  we discuss separately, positive and negative magnetoresistance, 
trivial and less-trivial magnetoresistance geometry. In all configurations, the most striking is  the hump of the negative magnetoresistance  (NMR) in the AFM state  that is the focus of our study.  In the AFM state, at $H<H_{s}$, as temperature increases, the NMR hump quickly gets narrower
whereas its  magnitude changes weaker (see Figs.~\ref{fig:flattened_MR}). Close to and above $T_N$, it starts decreasing in magnitude, and widens in field \cite{maltsev_tbp}. Finally, the NMR hump gradually vanishes  and transforms to the positive parabolic magnetoresistance.

\subsubsection{Negative and positive magnetoresistance}

\begin{figure}[h]
	\includegraphics[width=220pt]{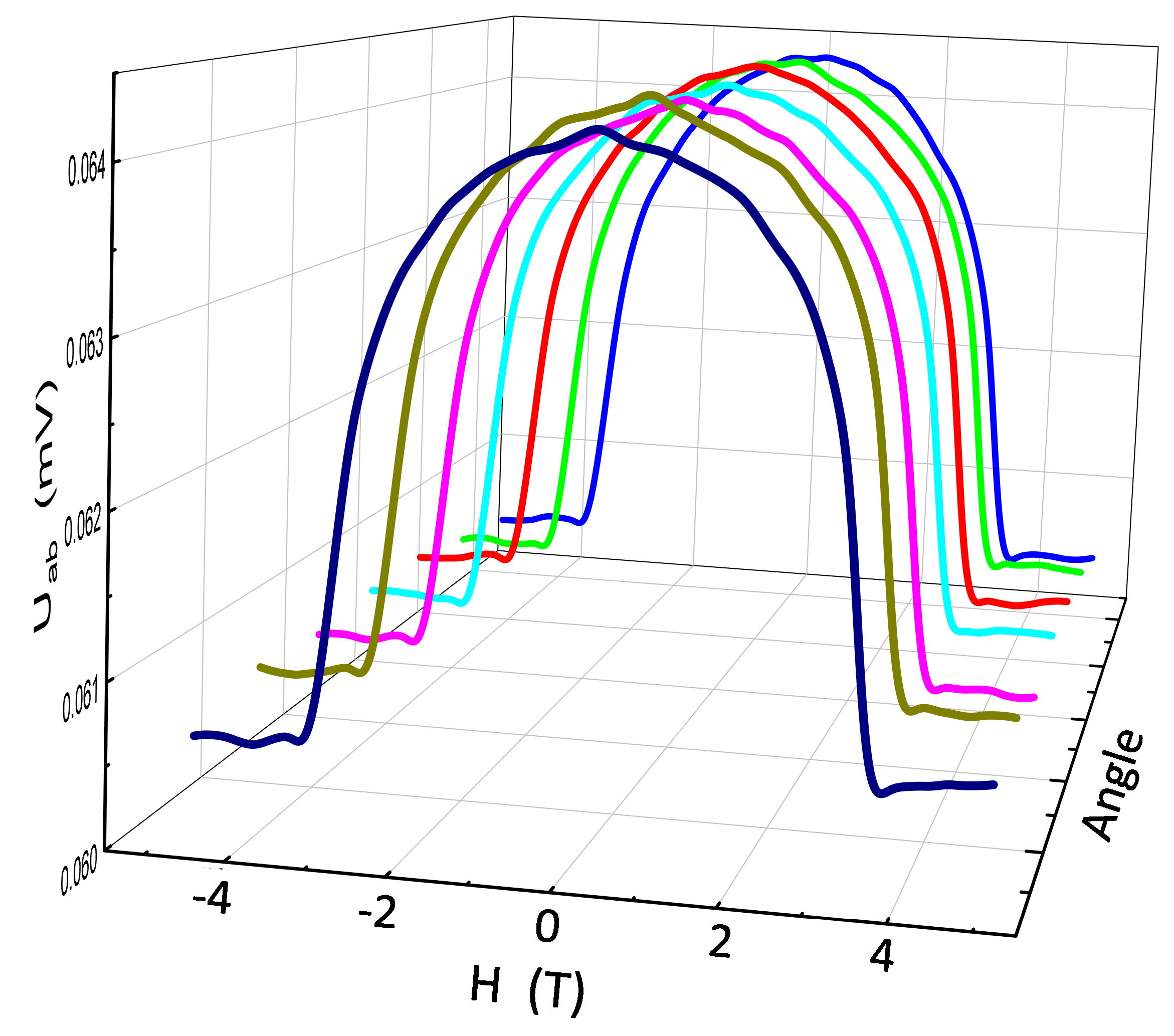}
	\caption{Magnetoresistance in the easy-plane $R_{ab}(H\|ab)$ measured for different angles (from 0 to $90^\circ$, with $15^\circ$ step) between the current and field directions. $z$-axis - diagonal voltage drop with 100$\mu$A bias current.  
		}
	\label{fig:isotropy}
\end{figure}

Figures~\ref{fig:flattened_MR} and \ref{fig:isotropy} show the $R(H)$ data for several field directions relative to the crystal axis, 
and for various angles between the bias current and applied field. 
We first describe the data on a qualitative level. 

The NMR hump has several remarkable features: (i) it is similar for  current applied  in the $ab$-plane and in perpendicular direction  (see  Figs.~\ref{fig:flattened_MR}, \ref{fig:CPMR}, and Appendix \ref{app:NMR anatomy}), (ii) it is independent of the  in-plane current direction relative to magnetic field (see Fig.~\ref{fig:isotropy}), (iii) it amounts up to about 6\% for all studied bulk samples and flakes, and (iv) it sharply terminates at the field of complete spin polarization $H_{s}$ (cf. Figs.~\ref{fig:M(HT)-main} and  \ref{fig:CPMR}).

\underline{NIMR isotropy in the $ab$-plane.}	Regarding this class of compounds, in literature an exciting axion insulator state was discussed, whose ``smoking gun'' evidence would be a sharp anisotropy of the magnetoresistance relative to the mutual orientation of the magnetic field and current.
	In order to test this possibility we measured the  magnetoresistance anisotropy 
	by rotating the sample with in-plane bias current in magnetic field aligned precisely in the $ab$-plane.
	Figure \ref{fig:isotropy} shows that   the magnetoresistance is  fully isotropic, i.e. independent of the angle between field and current directions, thus confirming (i) EuSn$_2$As$_2$ to be topologically trivial, (ii) the magnetoresistance is irrelevant to the chiral anomaly, contrary to the suggestion in  Ref.~\cite{li_PRX_2019}, and (iii)  the $ab$ plane to be the easy-magnetization plane with 	equivalent $a$   and $b$ axis.	

\underline{Sample- and configuration dependent diversity of the} \underline{NIMR shape.} At first sight, the negative magnetoresistance  looks 
similar for all  studied  samples. However, upon  closer examination, we find  two qualitatively different types of the functional  $R(H)$ dependence:
 the NIMR hump is almost {\em parabolic} for $R_{ab}(H_{ab})$ geometry in all studied samples (Fig.~\ref{fig:isotropy}), 
whereas  for  $R_{ab}(H_{c})$  in some samples NIMR  has a more {\em flattened} shape (Fig.~\ref{fig:flattened_MR})a,b,c.
Correspondingly, we classify  the results with the studied  EuSn$_2$As$_2$ samples 
 in two groups displaying either the ``parabolic''  or the ``flattened'' type $R(H)$ dependence (for more detail, see   Appendix \ref{app:NMR anatomy}). 
Below we argue that minor variations of the NIMR shape (sample-dependent, and the field-against-current orientation dependent)  come partly from the contribution of classical positive MR in a two-band model with different electron and hole mobilities.

At low temperatures $T \ll T_N$,  as field increases and exceeds $H_{s}$,   
the NIMR hump  changes sharply to the smooth positive parabolic rise.  
We highlight that the sharp transitions between the negative  and positive magnetoresistance 
(i.e. the $R(H)$ minima) coincide with the sharp magnetization saturation at $H=H_{\rm s}$ for the respective field direction (cf. Figs.~\ref{fig:M(HT)-main} and \ref{fig:CPMR}a). 
This is consistent also with previous studies \cite{arguilla_InChemFront_2017, chen_ChPhysLet_2020, li_PRB_2021, sun_SciChin_2021, gui_ASCCentrSci_2019}. 

\subsubsection{Classical positive magnetoresistance  at $H>H_s$.}
	
\begin{figure}[ht]
	\includegraphics[width=200pt]{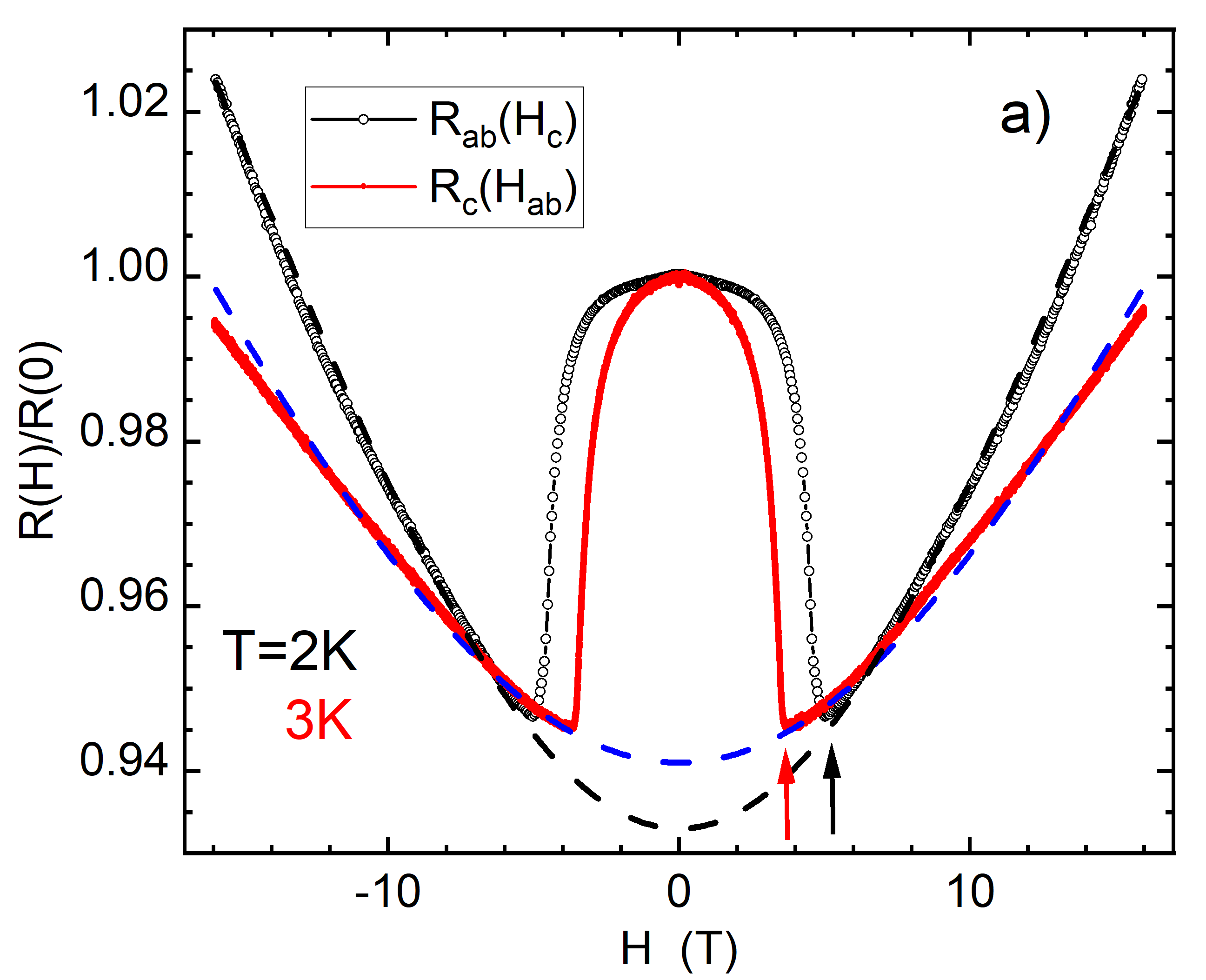}	
	\includegraphics[width=200pt]{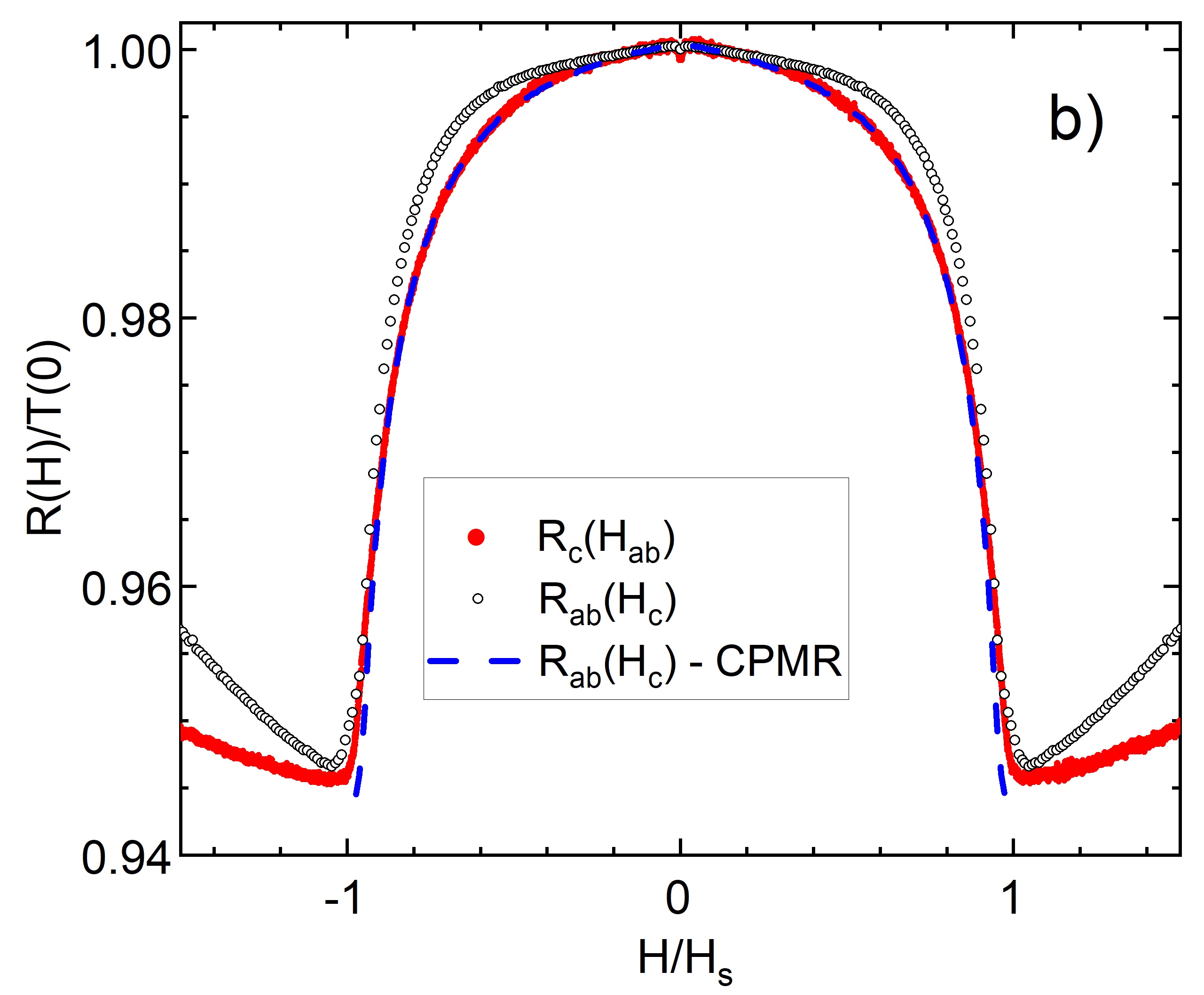}	
	\caption{(a) Normalized magnetoresitance for the  bulk crystal in various geometry: 
red  dots - 	$R_c(H\|{ab})$, black open circles - $R_{ab}(H\|c)$, dashed blue and black lines - the CPMR fitting  with a simplified 
parabolic $(\mu H)^2$	 dependences. 	Vertical arrows indicate $H_s$ for two field orientations.
(b) The same normalized $R_{c}(H_{ab})$ (open circles) and $R_{ab}(H_c)$ (red dots) dependences vs normalized field $H/H_s$. 
The blue dashed curve shows 	the latter dependence from which the fitting  PMR parabolic curve  is subtracted.
}
\label{fig:CPMR}
\end{figure}

Above the full spin polarization field $H_s$, the positive magnetoresistance (PMR) is observed in all geometries when the bias 
current  $\mathbf{j}$ is perpendicular to the field  (Figs.~\ref{fig:flattened_MR}c,d and \ref{fig:CPMR})a.  PMR is essentially weaker 
 when $\mathbf{j}$  and $\mathbf{H}$  lie  in the same $ab$ plane (see Fig.~\ref{fig:isotropy}). These facts indicate  that the 
 positive magnetoresistance has a classical orbital character, and we  further call it ``classical PMR''(CPMR).

At first sight, CPMR is reminiscent of the simple text-book dependence $\delta R(H)/R(0) = (\omega_c\tau)^2 \equiv (\mu H)^2$. 
However,  for the closed isotropic Fermi surface, within the single band transport, and neglecting  electron-electron interaction corrections \cite{kuntsevich_PRB_2013}, Eq.~(\ref{eq:(1)})  for electron-impurity scattering suggests $\rho_{xx}$ to be strictly independent of field. 
In order to make a more rigorous comparison  of CPMR  with theory, we invert again the conductivity tensor for the two-band system, 
Eq.~(\ref{eq:(1)}), and obtain:
\bea
\delta\rho_{xx}(H) &=& \rho_{xx}(H) - \rho_{xx}(0)\nonumber\\
\rho_{xx}(0) &=& \frac{1}{e(\mu_e n_e +\mu_h n_h)} \nonumber \\
\rho_{xx}(H) &=& \frac{1}{e}\frac{H^2(\mu_h \mu_e)(n_h\mu_e +  n_e \mu_h)+ n_h\mu_h+n_e \mu_e}{H^2(\mu_h \mu_e)^2(n_h-n_e)^2 +(n_e \mu_e + n_h\mu_h)^2}\nonumber
\\
\label{eq:CPMR}
\eea

The dashed curve in  Fig.~\ref{fig:flattened_MR}\,d shows that  the CPMR data $R(H)/R(0)$ for $H>H_s$ are well fitted by Eq.~(\ref{eq:CPMR}) with parameters listed in Table 1. In this comparison, we used a single variable prefactor: scaled the calculated $\rho(H)$ by about 10\% to fit the measured $\rho_{ab}(H=0)$ value because of a large uncertainty in the sample  sizes and intercontact distance (as was mentioned above) which affect the measured absolute resistivity value. The fitting curve described by Eq.~(\ref{eq:CPMR}) in low fields is parabolic,  but becomes less steep as field increases.
We note that  Eq.~(\ref{eq:CPMR}) is not applicable for transport across the layers,  $\mathbf{j} \|c$,  since  in the quasi two-dimensional systems it has a more complicated character, as  discussed below.

 To conclude this section: the two-band transport model of impurity scattering explains rather well the non-linear Hall resistance (at least, at low temperatures),   and the positive magnetoresistance data for in-plane transport (at $H>H_s$, when CPMR is not masked by NIMR). The same set of 
 parameters (Table 1) provides reasonable fitting of $R_{xy}(H)$, $\rho_{xx}(H=0)$, and  $\delta R_{xx}(H)$ data for  different samples  and for the exfoliated flake. This suggests that no other mechanisms are required  for describing, in the first approximation, the  in-plane transport behavior.

\underline{Unconventional charge transport in $c$-direction}, across the layers.
The above text-book formula (\ref{eq:CPMR}) derived for closed electron trajectories and for isotropic scattering is not applicable to the interlayer magnetotransport (i.e. $R_c$) in a field $H\| ab$. 
 Indeed, if one  applies formally the same Eq.~(\ref{eq:CPMR})  for CPMR in the $R_{ab}(H_c)$ and $R_{c}(H_{ab})$ configurations, the extracted values of an effective mobility $\mu^{\rm eff}$ would differ by 0.5\% only. This is clearly seen in Fig.~\ref{fig:CPMR}b, where the prefactors of the parabolic $(\mu H)^2$ dependence for $R_c(H)$ and $R_{ab}(H)$ differ  by 0.015\%. Such minor difference of mobilities (along and across the layers) is in a striking contrast with the factor of $130$ anisotropy in zero-field resistivity measured along and across the $ab$ plane (see Fig.~\ref{fig:R(T)}).
 
 For the $R_{c}(H_{ab})$ configuration one expects a crossover from a parabolic (in low field) to linear (in high field) MR \cite{MosesMcKenzie1999,Schofield2000,GrigPRB2014} and can approximately use Eq. (34) of Ref. \cite{Schofield2000}.
Even more surprising may look  Fig.~\ref{fig:flattened_MR}a that shows a non-zero CPMR  for bias current along the field, $\mathbf{j} \|\mathbf{ H} \|c$. Similar longitudinal interlayer magnetoresistance was also observed in many other quasi-2D metals being their general feature \cite{Wosnitza2007,Kartsovnik2009,Kang2009,GrigPRB2012}; it was explained and described a decade ago \cite{WIPRB2011,GrigPRB2012,GrigPRB2013}. These two facts indicate  that charge transport across the layers in crystals with highly anisotropic conduction 
has a more complex character, where the carriers, moving on average in the $c$ direction, propagate  also along the highly conducting $ab$ planes and experience therefore action of the perpendicular magnetic field. This interesting issue is however beyond the scope of the current paper.

\subsubsection{Negative magnetoresistance} 
As we mentioned above, there are two qualitatively different types of the  NIMR field dependence.
Data of the 1st type (see Figs.~\ref{fig:isotropy}, \ref{fig:CPMR}) is almost parabolic $\delta R(H)/R(0)\propto -a H^2$.
The 2nd type NIMR (call it ``flattened'') is more flat in low fields (Figs.~\ref{fig:flattened_MR})\,a,b,c. 

The  approximately parabolic shape of NIMR  in the 1st type samples is  similar to that observed  earlier on EuSn$_2$As$_2$ in Refs.~ \cite{arguilla_InChemFront_2017, li_PRB_2021}. On the other hand, the flattened type NIMR  for EuSn$_2$As$_2$ may be seen in Ref.~\cite{chen_ChPhysLet_2020}. Comparing our results on EuSn$_2$As$_2$ with those measured  on vdW AFM sister  materials EuSn$_2$P$_2$ \cite{gui_ASCCentrSci_2019} and  EuFe$_2$As$_2$ \cite{jiang_NJP_2009, sanchez_PRB_2021}, we also find  qualitative similarity between NIMR in all these materials. 

More detailed  measurements \cite{maltsev_tbp} of magnetoresistance performed with various exfoliated flakes confirmed that NIMR is independent of the  sample thickness. This evidences  that NIMR does not originate from  large-scale macroscopic lattice defects (such as misfit dislocations, domains, misoriented grains), and on the sample $ab$-plane bending. Indeed,   these imperfections would be  different for the bulk crystal and for exfoliated  flakes. More detail on the XRD structural sample investigation is given in Supplemental materials \cite{SM}.

\underline{Influence of  the geometry of mesurements on NIMR.}
One of the  reasons for the  non-universality of NIMR is the obvious dependence of CPMR  on the sample mobility $\mu$ that is anisotropic in layered crystals. The flattening, therefore, depends on the charge transport and field directions: 
the flattening is stronger for  $R_{ab}(H_c)$,  weaker for $R_c(H_{ab})$ - see Fig.~\ref{fig:CPMR},  and is the weakest for $R_{ab}(H_{ab})$ - see Fig.~\ref{fig:isotropy}, because CPMR is missing for $\mathbf{j}\|\mathbf{H}\| ab$. 

In Fig.~\ref{fig:CPMR}a we compare magnetoresistance measured on the same sample in different configurations:  $R_{ab}(H\|c)$  (showing  the most strong flattening)  and $R_c(H\|ab)$. After subtracting CPMR from the original data, and normalizing magnetic field $H/H_s$ (note, $H_s^{(ab)}$ and $H_s^{c}$ are different- see Fig.~\ref{fig:M(HT)-main}) the two resulting curves in Fig.~\ref{fig:CPMR}b almost coincide; their proximity 
demonstrates that the configuration-dependent part of the flattening is mainly caused by a different classical parabolic CPMR (Eq.~\ref{eq:CPMR}). In other words, NIMR develops on top of the  background CPMR. In the configuration  $R_{ab}(H\|ab)$, where  the positive MR is missing,  NIMR looks closest to parabolic. 

In agreement with the above arguments, we conclude   that the parabolic-type NIMR, 
$\delta R(H)\propto - [M(H)]^2 \propto -H^2$,  to be {\em the generic type dependence} for the layered AFM crystals. 
In Appendix \ref{app:NMR anatomy} we present scaling analysis of the two types of NIMR and provide  additional arguments in favor of this conclusion. 
We also discuss  potential mechanisms for a sample-dependent departure of  NIMR from the generic parabolic dependence. 
Searching for the potential origin of different types of NIMR, we  performed detailed XRD, and transmission electron microscopy (TEM) examination of our samples, and  precise magnetization measurements in low-field. For testing potential non-trivial topological order, suggested in \cite{li_PRX_2019}, we also performed ARPES measurements supplemented with DFT band structure calculations. 
This information is presented in Appendices to our paper.

\section{Theoretical description of Negative isotropic magnetoresistance in layered conductors with AFM order}
\label{sec:theory} Searching for an explanation for the observed NIMR, we analyzed the known MR mechanisms which might be potentially relevant to the
studied case and found none to be able to explain the observed NIMR (see Ref.~\cite{PRL_tbp} and also Appendix \ref{app:mechanisms}).

As we demonstrated in Fig.~\ref{fig:NMR-deviation_parabolic}, NIMR is universal when plotted 
versus normalized magnetic field $\delta R(H/H_s)$; in other words, $\delta \rho(H)=  \delta \rho[M(H)]$.
Furthermore, $\delta \rho(M)$ is fully {\em isotropic} with respect to the field and current direction. This is in a sharp contrast with such well known effects as giant magnetoresistance 
(GMR), where the negative magnetoresistance is {\em anisotropic} (for more detail, see Appendix \ref{app:mechanisms}).
The observed isotropy of NIMR suggests the conventional short-range scattering by  point-type defects \cite{Abrik}.  
From the fact that NIMR develops in the AFM state solely and is tied to magnetization field dependence,  
it follows that the electron scattering rate diminishes as spin polarization of Eu-atom layers changes from AFM- to FM-type.
In contrast, ordinary scattering on magnetic impurities and magnons in the AFM metals and in AFM structures results in the positive MR 
(for more detail, see Appendix \ref{app:mechanisms}).

\subsection{Theoretical model and results}

Very recently, the novel so-called exchange splitting mechanism of negative
magnetoresistance in layered antiferromagnetic semimetals has been proposed 
\cite{PRL_tbp} to explain this effect in EuSn$_{2}$As$_{2}$. Qualitatively,
this mechanism can be understood as follows. 
\begin{figure}
	\begin{center}
		\includegraphics[width=220pt]{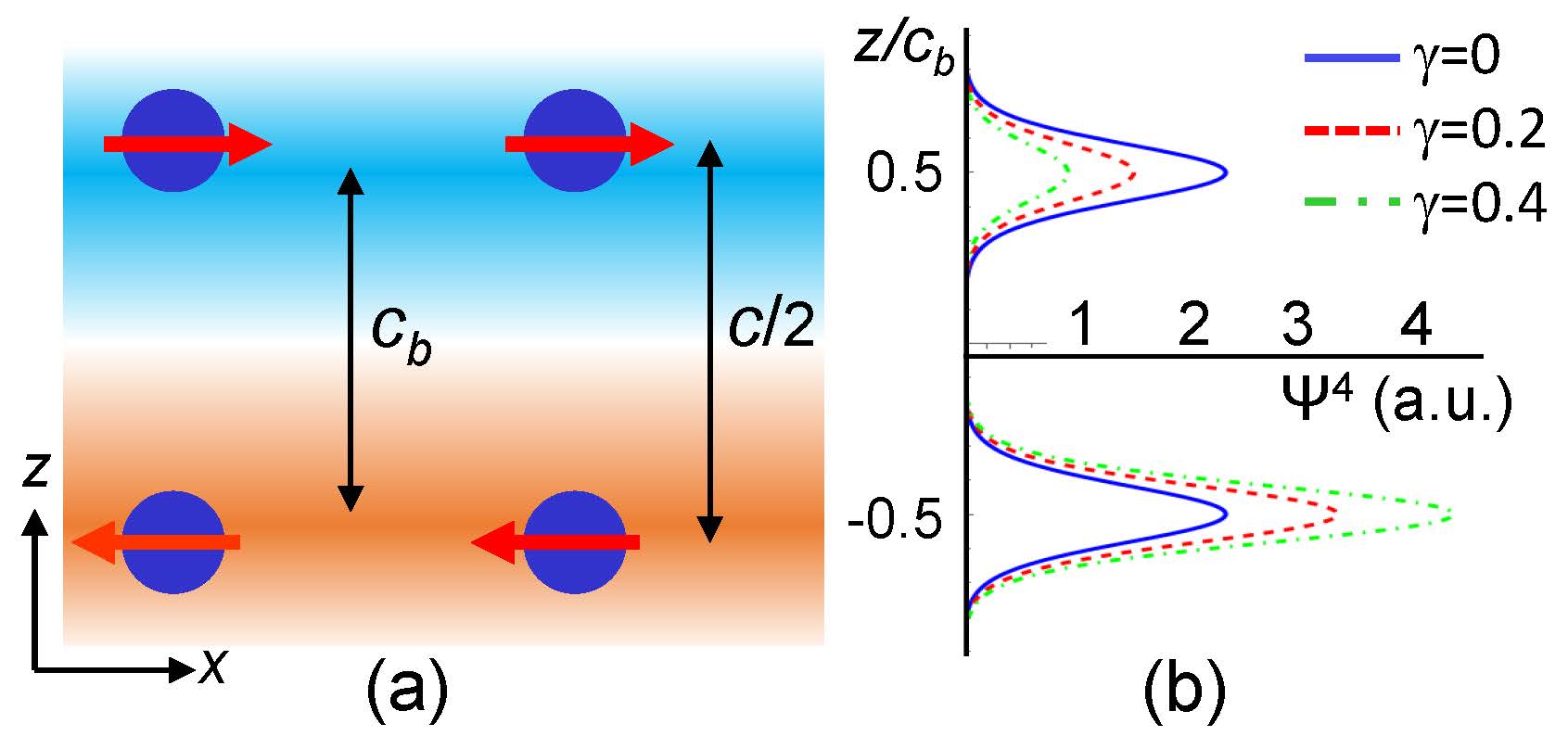}
		\caption{Schematic illustration of the difference between spin-up and spin-down
		electron density distribution along the $z$-axis, represented by color intensity (orange -- positive, blue -- negative), in the half-cell of the $c/2$-size. $c_b\approx c/2$ denotes the distance between two maxima of electron spin density. Dark blue circles show the Eu-atoms, and arrows - their magnetization direction in the AFM state.}
	\end{center}
			\label{fig:WF-cartoon}
\end{figure}

The conduction electrons originate from double layers of SnAs confined 
between the layers of Eu atoms (see Figs.~\ref{fig:crystal_structure}a and \ref{fig:WF-cartoon}).
The A-type AFM ordering of Eu spins doubles the lattice constant along the interlayer $z$-direction and 
violates the binary $\hat{Z}_{2}$ symmetry between two adjacent Eu layers for
each spin component $\sigma $ of conducting electrons. 
This $\hat{Z}_{2}$ symmetry violation changes the $z$-dependence of the electron wave function 
$\Psi _{\sigma }\left( z\right) $ in bilayers containing two Eu layers 
of opposite spin polarization or AFM magnetization $\boldsymbol{L}_{AFM}$.
The schematic picture of the electron wave functions is shown in Fig.~\ref{fig:WF-cartoon}.
 In the paramagnetic state the electron wave function $\Psi _{\sigma }\left( z\right) $ is symmetric to the
permutation $1\leftrightarrow 2$ of two layers within the bilayer for each
spin component $\sigma $. The AFM order reduces the electron energy of
exchange interaction with Eu magnetic moment on one of these two layers and
increases this energy on the other. It squeezes the electron wave function $\Psi
_{\sigma }\left( z\right) $ toward the layer of lower energy. The electron
scattering rate by short-range impurities or other crystal defects in the
Born approximation is proportional to the fourth power of electron wave
function \cite{Abrik}  
\begin{equation}
	1/\tau \propto \int \left\vert \Psi _{\sigma }\left( z\right) \right\vert
	^{4}dz , \label{ScRate}
\end{equation}%
because it contains the square of matrix element $T_{\boldsymbol{n}^{\prime }\boldsymbol{n}}^{(i)}=\int d^{3}\boldsymbol{r}
\Psi _{\boldsymbol{n}^{\prime }}^{\ast }\left( \boldsymbol{r}\right)
V\left( \boldsymbol{r}\right) \Psi _{\boldsymbol{n}}\left( \boldsymbol{r}
\right) $ of electron scattering by disorder potential $V\left( \boldsymbol{r}\right) $ between the quantum states $\boldsymbol{n}$ and $\boldsymbol{n}^{\prime }$.
The compression of electron wave function $\Psi_{\sigma }\left( z\right) $ induced by 
bilayer $\hat{Z}_{2}$ symmetry violation increases the integral in Eq. (\ref{ScRate}), proportional to the electron scattering rate and metallic resistivity. Hence, the electric resistivity is larger in the AFM state than in paramagnetic. This enhancement requires a short correlation length of the scattering potential $V\left( \boldsymbol{r}\right) $, shorter than the lattice constant in $z$-direction doubled by AFM order, which is typical for impurities and other point crystal defects responsible for residual resistivity. 

The negative magnetoresistance in the AFM state appears because of the field dependence of AFM order parameter $\boldsymbol{L}_{AFM}$. 
An external magnetic field $\boldsymbol{H}$ tilts the spins of neighboring Eu
layers  towards $\boldsymbol{H}$,
giving the field-induced magnetization $\boldsymbol{L}_{\|}=\chi \boldsymbol{H}$ along the magnetic
field  $\boldsymbol{L}_{AFM}(0)$. Hence, it reduces the antiferromagnetic vector $\boldsymbol{L}_{AFM}$ approximately  according to the	classical relation  
	\begin{equation}
		L_{AFM}\left( H\right) \approx \sqrt{L_{AFM}^{2}\left( 0\right) -\left(\chi
			H\right) ^{2}}.  
		\label{M}
	\end{equation}

Taking the magnetic susceptibility in AFM state 	$\chi(H) 	\approx const$, 
	in agreement 	with the textbook behavior \cite{kittel}  and with experimental data for EuSn$_2$As$_2$ 
	in our Fig. \ref{fig:M(HT)-main} and in other papers,  we obtain:\\
	\begin{equation}
		L_{AFM}\left( H\right) \approx L_{AFM}\left( 0\right) \sqrt{1 -\left(
			H/H_s\right) ^{2}},  \label{M1}.
	\end{equation}
Here $H_s$ is the field of complete spin polarization, $H_s\approx 4$T  in EuSn$_{2}$As$_{2}$.
	Since the exchange term in electron energy $\Delta E_{ex}\propto L_{AFM}$,  
	\begin{equation}
		\Delta E_{ex}(H)\approx \Delta E_{ex}(0) L_{AFM}(H)/L_{AFM}(0) , 
		\label{EexB}
	\end{equation}
the bilayer wave-function asymmetry and spatial confinement
also decrease quadratically in a low 	magnetic field $H<H_s$. 
This leads to a parabolic negative 	magnetoresistance  almost up to the full spin polarization point $H_s$.  
At $H\to H_s$ the enhanced critical fluctuations are expected to make the second-order phase transition sharper and 
	lead to a faster decrease of $L_{AFM}\left( H\right) $ than in the mean-field approximation described by Eq. (\ref{M1}).  

The distinct feature of this mechanism of negative magnetoresistance is its almost complete isotropy to the magnetic field and electric current directions. The minor anisotropy to the field directions appears because of the anisotropic field dependence of AFM order parameter $L_{AFM}\left( \boldsymbol{H}\right) $ due to magnetic anisotropy and the spin-flop transition. As shown in Ref. \cite{PRL_tbp}, the relative increase of resistivity due to the AFM ordering and NIMR effect is 

\begin{equation}
	\frac{\delta \rho (H)}{\rho (0)}
	\approx \frac{\gamma^{2}}{1+2\gamma^{2}} ,  
	\label{drhoR}
\end{equation}
where the field-dependent ratio 
\begin{equation}
\gamma =\gamma (H) =\frac{\Delta E_{ex}(H)}{2t_{0}}  \approx \frac{\Delta E_{ex}(0)}{2t_{0}} 
	\sqrt{1 -\frac{H^2}{H^2_s} }
	\label{gamma}
\end{equation}%
of the exchange splitting energy $\Delta E_{ex}$ of conducting electrons to their
hopping amplitude $t_{0}$ between the opposite AFM sublattices describes the degree of $\hat{Z}_{2}$-symmetry 
violation between AFM sublattices and, hence, can be called the asymmetry parameter.
From Eqs. (\ref{drhoR}) and 
 (\ref{gamma}) we see that at small $\gamma \ll 1$ and not very strong field $
H \ll H_s $ the NIMR correction is quadratic in magnetic field.
	\begin{equation}
		\frac{\rho (H)-\rho (0)}{\rho (0)}
		\approx \frac{\gamma^{2}(H)-\gamma^{2}(0)}{1+2\gamma^{2}(0)} \approx - \frac{H^2}{H^2_s}\left( \frac{\Delta E_{ex}}{2t_{0}} \right) ^2.  
		\label{drhoRm}
	\end{equation}

At $H>H_{\mathrm{s}}$ the NIMR correction (\ref{drhoR})-(\ref{gamma}) disappears, and
one returns to the usual classical positive magnetoresistance in multiband conductors due to
impurity scattering, which is parabolic at low field when $\omega_c \tau \ll 1 $: 
\begin{equation}
	\rho_{zz}^{m}\left( H\right) /\rho_{zz}^{m}\left( 0\right)= 1+\omega_c
	^{2}\tau ^{2},\ \ \omega_c \tau \ll 1,  \label{RMzB1}
\end{equation}
where $\omega_c =eH/(m^{\ast}c)$ is the cyclotron frequency. Combining Eqs. (\ref{drhoR}) and (\ref{RMzB1}) 
gives the schematic MR curve illustrated in Fig.~\ref{fig:comparison} in agreement with experimental data.

Note that the AFM does not violate the symmetry $\hat{Z}_{2}\cdot \hat{%
	\theta}$, where $\hat{\theta}$ is the time reversal, which changes the spin
projection $\sigma $. However, there is a protection of electron scattering
to the opposite-spin states by a potential disorder, similar to the
protection of chiral edge states to backscattering in topological insulators.

\subsection{Rough theoretical estimates of maximal NIMR effect in EuSn$_2$As$_2$}
The conduction electrons in EuSn$_{2}$As$_{2}$ come from SnAs bilayers.
Below the Neel temperature $T_{N}$ in A-type antiferromagnet, the two layers
within one bilayer feel opposite
magnetization of Eu spins from the nearest spin-polarized 2D Eu planes. The
corresponding Zeeman-exchange spin-splitting in EuSn$_{2}$As$_{2}$ is 
$\Delta E_{ex}\approx 30-40$\,meV, as follows from the estimate
	based on our band-structure calculations in a ferromagnetic state \cite{degtyarenko_tbp}.

The interlayer transfer integral $t_{0}$ in EuSn$_{2}$As$_{2}$ is more difficult to determine.  Unfortunately, the 
	ARPES data \cite{li_PRX_2019} do not have sufficient energy resolution to measure the bilayer 
	splitting and $t_{0}$ directly. The $t_{0}$ value can only be roughly estimated from the observed resistivity anisotropy at low temperature $\rho _{zz}/\rho _{xx}\approx 130$ (see Fig. \ref{fig:R(T)}) and the width of in-plane energy band $4t_{x}\approx 1.9$eV, which is taken from our ARPES data and from the DFT calculation. As a
	result, we obtain the estimate  $t_{0}\approx t_{x}\sqrt{\rho _{zz}/\rho _{xx}}\approx 42$meV. Hence, the 
	parameter $\gamma (0) =\Delta E_{ex}/(2t_{0})\approx 0.36$,  which according to Eq. (\ref{drhoR}) gives the maximal NIMR effect $\delta \rho (H)/\rho (0)\approx 0.1$ in EuSn$_{2}$As$_{2}$. 
	
	The observed NIMR effect is somewhat smaller than our theoretical prediction. For the longitudinal geometry $R_{ab}(H_{ab})$, which does not have a large contribution from positive MR,  $\delta \rho (H)/\rho (0)\approx 0.06$ (see Fig. \ref{fig:comparison}). The difference $\approx 4\%$  between the experiment and the above theoretical estimate comes because of (i) too rough theoretical estimates of $\Delta E_{ex}$ and $t_0$ and (ii) idealized theoretical model leading to Eqs. (\ref{drhoR})-(\ref{gamma}). 

	According to Eqs. (\ref{drhoR})-(\ref{gamma}) the NIMR effect depends quadratically 
	on $\gamma (0) =\Delta E_{ex}/(2t_{0})$, and reducing this ratio by $15-20$\%  gives a  good agreement between our theoretical model and experiment. Another possible source of error in our calculations comes from the assumption of negligible overlap of wave functions $\psi_{1}$ and $\psi_{2}$  (see  
	Eqs. (\ref{psipmWF}) and (\ref{psipmS}) in Appendix \ref{app:theory details}) and assumed in the derivation of Eq. (\ref{dI}). A finite value at $z=0$ of the wave function in Fig. \ref{FigWF} indicates that this overlap is nonzero. This assumption also leads to the overestimation of NIMR effect in Eq. (\ref{drhoR}) and its deviation from experimental data. 

	The exchange splitting mechanism of negative magnetoresistance corresponds to the energy scale $\Delta E_{ex} \approx 30-40$\,meV 
that is	much larger than the usual Zeeman splitting and temperature $T$. Hence, it is rather robust and should appear in many other compounds. However, in most AFM metals the effect is small by the parameter $\gamma^2$ as described in Eq. (\ref{drhoR}).

	\subsection{On the temperature dependence of NIMR}
	\label{sec:T-dependence}
	At low temperatures $T<T_N\ll \Delta E_{ex}$ in the AFM state, the main effect of finite temperature on the proposed NIMR mechanism is not on its magnitude but on the field $H_s$ of complete spin polarization and, hence, on the magnetic-field 
width  of the NIMR hump. According to the mean-field theory, one expects that the NIMR interval $H<H_s$ shrinks quadratically as temperature increases, $\Delta H_s(T)/H_s(0) =H_s(T)/H_s(0)-1 \propto -(T/T_N)^2$. However, the fluctuations may modify this oversimplified $\Delta H_s(T)$ dependence, and their theoretical study is beyond the scope of our paper.

\section{Quantitative comparison of the measured magnetoresistance with theory}
\label{sec:data comparison}

\begin{figure}[tbh]
		\includegraphics[width=120pt]{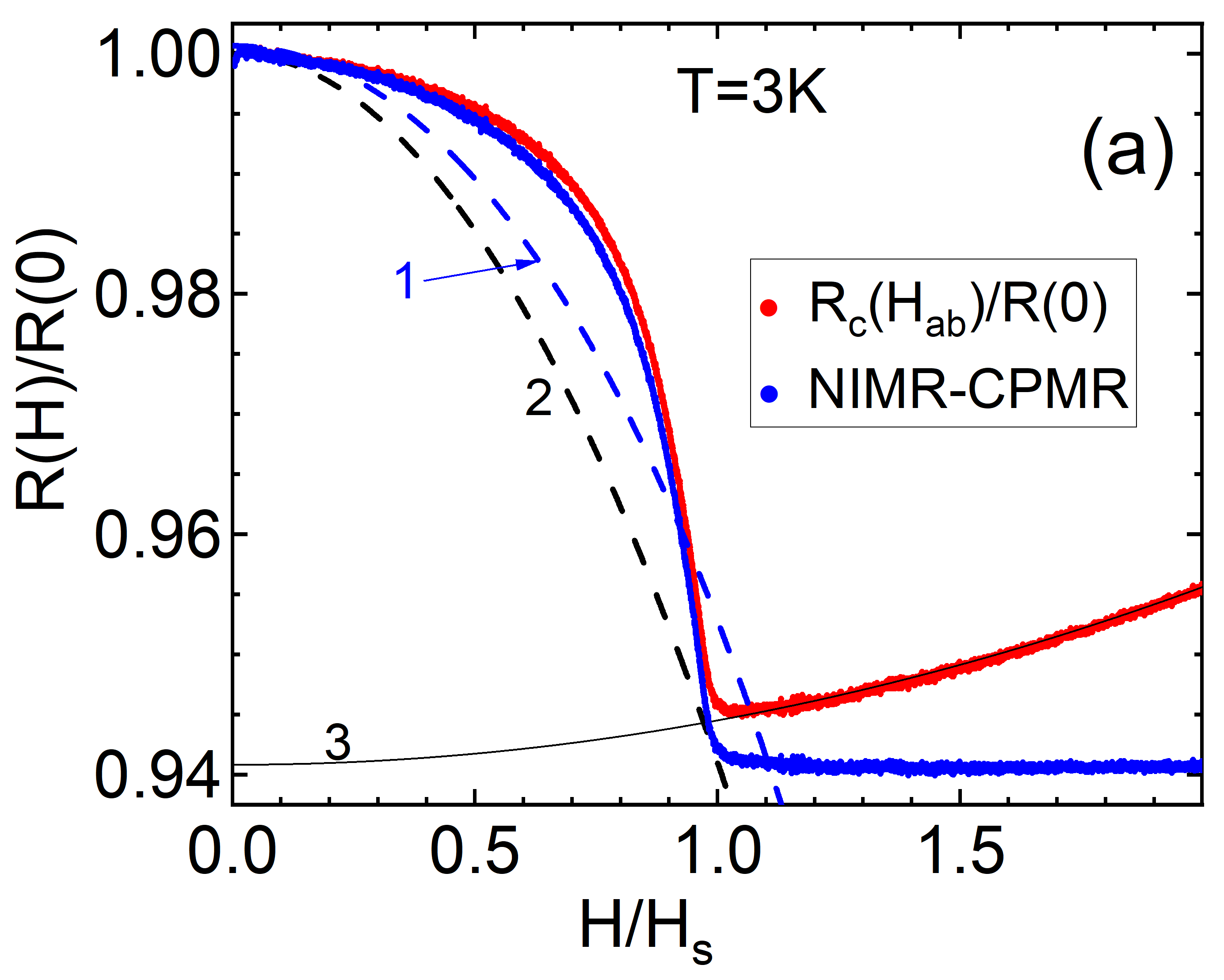}
		\includegraphics[width=120pt]{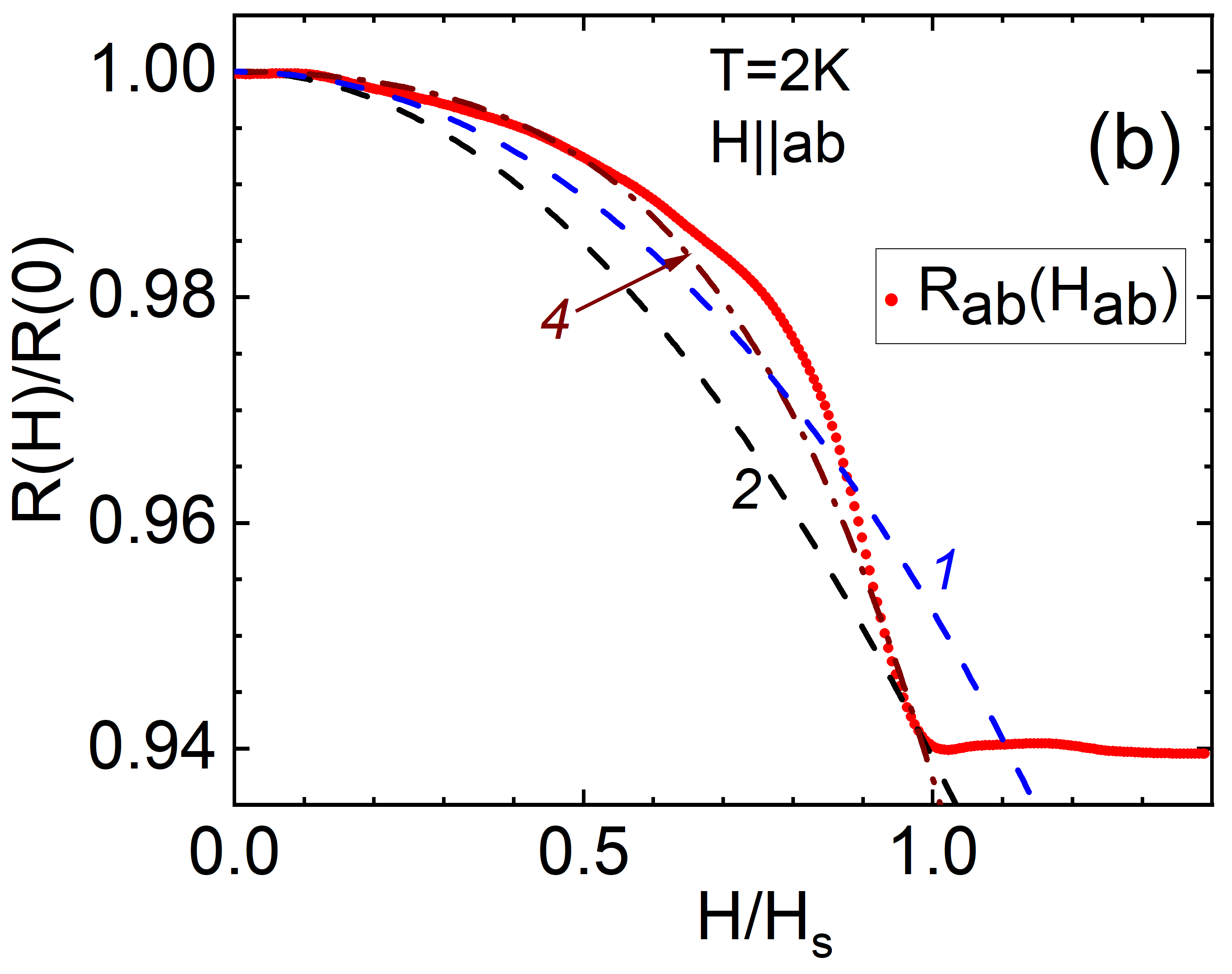}
\caption{Normalized magnetoresistance $R(H)/R(0)$ vs normalized magnetic field $H/H_s$. (a) Red dots - $R_c (H_{ab})$  at $T=3$K, blue dots - same data with subtracted  CPMR. 
	(b) Red dots $R_{ab}(H_{ab})$ at $T=2$K.
	Solid curves {\it 1} - theoretical dependence with   $\gamma(0)=0.23$ evaluated as described in the text.
Parabolas {\it 2}  are guide to the eye, matching the data at $H=H_s$. 
Solid line {\it 3} --  CPMR. Curve {\it 4} - the 4th order fitting polynomial.
}
\label{fig:comparison}
\end{figure}
\subsection{Negative  magnetoresistance}
In Figure~\ref{fig:comparison} we compare  the calculated negative magnetoresistance $R(H)/R(0)$ with the experimental data from Figs.~\ref{fig:CPMR}, and \ref{fig:isotropy}  at the lowest temperature.  
One can see that the theoretical dependence (Eqs.~(\ref{drhoR}) and (\ref{gamma})) reasonably well corresponds 
the measured NIMR  after subtracting PMR  from the row $R(H)/R(0)$ data.
The theoretically estimated prefactor value $\gamma(H=0)=0.23$ is consistent with a sample parabola (curve {\em 1} in Fig.~\ref{fig:comparison}a) within 4\%.
Taking into account the large uncertainty in the $\Delta E_{ex}$ and $t_0$ values, 
the agreement between the theoretical and experimental 
 parabolic prefactor values seems satisfactory. 

Three notes should be made here. Firstly, as we mentioned above, in the vicinity of the phase transition (AFM-PM), the $R(H)$ dependence is expected to become steeper than Eq.~(\ref{gamma}) suggests. The  sample curve {\it 4} illustrates that the agreement may be improved 
when the 4th order terms are included.

Secondly, in Appendix~\ref{app:NMR anatomy} we show that flattening of the low field NIMR data (i.e., deviation from parabola in low fields) 
may be caused by magnon scattering. The latter produces an extra PMR term that quickly decays as $H\rightarrow H_s$, because magnons develop in the AFM state solely.

Thirdly, our scaling analysis of NIMR   (see Appendix \ref{scaling}) reveals a universal temperature dependence of $R(H/H_s)$
 via  $[H_s(0)/H(T) -1 ]\propto T^\xi$  with $\xi\approx 2.5$. The simple mean-field arguments suggest a $H_s\propto T^2$ dependence.   The empirical result might therefore be useful for future refined  theoretical consideration.

\section{Conclusions}
\label{sec:conclusion}

The major aim of our studies was to	clarify the origin of the {\em negative} magnetoresistance observed in   layered non-Weyl AFM semimetals. 

(1) In this paper we presented results of 	comprehensive magnetotransport measurements with   EuSn$_2$As$_2$, a typical representative 
material of this class.
The  considered materials demonstrate a negative  magnetoresistance $R(H)$, firmly linked with magnetization $M(H)$ changes. 
The distinctive feature of  NMR in these materials  is its full isotropy with respect to the magnetic field 
and current directions.   In order to distinguish this NMR from other known types of NMR (such as GMR and CMR), we  called it ``negative isotropic magnetoresistance'' (NIMR). Our results give a strong evidence that NIMR in the layered AFM semimetals is an intrinsic property, almost irrelevant to  defects, domains, and other sample-specific disorder. 

(2) We explored in detail  the  novel  magnetoresistance mechanism suggested in \cite{PRL_tbp}, and compared it with experimental data.
In the proposed theory, the magnetoresistance  primarily originates from  exchange splitting of the 
energy levels of  charge carriers
confined between differently polarized layers of Eu  magnetic ions  in the host lattice. 
Other, more  conventional mechanisms, such as electron  scattering by magnons and by staggered layered magnetization, play  a secondary role. The combination of these mechanisms may explain a certain sample-dependent variation of the measured NIMR in the AFM state. 

(3) Studying  temperature dependence of NIMR,  we have shown that whereas the magnitude of the NIMR hump is roughly $T-$ independent in the AFM-domain of temperatures, the width of the $R(H)$-hump in magnetic field, $\Delta H(T) \approx H_s(T)$,  shrinks with temperature as $[H_s(0)/H_s(T)] -1 \propto  T^\xi$ with $\xi = 2.5$. This exponent differs  somewhat  from the above theoretical estimate  $\xi = 2$ (sec. ~\ref{sec:T-dependence}), due to the oversimplified character of the mean mean-field model.

 It might be that  taking account of electron-electron interaction in short-range scattering  may improve agreement with theory. Indeed,  this mechanism   produces the square root temperature dependence of the {\em positive}  MR, $\delta\rho(H,T) \propto (\omega_c\tau)^2 (T\tau)^{-1/2}$,    in the ballistic interaction regime $T\tau\gg 1$ \cite{interaction-MR, MR_ee-corrections_exper}); the latter leads to broadening of the observed hump with temperature increasing. Detailed  investigation of this issue is however beyond the scope of our paper.

(4) We have also shown that the  {\em positive} magnetoresistance which is always observed (for current perpendicular to magnetic field direction) in layered AFM materials above the field of complete polarization, is 
described by conventional impurity scattering in a two-band system,  
$\delta R(H)/R(0) \propto (\mu^{\rm eff} H)^2$. 
We have shown that the two-band model with the same parameters describes also quantitatively the non-linear field dependence of the Hall resistance. 
(For transport across the layers the classical positive MR may be described by Eq. (34) of Ref. \cite{Schofield2000}.)

 (5) The proposed mechanism of negative magnetoresistance is isotropic with respect to the directions of both the magnetic field and electric current, in agreement with our experiment. The magnitude of NIMR effect, estimated in Eqs. (\ref{drhoR})-(\ref{gamma}), is in a good agreement with experiment either.

(6) The presented ARPES data and DFT calculations confirm the multi-band energy spectrum  with major contribution from the hole  and electron bands around the $\Gamma$ point.This    agrees with the measured nonlinear magnetic field dependence of the Hall voltage.

(7) Our  DFT calculations reveal the exchange splitting  of the Sn-5 states in the AFM-ordered phase, and provide 
  solid ground for our theory of the negative isotropic  magnetoresistance \cite{PRL_tbp}. The calculated exchange splitting 
  of the energy levels enables to estimate the prefactor of the theoretical field dependence of NIMR, which appears to be in a good agreement with experimental data.

\section{Acknowledgements}
AVS, OAS, KSP, and VMP acknowledge support from  RSCF grant \#23-12-00307.  

\appendix
\section{Anatomy of the Magnetoresistance}
\label{app:NMR anatomy}

The normalized negative magnetoresistance $\delta R(H)/R(0)$ has the same magnitude for the samples showing ``parabolic'' and ``flattened'' type magnetoresistance.
The following pattern is worth noting:
 the observed flattening is stronger  for $R_{ab}(H_c)$ and for $R_c(H_c)$  than for $R_{ab}(H_{ab})$. 
In Figure \ref{fig:CPMR}b we demonstrate that  this difference is mainly due to
the mobility anisotropy, and after subtracting the corresponding CPMR contribution, the $R(H)$ dependence becomes isotropic for one and the same crystal.

Besides this configuration-dependent regularity, there is also a sample-dependent flattening, which  
we associate with magnon scattering. 

\subsection{``Parabolic'' and ``Flattened'' -type negative magnetoresistance in the AFM state}
The difference in functional dependences of NIMR  is most clearly pronounced at the lowest temperatures (cf Figs.~ ~\ref{fig:flattened_MR} and \ref{fig:isotropy}). 
The measured nearly-parabolic NIMR  shown in 
Fig.~\ref{fig:comparison} in low fields $H\ll H_s$  looks  qualitatively similar to the theoretical  
dependence $\delta R(H)\propto -  H^2$   \cite{PRL_tbp}.
Deviations from the parabolic dependence were mentioned in  Sections ~\ref{sec:theory}, \ref{sec:data comparison})
and might be a consequence of the proximity to the 2nd order phase transition at $H=H_s$. 
Below, we analyze a potential alternative origin of the $R(H)$ shape flattening.

\begin{figure}[ht]
	\includegraphics[width=118pt]{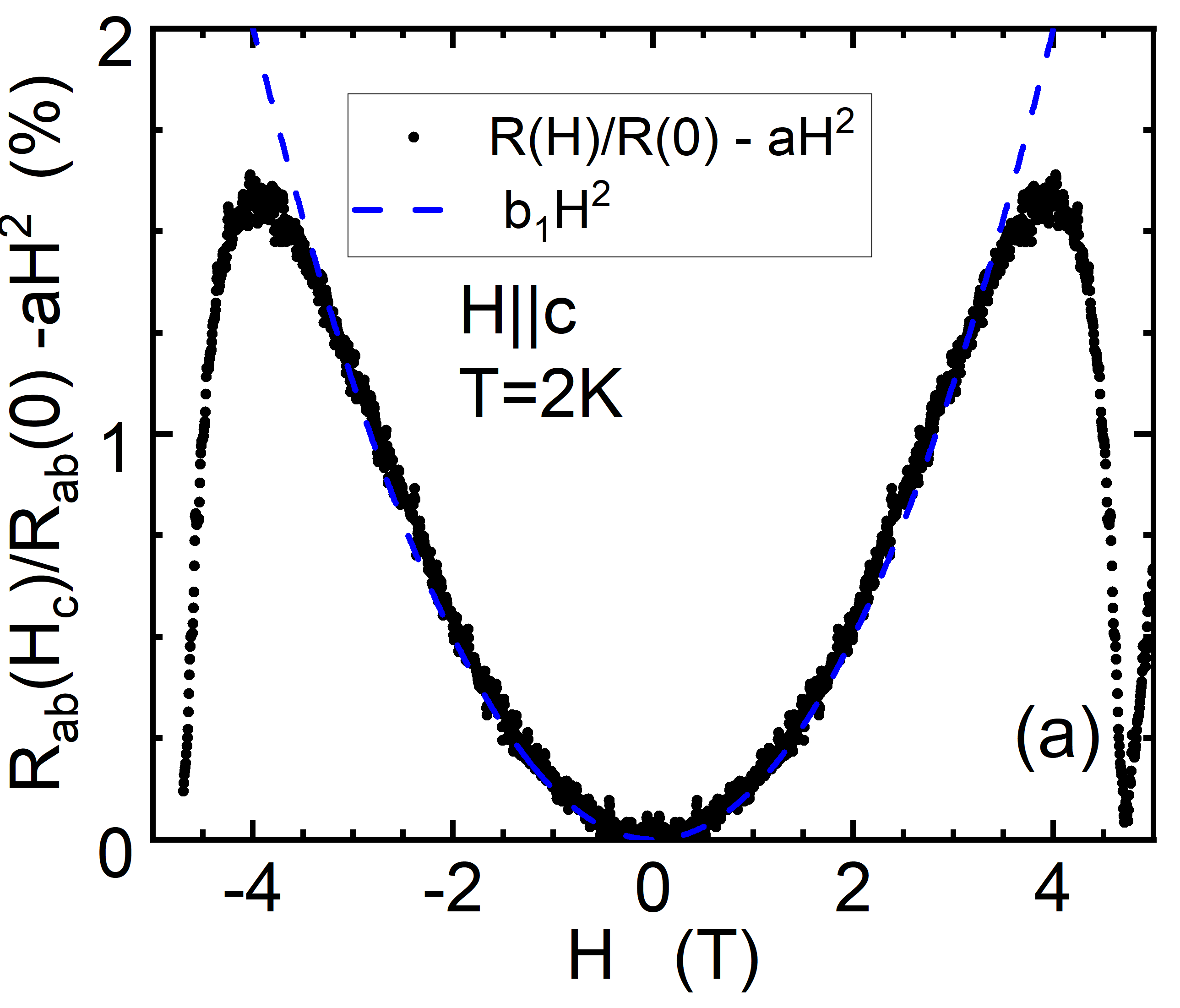}
	\includegraphics[width=122pt]{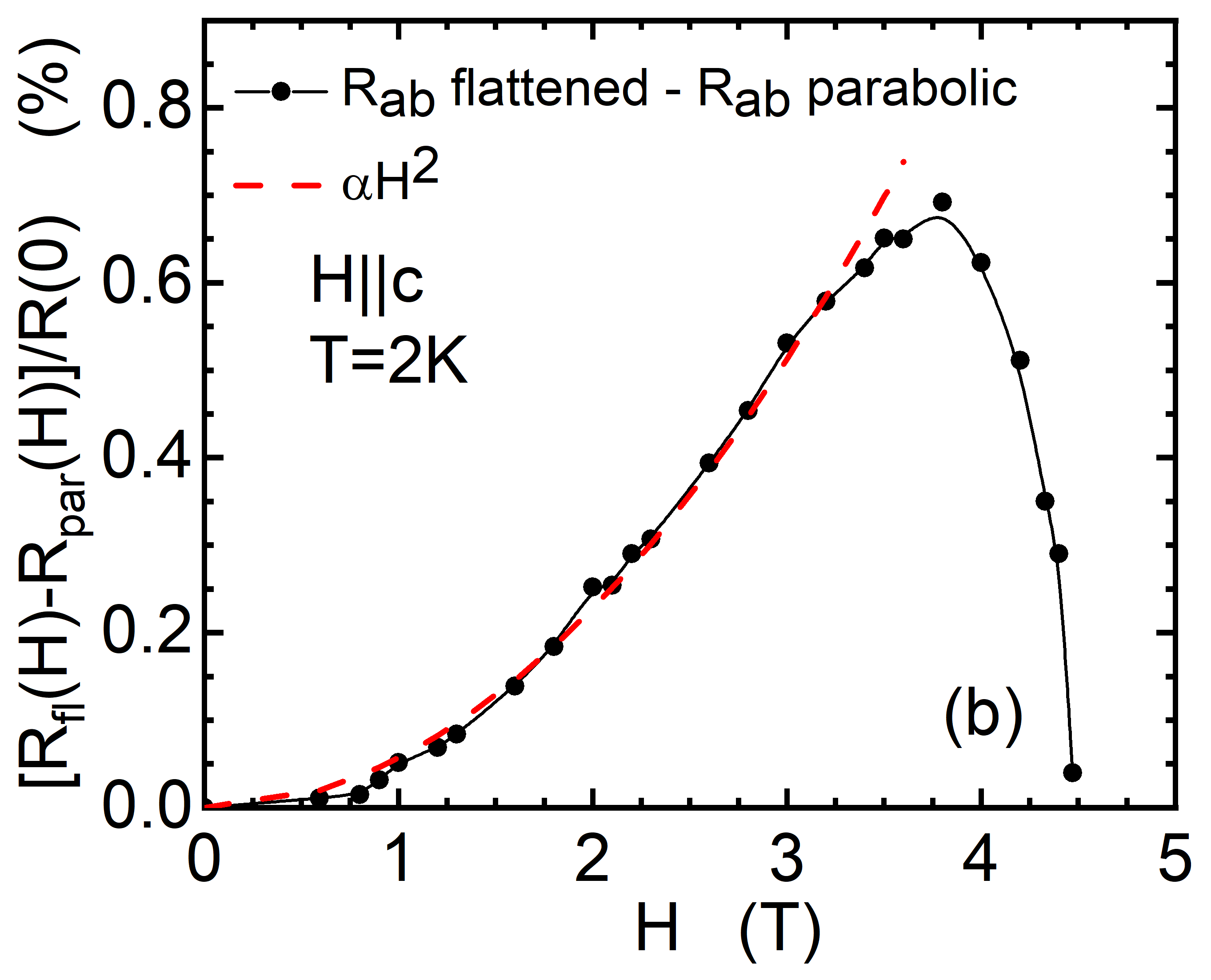}
	\caption{(a) Example of the deviation of the measured $R_{ab}(H\|c$) dependence 
		from the true parabola $(-a H^2)$ with the prefactor 	$a$  adjusted to fit the data at $H=H_{s}$ (see Fig.~\ref{fig:comparison}). 
(b) Difference in the magnetoresistance $R_{ab}(H\|c)$ between the ``flattened'' and ``parabolic'' type samples. Dashed curve - $(b_2 H^2)$ with adjustable prefactor $b_2$.
}
	\label{fig:NMR-deviation_parabolic}
\end{figure}

An example of the deviation $(R_{ab}(H) - a H^2)$ for the ``parabolic-type'' sample is shown in  Fig.~\ref{fig:NMR-deviation_parabolic}a. For this analysis, the prefactor $a$ is chosen to fit the data at $H=H_s$. Above 
$\approx 0.85 H_s$ the deviation sharply vanishes.

In Figure ~\ref{fig:NMR-deviation_parabolic}b we plotted the difference between $R_{ab}(H)/R(0)$ 
for the representative 1st and 2nd type samples in the $H\|c$ geometry where the difference is most pronounced.

One can see that the ``flattened'' type NIMR differs from the``parabolic'' NIMR also by an extra positive MR term: 
 the latter is two times less, it is very similar to the  one shown in Fig,~\ref{fig:NMR-deviation_parabolic}a. 
  We conjecture that all  positive contributions to MR in the AFM state originate from carrier 
scattering by magnons in the AFM crystal, and/or  
by randomly staggering layers magnetization. The former mechanism is universal, whereas the latter - sample dependent.
Both PMR terms in the AFM state vanish sharply in fields above $0.8\times H_s$. 

The most important result is that in lower fields, the deviation by itself is parabolic in field as Figs.~\ref{fig:NMR-deviation_parabolic}a,b show. 
This  means, in particular, that  there are no exponential-type hopping or temperature activated contributions to NIMR, and, respectively, no over-a-barrier type potential scattering mechanisms, like Kondo-scattering.

\subsection{Scaling analysis of the negative magnetoresistance  in the AFM state}
\label{scaling}
\begin{figure*}[ht]
\includegraphics[width=160pt]{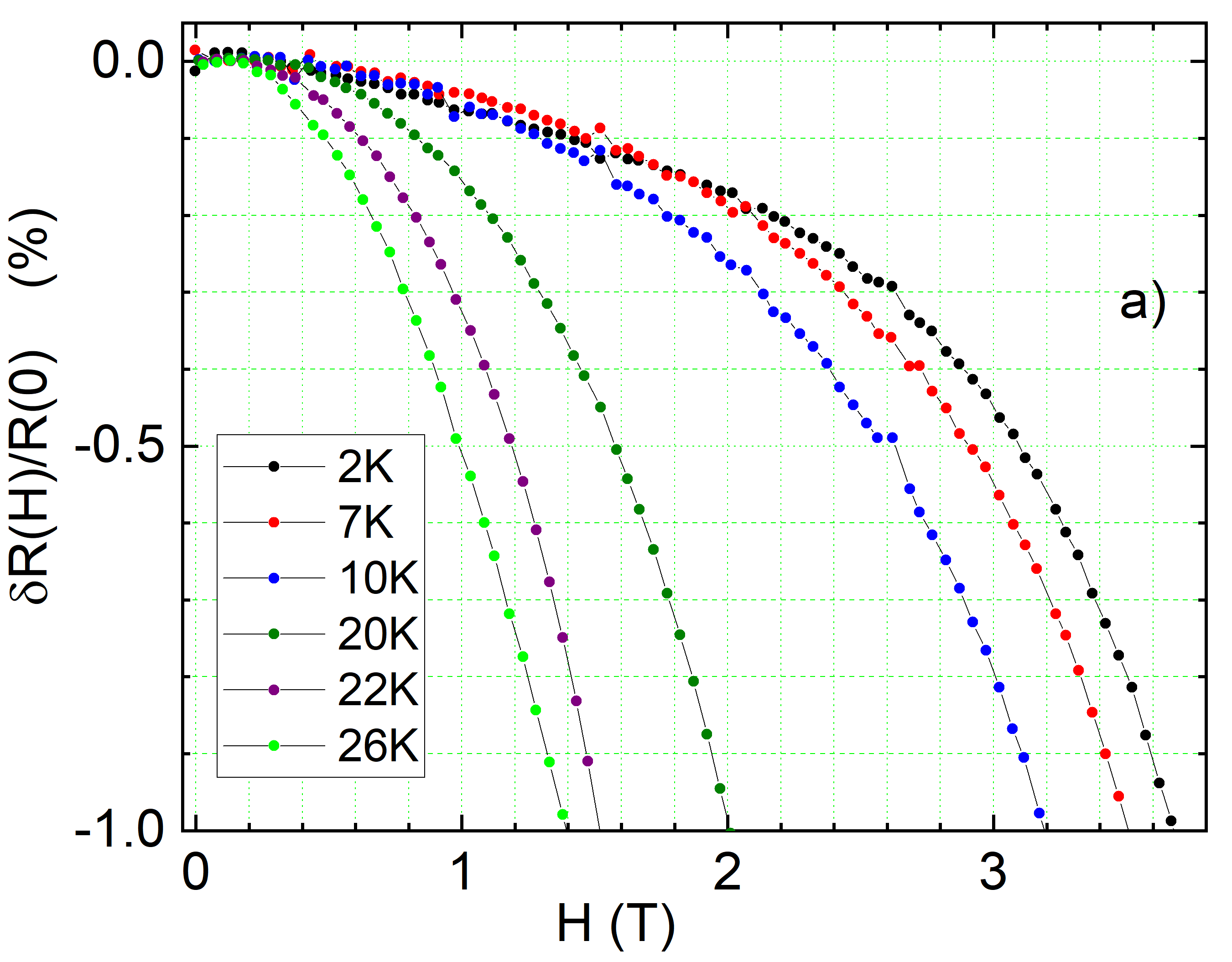}
\includegraphics[width=160pt]{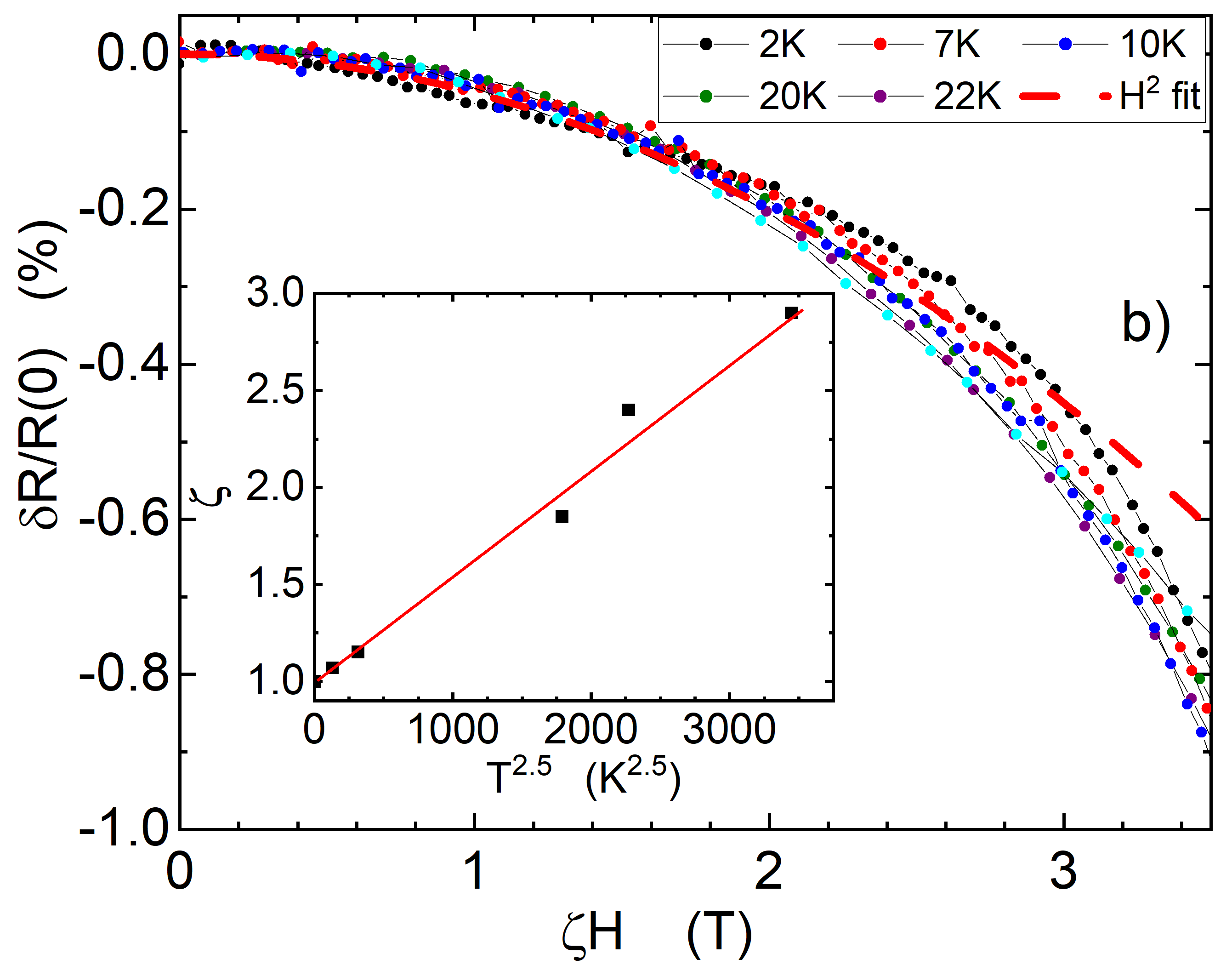}
			\includegraphics[width=160pt]{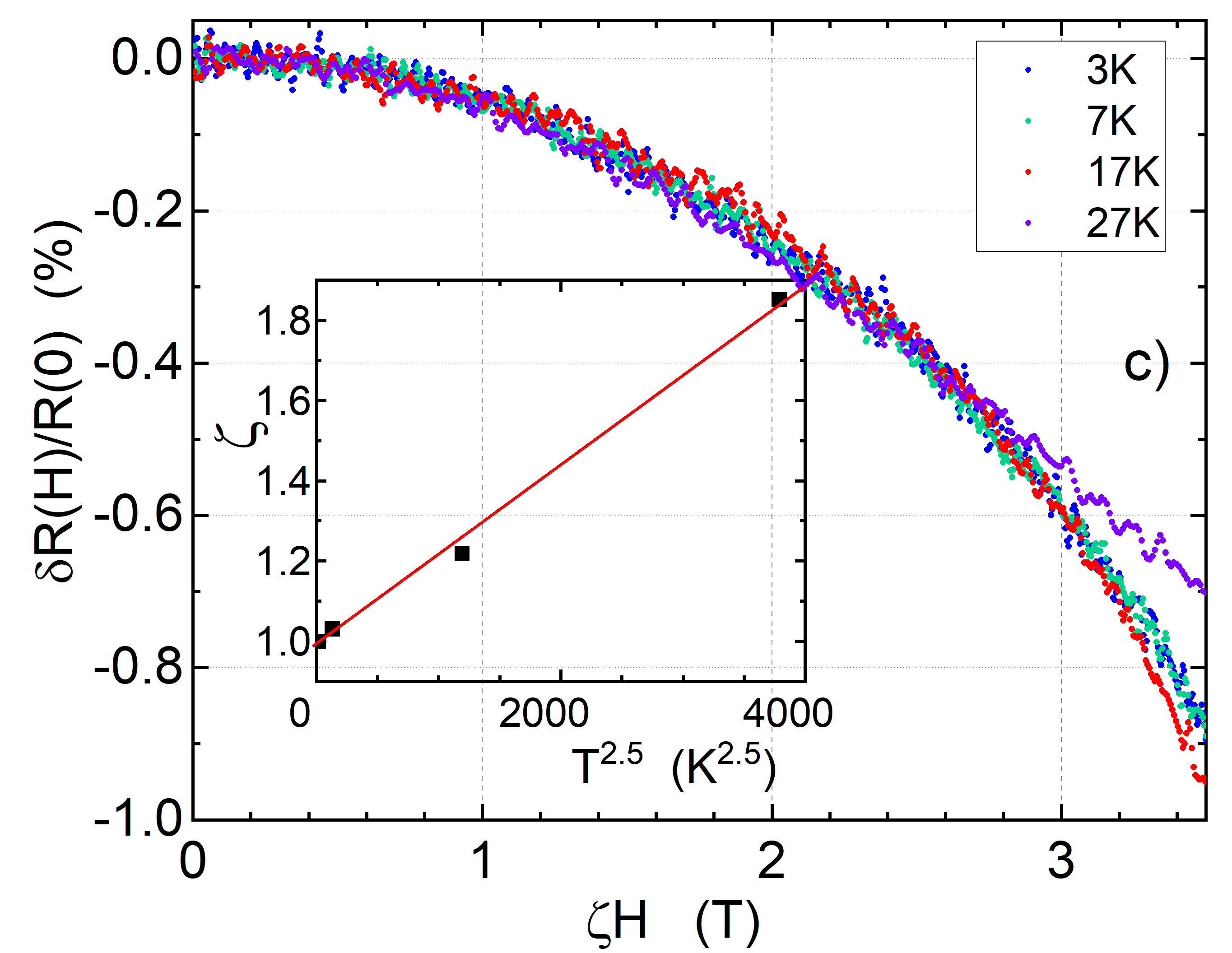}
		\caption{Normalized  negative magnetoresistance $\delta R(H,T)/R(0,0)$: 
			(a) Initial ``flattened'' type NIMR data from Fig.~\ref{fig:flattened_MR}c. 
			(b) Same  ``flattened'' type NIMR data  scaled
						versus  $H^*= H(0)/\zeta(T)$ at various temperatures. 
			(c)  ``Parabolic'' type NIMR data scaled similarly.
			The insets show the scaling factor $\zeta(T)$ that is $\propto T^{2.5}$ for both plots. 
}
	\label{fig:R(HT)-scaling}
\end{figure*}

In order to test whether the two different types of NIMR have  a common or different physics behind, we performed their scaling analysis. Figures \ref{fig:R(HT)-scaling}a and \ref{fig:R(HT)-scaling}b  show that the normalized magnetoresistance data taken at various temperatures for either type NIMR  may be collapsed onto a single curve when field is scaled with a single scaling parameter $\zeta(T)=H(T=0)/H(T)$. The curves at the lowest temperature, $T=3$\,K and 2\,K for panels (a) and (b), respectively, have been taken as a reference curve  $R(H,T=0)$. 

Remarkably, in both cases the  scaling parameter has the same temperature dependence: $\zeta \propto (1+ T^\xi)$; in other words, the width of the NIMR hump shrinks with temperature  as $\Delta H \propto  1/(1+T^{\xi})$ where $\xi= 2.5\pm 0.2$.
The scaling holds in the temperature range up to $T_N$.
The similarity of the two scaling curves and identical $\zeta(T)$ critical indices for two types samples confirm that (i) the physics behind two types of NIMR is the same,  and (ii) the difference between the two types of NIMR is due to a 
larger positive magnetoresistance contribution in the ``flattened''-type, than in the ``parabolic''-type NIMR.

To conclude this section:  our scaling analysis demonstrates that 
negative magnetoresistance in the AFM A-type semimetals is composed of {\em two parabolic contributions of the opposite sign}. The major universal negative contribution is described by our theory (Eqs.~(\ref{drhoR}), (\ref{gamma}) and Ref.~\cite{PRL_tbp}), the  minor sample-dependent positive contribution  
vanishes sharply when field approaches $H_s$. 

\section{Overview of the known mechanisms of MR in magnetically ordered systems.}
\label{app:mechanisms}
In search of an alternative  explanation for the observed NIMR, we  tried on well known MR mechanisms
which might be potentially relevant to the studied case and found none to be able to explain it.

\begin{enumerate} 
\item   
In ferromagnetic oxides, such as  Ln$_{1-x}$Ca$_x$MnO$_3$, the magnetoresistance is negative  and also correlates with magnetization 
\cite{hundley_APL_1995}:
the magnetoresistance in this case is phenomenologically described as $R(H,T)\propto \exp[-M(H,T)/M_0]$ 
and  is explained by polaronic hopping  transport below the ferromagnetic ordering temperature $T_c$. 
The model of polaron hopping transport in insulating oxides is evidently  inapplicable to the semimetallic and well conducting layered AFM semimetals (and particularly to  EuSn$_2$As$_2$), and to the diffusive- rather than hopping-type transport.

\item
Giant magnetoresistance.	The observed parabolic NIMR, at first glance, is reminiscent of a giant magnetoresistance (GMR)  \cite{dagotto_book} in Fe/Cr superlattices where resistance decreases as magnetization in the neighbouring Fe-layers turns with external field from antiparallel to parallel \cite{baibich_PRL(1988)}.
For a single Fe-Cr-Fe sandwich, GMR amounts to $\Delta R/R \sim 1.5$\%  
\cite{camley-barnas_PRL(1989)}, whereas for the multilayered superlattice it reaches about 50\% \cite{baibich_PRL(1988)}.
For  ideally smooth superlattices, GMR should develop in transport across the layers (CPP); 
its presence in real structures for transport along the layers (CIP) have been attributed to  diffusive scattering of
electrons due to interface roughness \cite{camley-barnas_PRL(1989), baibich_PRL(1988)},  
As result, GMR is anisotropic \cite{barthelemy_enciclopedia}.
 In contrast, in EuSn$_2$As$_2$ under study, and in other similar AFM semimetallic crystals, the
layered structure consists of  monolayers of Eu atoms located in the correct lattice sites whose ``roughness'' is 
absent, and the magnetoresistance is fully isotropic. 
  
The key-experimental facts suggesting  irrelevance of GMR/CMR phenomenon to the NIMR under study  are as follow:\\
- fully isotropic NIMR (for current in-plane CIP, and perpendicular to the plane CPP, see Figs.~\ref{fig:isotropy}, \ref{fig:CPMR} and Fig.~4 of Ref.~\cite{PRL_tbp}). This fact immediately prompts the point-like scattering to be responsible, rather than scattering by anisotropic extended defects (phase separation, domain boundaries, etc.). In contrast, GMR is usually lager in CPP geometry than in CIP (see e.g., Fig.6 of Ref. \cite{barthelemy_enciclopedia});\\
-independence of NIMR on the crystal thickness (60nm-0.5mm) (see Ref. \cite{maltsev_tbp}); \\
-the obvious absence of the interface roughness of atomic layers in single crystals (NB: the roughness is  introduced to explain the in-plane magnetoresistance within GMR)  \cite{baibich_PRL(1988), barthelemy_enciclopedia};\\
- high quality lattice structure of the studied single crystals, the absence of grains and boundaries (confirmed by the precise XRD examination, and by the ARPES measurements of the energy band structure), and the absence of physical ground for phase separation effects. \\
- rather different temperature dependence of the width of NIMR hump. In the resistor network model for GMR and CMR \cite{yin_PRB_2000} NMR scales as $(H/T)$.  Hence, the GMR and CMR hump broadens as $T$ increases,  that is confirmed by numerous measurements (e.g., see Fig.~1 of \cite{xu_JPCM_1999}, Fig.~4 of \cite{yin_PRB_2000},  and Fig.~3 of \cite{fontcuberta_PRL_1996}).

In contrast, all experiments with layered AFM crystals show that the NIMR hump shrinks monotonically as temperature increases up to $T_N$ (see Fig.~\ref{fig:R(HT)-scaling}a and 
Figs.~4 of \cite{ying_PRB_2012}, Fig.~5 of \cite{gui_ASCCentrSci_2019}, Fig.~3b of \cite{chen_ChPhysLet_2020}, and Fig.~6b of  \cite{jiang_NJP_2009}). Furthermore, the width of the NIMR hump scales as $H T^\zeta$ with $\zeta>0$ 
(see Fig.~\ref{fig:R(HT)-scaling}).\\
-rather different shape of the magnetoresistance. NIMR for all studied compounds is convex in the  range of  temperatures up to $T_N$ and fields up to $H_s$. By contrast, in the
effective medium theory for doped perovskites \cite{yin_PRB_2000},  the curvature of the magnetoresistance is  $(H/T)$-dependent and becomes concave-upward, concave-downward, and linear as $H/T$ increases. 
\item
Anisotropic interlayer exchange between Eu moments has been conjectured for a sister compound, EuFe$_2$As$_2$. 
There, the Fe moments are ordered in the striped SDW state and there is a 
biquadratic coupling between the Eu and Fe systems. In contrast,
for EuSn$_2$As$_2$, the magnetic system consists of Eu magnetic  moments solely. -
The half-filled $4f$  orbital of Eu$^{2+}$ has
zero orbital angular momentum ($L = 0$) and negligible single-ion
anisotropy. For these reasons, the emergence of the in-plane anisotropy could not
be expected and the anisotrtopy can hardly be an origin of NIMR for the wide class of materials listed above.
\item 
Electron-magnon scattering in antiferromagnets was considered theoretically in Refs.~\cite{yamada-takada_1973-1, yamada-takada_1973-2}, 
and for all  relevant  cases in the AFM state (i.e., at $T<T_N$ and $H<H_s$) was found to be {\em positive}. 
In particular, for $H\|$  easy axes (i.e. in our case, $H\| ab$ along the antiferromagnetic vector $\mathbf{L}_{AFM}$), 
in very low field below the spin flop field $H<H_{c1}$, and at low temperatures  $\mu_B H_{c1} \ll T\ll T_N$:
\beq
\frac{\delta\rho(H)}{\rho(0)} \propto
\left( \frac{H}{T} \right)^2, 
\label{eq:3.18-Yamada}
\eeq
where  the spin-flop field $H_{c1}$  is $\ll H_s$. 

For another geometry, $H\perp $ easy axes (i.e., in our case, $H\perp$ easy magnetization $ab$ plane), 
as temperature raises, the magnetoresistance changes from
\beq
\frac{\delta\rho(H)}{\rho(0)} \propto \left(\frac{T}{T_N}\right)^2\left(\frac{H}{H_s} \right)^2 \quad \textrm{at 
$T  \ll H \ll T_N$}, 
\label{eq:3.30-Yamada}
\eeq
to
\beq
\frac{\delta\rho(H)}{\rho(0)} \propto  \left(\frac{T|H|}{T_N^2}  \right)  \quad \textrm{at $ H\ll T <T_N$}.
\nonumber
\label{eq:3.29-Yamada}
\eeq

Since $g\mu_B\approx 1.35$\,K/T, at the lowest temperature of our measurements $T\approx 2$K,
Eq. (\ref{eq:3.29-Yamada})  is applicable in the  field range  from  $\approx 2$T to $H_s$.
In any case, in the AFM state, resistance  always {\em increases} with field because of a field-induced
increase in  spin fluctuations; this is in contrast to a 
suppression of the spin fluctuations in an external magnetic field for a ferromagnetic metal (see further).

The major theoretical results are confirmed by a large  number of experimental studies, e.g. on antiferromagnetic Mn$_2$Au \cite{bodnar_PRA_2020} and PrB$_6$ \cite{ali_JAP_1987}. Since EuSn$_2$As$_2$ in the AFM state exhibits  NIMR rather than PMR,
we conclude this mechanism, at least,  doesn't play a major role  in NIMR.  

\item
Electron-magnon scattering in ferromagnets.\\
The  low temperature NMR was reported for ferromagnetic DyNiBC, GdNiBC, HoNiBC, and TbNiBC \cite{fontes_PRB_1999}.
A strongly anisotropic NMR was  also observed in semiconductor  GaAs nanowires ``wrapped'' in a ferromagnetic (Ga,Mn)As shell \cite{butschkow_PRB_2013}.
NMR in ferromagnets originates from the  magnetic-field-induced suppression of magnon
scattering  that causes a sizable resistance decrease of several 10\%.  Functionally, 
at  low temperatures  $\Delta R(H)$ is almost linear for ferromagnetic boron carbides,  $\Delta R(H) \propto -H$
\cite{fontes_PRB_1999}, or   concave  for ferromagnetic   GaAs/(Ga,Mn)As core-shell nanowires \cite{butschkow_PRB_2013}.
In all these cases of ferromagnet compounds, the $R(H)$  functional dependence is rather different from our case, Fig.~\ref{fig:comparison}, where it is always convex.

\item 
Kondo-scattering, i.e. the scattering of conduction electrons  due to randomly located  magnetic impurities. 
This scattering results in a characteristic minimum in electrical resistivity with temperature \cite{anderson_PR_1961},
due to an additional term $ \rho(T) \propto \ln(\mu/T)$.
In our case, firstly, no such a minimum or even a tendency is observed in Fig. \ref{fig:R(T)}. Secondly, there are no magnetic impurities in the stoichiometric EuSn$_2$As$_2$ compound grown from high purity raw materials.

\item 
 Electron scattering by staggered magnetization. The XRD examination of the studied crystals (Supplemental materials \cite{SM}) revealed that the major type of the lattice defects is  twisting the layers around the $c$-axis within $\pm 7^\circ$. We speculate that  the intralayer FM-magnetization direction may somewhat stagger from layer to layer around the $c$ axis, because  magnetization may be tied to the crystal axes ($a$ or $b$) by a second order in-plane anisotropy. This reason will cause a sample-dependent  electron scattering. 

\end{enumerate}

Summarizing the above brief review, we conclude that none of the previously known mechanisms can explain the observed NIMR
in the layered AFM  semimetals. However, a minor sample-dependent  positive contribution to MR in the AFM state 
might be related with  either AFM magnon scattering, or with FM-static disorder.  
	
The  AFM order in defect-free samples and, hence, the magnon scattering are evidently universal.
On the other hand, disorder pinning of the acoustic  magnons  is sample-dependent and may be a reason of the sample-dependent NIMR diversity. Alternatively, 
static spatial fluctuations of the magnetization direction due to the layers misorientation  might be larger for samples with  flattened-type NIMR. 
At the moment, however, we don't have either experimental verification of these possibilities, or a detailed theoretical description of the temperature and magnetic field-dependence of the scattering mechanism (and PMR).

\section{Theoretical derivation of the negative isotropic magnetoresistance}
\label{app:theory details}

\subsection{Electron quantum states and wave functions in AFM metal}

\begin{figure}[tbh]
	\includegraphics[width=0.4\textwidth]{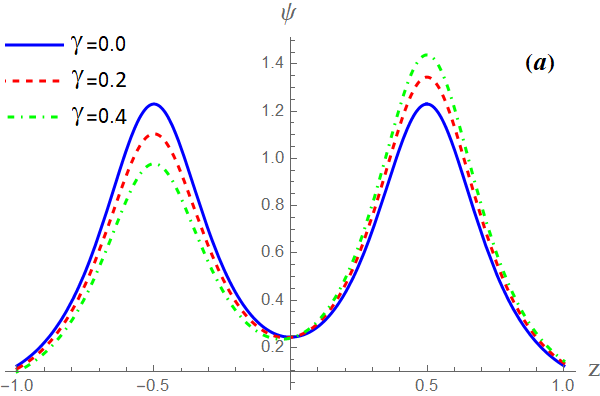}\\
	\medskip
	\includegraphics[width=0.4\textwidth]{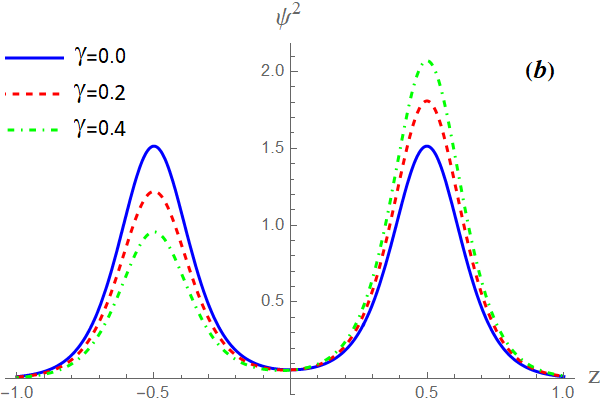}\\
	\medskip
	\includegraphics[width=0.4\textwidth]{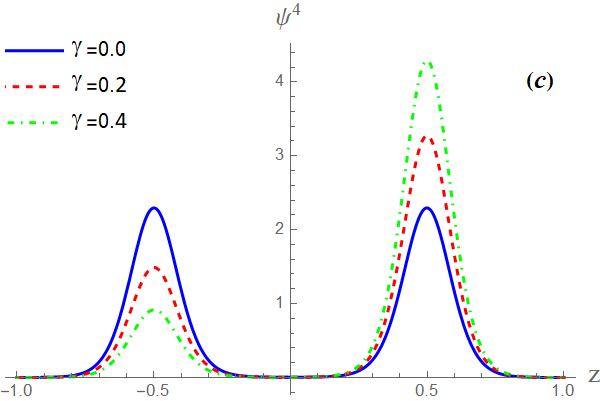}
	\caption{Normalized electron wave function $\psi (z)$ (a) of the lowest-energy quantum state, given by Eq. (\ref{psipmWF}), in a bilayer, modeled by a double-well potential, its square $\psi^2 (z)$ (b) and fourth power $\psi^4 (z)$ (c). The asymmetry parameter $\gamma =\Delta E_{ex}/2t_z =0$ (solid blue line), $\gamma =0.2$ (dashed red line) and $\gamma =0.4$ (dot-dashed green line).
		This figure illustrates the violation of $\hat{Z}_2$ symmetry of the electron wave function in bilayers due to the A-type AFM order and the enhancement of electron scattering rate proportional to $\psi^4 (z)$ in Eq. (\ref{ScRate}). 
	}
	\label{FigWF}
\end{figure}

Let us consider the electron wave functions squeezing in the AFM state with
a collinear and commensurate with lattice AFM order (see Figs. \ref{fig:WF-cartoon} and \ref{FigWF}
for illustration). The two AFM sublattices are numerated by the index $i=1,2$
or by the \textquotedblright pseudospin\textquotedblright\ $\lambda =\lambda
\left( i\right) =3/2-i$, while the projection of spin on the magnetization
axis $\boldsymbol{M}_{AFM}$ is denoted by index $\sigma =\pm 1\equiv
\uparrow ,\downarrow $. The translation $\hat{T}_{\boldsymbol{A}}$ by the
vector $\boldsymbol{A}$ connects opposite AFM sublattices, and the product $%
\hat{T}_{\boldsymbol{A}}\hat{\theta}$ keeps the AFM crystal unchanged, where 
$\hat{\theta}$ is the time reversal operator. The quantum basis consists of
four states, $\left\vert i,\sigma \right\rangle =\left\{ 1\uparrow
,1\downarrow ,2\uparrow ,2\downarrow \right\} $, corresponding to wave
functions $\psi _{i,\sigma }=\left\{ \psi _{1\uparrow },\psi _{1\downarrow
},\psi _{2\uparrow },\psi _{2\downarrow }\right\} =\psi _{i,\sigma }\left( 
\boldsymbol{r}\right) $. By $\hat{T}_{\boldsymbol{A}}\hat{\theta}$ symmetry $%
\psi _{1\uparrow }\left( \boldsymbol{r}\right) =\psi _{2\downarrow }\left( 
\boldsymbol{r}\right) $ and $\psi _{2\uparrow }\left( \boldsymbol{r}\right)
=\psi _{1\downarrow }\left( \boldsymbol{r}\right) $. 

The usual Zeeman splitting $\Delta E_{Z}(\boldsymbol{H})=\left( \boldsymbol{%
	\vec{\sigma}\cdot \vec{H}}\right) g\mu _{H}/2$ in a relevant external
magnetic field $H\lesssim 5$T is much smaller than the exchange splitting $%
\Delta E_{ex}\gtrsim 10$meV and will be neglected below. Then the
AFM-sublattice part of the electron Hamiltonian for each electron
quasi-momentum $\boldsymbol{k}$ is given by the $4\times 4$ matrix in the
basis $\left\vert i,\sigma \right\rangle $, which decouples into two $%
2\times 2$ matrices: 
\begin{equation}
	\hat{H}_{\sigma }=\left( 
	\begin{array}{cc}
		\Delta E_{ex}\sigma /2 & t_{0} \\ 
		t_{0}^{\ast } & -\Delta E_{ex}\sigma /2%
	\end{array}%
	\right) .  \label{HBLs}
\end{equation}%
Here the non-diagonal term $t_{0}=t_{0}^{\ast }$\ is the intersublattice
electron transfer integral. The diagonalization of Hamiltonian (\ref{HBLs})
gives two eigenvalues 
\begin{equation}
	E_{\pm ,\sigma }=\mp \sqrt{\Delta E_{ex}^{2}/4+t_{0}^{2}}  \label{Epm}
\end{equation}%
and the corresponding wave functions (WF) 
\begin{equation}
	\psi _{\pm ,\sigma }=\frac{\psi _{1}\left( \sigma \gamma \pm \sqrt{\gamma
			^{2}+1}\right) +\psi _{2}}{\sqrt{1+\left( \sigma \gamma \pm \sqrt{\gamma
				^{2}+1}\right) ^{2}}},  \label{psipmWF}
\end{equation}%
where $\psi _{1},_{2}$ are the electron wave functions \textquotedblleft
localized\textquotedblright\ mostly on the first and second AFM sublattices
and 
\begin{equation}
	\gamma =\Delta E_{ex}/2t_{0}  \label{gammaS}
\end{equation}%
the ratio of the exchange splitting $\Delta E_{ex}$ of conduction electron
bands to their hopping amplitude $t_{0}$ between the opposite AFM
sublattices, illustrated in Fig.~\ref{FigScheme}.

Without the AFM order, i.e. at $\Delta E_{ex}=0$, one gets the electron spectrum $E_{\pm ,\sigma }=\mp t_{0}$ and the corresponding normalized
eigenstates 
\begin{equation}
	\psi _{\pm ,\sigma }^{0}=\left( \psi _{1}\pm \psi _{2}\right) /\sqrt{2},
	\label{psipmS}
\end{equation}%
which are symmetric and antisymmetric superpositions of electron states on  sublattices $1$ and $2$, as it should be when the $\hat{Z}_{2}$ symmetry is
conserved. From Eqs. (\ref{psipmWF}) we see that AFM lifts this symmetry, making the eigenfunction amplitude larger on one of two sublattices, as
illustrated in Fig. \ref{FigWF}. This enhances the integral in Eq. (\ref{ScRate}),  as seen already from Figs. \ref{FigWF} and shown analytically below.

\subsection{Mean free time in the Born approximation}

In the Born approximation, i.e. the second-order perturbation theory in the
impurity potential, the electron scattering rate is given by the Fermi's
golden rule \cite{Abrik}:%
\begin{equation}
	\frac{1}{\tau }=\frac{2\pi }{\hbar }\sum_{n^{\prime },i}\left\vert T_{ 
		\boldsymbol{n}^{\prime }\boldsymbol{n}}^{(i)}\right\vert ^{2}\delta\left(
	\varepsilon _{\boldsymbol{n}}-\varepsilon _{\boldsymbol{n}^{\prime }}\right)
	.  \label{TauGFR}
\end{equation}%
where the index $n\equiv \left\{ \boldsymbol{k},\zeta ,\sigma \right\} $
numerates quantum states, $\zeta $ and $\sigma $ denote the electron subband
and spin projection, 
$\varepsilon _{\boldsymbol{n}}$ is the electron energy in state $n$, and $%
\delta (x)$ is the Dirac delta-function. The short-range impurities or other
crystal defects in solids are, usually, approximated by the point-like
potential 
$V_{i}\left( \boldsymbol{r}\right) =U\delta\left( \boldsymbol{r}-\boldsymbol{%
	r}_{i}\right) $. Here we omit the spin index $\sigma $ because it is
conserved by the potential scattering. The corresponding matrix element of
electron scattering by this impurity potential is $T_{\boldsymbol{n}^{\prime
	}\boldsymbol{n}}^{(i)}=U \Psi^{\ast } _{\boldsymbol{n}^{\prime }}\left( 
\boldsymbol{r}_{i} \right) \Psi _{\boldsymbol{n}}\left( \boldsymbol{r}_{i}
\right) $, where $\Psi _{\boldsymbol{n}}\left( \boldsymbol{r}\right) $ is
the electron wave function in the state $\boldsymbol{n}$.

For short-range impurities the matrix element does not depend on electron
momentum $\boldsymbol{k}^{\prime }$. Then the summation over $\boldsymbol{k}%
^{\prime }$ with the $\delta $-function in Eq. (\ref{TauGFR}) gives the
electron density of states (DoS) $\nu _{F\zeta }\equiv \nu _{\zeta }\left(
\varepsilon _{F}\right) $ at Fermi energy $\varepsilon _{F}$ per one spin
component and per one subband $\zeta $. The total DoS per one spin $\nu
_{F}=\sum_{\zeta }\nu _{F\zeta }$. If the impurities are uniformly and
randomly distributed in space, the sum over impurities rewrites as an
integral over impurity coordinate: $\sum_{i}\rightarrow n_{i}\int d^{3}%
\boldsymbol{r}_{i}$, where $n_{i}$ is the impurity concentration. Then Eq. (%
\ref{TauGFR}) becomes 
\begin{equation}
	\frac{1}{\tau }=\frac{2\pi }{\hbar }n_{i}U^{2}\int d^{3}\boldsymbol{r}%
	_{i}\sum_{\zeta ^{\prime }}\nu _{F{\zeta }^{\prime }}\left\vert \Psi _{%
		\boldsymbol{k}^{\prime }\zeta ^{\prime }}^{\ast }\left( \boldsymbol{r}%
	_{i}\right) \Psi _{\boldsymbol{k}\zeta }\left( \boldsymbol{r}_{i}\right)
	\right\vert ^{2}.  \label{Tau1}
\end{equation}%
The Bloch wave function is $\Psi _{\boldsymbol{k}\zeta }\left( \boldsymbol{r}%
\right) =\psi _{\zeta }\left( \boldsymbol{r}\right) \exp \left( i\boldsymbol{%
	kr}\right) $, where $\psi _{\zeta }\left( \boldsymbol{r}\right) $ is the
periodic function. Then Eq. (\ref{Tau1}) rewrites as%
\begin{equation}
	\frac{1}{\tau }=\frac{2\pi }{\hbar }n_{i}U^{2}\nu _{F}\,I,  \label{Tau2}
\end{equation}%
where the integral over one elementary cell 
\begin{equation}
	I\equiv \int d^{3}\boldsymbol{r}\left\vert \psi _{\zeta }\left( \boldsymbol{r%
	}\right) \right\vert ^{2}\sum_{\zeta ^{\prime }}\frac{\nu _{F\zeta ^{\prime
	}}}{\nu _{F}}\left\vert \psi _{\zeta ^{\prime }}\left( \boldsymbol{r}\right)
	\right\vert ^{2} .  \label{I}
\end{equation}%
For a single band $\zeta $ at the Fermi level, i.e. when $\nu _{F\zeta
	^{\prime }}=0$ for $\zeta ^{\prime }\neq \zeta $, this confirms Eq. (\ref%
{ScRate}). Eqs. (\ref{Tau2}) and (\ref{I}) also result to Eq. (\ref{ScRate}) if
only the scattering within the same band is allowed, e.g., due to the spin
conservation during the potential scattering. Eqs. (\ref{Tau2}),(\ref{I})
approximately give Eq. (\ref{ScRate}) also when there are several bands, but
the DoS $\nu _{F\boldsymbol{\zeta }}$ for one band $\zeta $ is much larger
than that for the others. The latter happens in Dirac semimetals, where the
DoS $\nu _{\zeta }\left( \varepsilon _{F}\right) \propto \varepsilon
_{F}^{d-1}$ strongly depends on the Fermi energy $\varepsilon _{F}$ and
where the energy difference $\varepsilon _{\zeta ^{\prime }}\left( 
\boldsymbol{k}\right) -\varepsilon _{\zeta }\left( \boldsymbol{k}\right)
\gtrsim \varepsilon _{F}$ for $\zeta ^{\prime }\neq \zeta $.

Now we estimate the difference $\delta I$ of two integrals (\ref{I}) for the
asymmetric and symmetric wave functions, given by Eqs. (\ref{psipmWF}) and (%
\ref{psipmS}). The host crystal lattice has the $\hat{Z}_2$ symmetry,
therefore $\int \psi _{1}^{4}\left( z\right) dz=\int \psi _{2}^{4}\left(
z\right) dz$. For simplicity, we also assume that the overlap of the wave
functions on different AFM sublattices is negligible, i.e. $\psi _{1}\psi
_{2}\ll \left\vert \psi _{1}\right\vert ^{2}$, and we neglect the products $%
\psi _{1}\psi _{2}\approx 0$. The main conclusion remains valid also when $%
\psi _{1}\psi _{2}\sim \left\vert \psi _{1}\right\vert ^{2}$, but the
calculation are more cumbersome.

First we consider the completely subband-polarized case $\nu _{F\boldsymbol{%
		\zeta }}/\nu _{F}=1$ for $\zeta =1$ and $\nu _{F\boldsymbol{\zeta }}/\nu
_{F}=0$ for $\zeta =2$. Then for the symmetric wave function (\ref{psipmS}),
corresponding to the lowest subband $\zeta $ without the AFM order and
bilayer asymmetry, the integral (\ref{I}) is 
\begin{equation}
	I_{0}=\int d^{3}\boldsymbol{r\,}\left\vert \psi _{+}^{s}\right\vert
	^{4}dz\approx \int d^{3}\boldsymbol{r\,}\psi _{1}^{4}/2.  \label{I0}
\end{equation}%
We now calculate the difference 
\begin{equation}
	\delta I\equiv I-I_{0}=\int d^{3}\boldsymbol{r\,}\left( \left\vert \psi
	_{+}\right\vert ^{4}-\left\vert \psi _{+}^{s}\right\vert ^{4}\right) ,
	\label{dI0}
\end{equation}%
which gives the correction to mean free time $\tau $ according to Eq. (\ref{ScRate}) or (\ref{Tau2}). After the substitution of Eqs. (\ref{psipmWF}) and (%
\ref{psipmS}), at $\psi _{1}\psi _{2}\ll \left\vert \psi _{1}\right\vert ^{2}
$ this simplifies to 
\begin{equation}
	\delta I\approx \boldsymbol{\,}\frac{\gamma ^{2}}{1+\gamma ^{2}}\int d^{3}%
	\boldsymbol{r}\frac{\psi _{1}^{4}\left( z\right) }{2}=\frac{\gamma
		^{2}\,I_{0}}{1+\gamma ^{2}}=\frac{\gamma ^{2}\,I}{1+2\gamma ^{2}}.
	\label{dI}
\end{equation}%
Combining Eqs. (\ref{dI}) and (\ref{Tau2}) gives the decrease of the
electron mean free time from the squeezing of electron  wave function by AFM
order and the corresponding enhancement of impurity scattering: 
\begin{equation}
	\frac{\delta \tau }{\tau_0 }\approx -\frac{\delta I}{I_0}\approx \frac{-\gamma
		^{2}\,}{1+\gamma ^{2}}.  \label{dt}
\end{equation}

\subsection{Negative isotropic magnetoresistance}

Substituting Eqs. (\ref{dI}), (\ref{gammaS}) and (\ref{EexB}) to (\ref{Tau2}) we obtain the relative increase
of resistivity due to the AFM ordering%
\begin{eqnarray}
	\frac{\delta \rho (H)}{\rho (0)}&\approx & -\frac{\delta \tau }{\tau }\approx \frac{\delta I}{I}\approx \frac{%
		\gamma ^{2}\,}{1+2\gamma ^{2}}\notag \\
	&\approx &\frac{\left[\Delta E_{ex}/(2t_{0})%
		\right] ^{2}\left( 1-H^{2}/H_{\mathrm{s}}^{2}\right) }{1+2\left[\Delta E_{ex}/(2t_{0})%
		\right] ^{2}\left( 1-H^{2}/H_{\mathrm{s}}^{2}\right) }.  \label{drhoG}
\end{eqnarray}%

At $\gamma ^{2}\ll 1$ Eq. (\ref{dI}) simplifies to $\delta I\approx \gamma
^{2}I_{0}\approx \gamma ^{2}I$, and Eq. (\ref{drhoG}) simplifies to
\begin{equation}
	\frac{\delta \rho (H)}{\rho (0)}\approx \gamma ^{2}\approx \left( \frac{\Delta E_{ex}}{2t_{0}}%
	\right) ^{2}\left( 1-\frac{H^{2}}{H_{\mathrm{s}}^{2}}\right) .  \label{drho}
\end{equation}%
Contrary to the NMR caused by chiral anomaly in Weyl semimetals, this
increase of resistivity is \emph{isotropic}. For example, our NIMR mechanism
applies both for the interlayer and in-plane current directions in layered
conductors, and it only slightly depends on the magnetic field direction due
to a magnetic anisotropy solely.

In usual 3D AFM metals the proposed NIMR mechanism is very weak, because the
ratio $\gamma =\Delta E_{ex}/2t_{0}\ll 1$. Indeed, $\Delta E_{ex}\lesssim 0.1
$eV, while $2t_{0}\sim 1$eV is comparable to the bandwidth. However, in
strongly anisotropic layered materials with A-type AFM order the ratio $%
\gamma =\Delta E_{ex}/2t_{0}\sim 1$, because the interlayer transfer
integral $t_{0}\lesssim 0.1$eV 
in van-der-Waals or other layered compounds is also small.{}

In strong magnetic field approaching the complete spin polarization (or spin-flip transition), at $H/H_s\approx 1$, the field dependence of
magnetization $L$ and exchange energy $\Delta E_{Zex}$ may differ from Eqs. (\ref{M}) and (\ref{EexB}). Because of fluctuations, the actual spin-flip transition is often sharper and is shifted to smaller fields than that given by the mean-field theory and Eq. (\ref{M}).
Then the magnetoresistance at $H\approx H_{\mathrm{s}}$ also develops sharper than that described by Eqs. (\ref{drhoG}) and (\ref{drho}).

\subsection{Contribution of the classical positive magnetoresistance: qualitative consideration}
\label{sec:PMR-classic} 

At $H>H_{\mathrm{s}}$ the obtained correction (\ref{drho}) disappears, and
one returns to the usual positive magnetoresistance in multiband conductors
due to impurity scattering, which is parabolic at low field when $\omega_c \tau \ll 1 $ \cite{Abrik}: 
\begin{equation}
	\rho_{zz}^{m}\left( H\right) /\rho_{zz}^{m}\left( 0\right)= 1+\omega_c
	^{2}\tau ^{2},\ \ \omega_c \tau \ll 1,  \label{RMzB}
\end{equation}
where $\omega_c =eH/(m^{\ast}c)$ is the cyclotron frequency. 
Here $\omega _{c}$ and $\tau $ refer to the carriers of the dominating band $%
m$. We highlight that the positive magnetoresistance $\gamma \rho
_{ii}\propto (\omega _{c}\tau )^{2}$ is intrinsic to the multicomponent
carrier system solely, since for the single-component system with isotropic
spectrum $\rho _{ii}$ does not depend on magnetic field up to $(\omega
_{c}\tau )^{4}$. Therefore, the existence of the parabolic PMR by itself is
in accord with our ARPES data and band structure calculations, which confirm
the presence of the hole band at $\Gamma $ point and the electron band at $M$
point. 

At $H<H_{\mathrm{s}}$ both Eqs. (\ref{drho}) and (\ref{RMzB}) contribute to magnetoresistance.  
Combining Eqs. (\ref{drho}) and (\ref{RMzB}) gives the schematic MR curve at $H<H_{\mathrm{s}}$:
\begin{equation}
	\frac{\rho \left( H\right) }{\rho \left( 0\right) }\approx 1-\frac{\left(
		\Delta E_{ex}/2t_{0}\right) ^{2}}{1+2\left( \Delta E_{ex}/2t_{0}\right) ^{2}}
	\frac{H^{2}}{H_{\mathrm{s}}^{2}}+\left[ \omega _{c}(H)\tau \right] ^{2},
	\label{RMzF}
\end{equation}
illustrated in Fig.~\ref{fig:comparison}, which resembles very much the experimental
observations. 

Below $H<H_s$ both the second and third terms are at work and compete, because both give the quadratic MR but of opposite signs. This is seen from the experimental data in Fig. \ref{fig:comparison}, where the positive term of magnetoresistance depends on mutual orientation of the resistivity component $R$ and magnetic field $\boldsymbol{H}$. The positive classical MR term is absent for longitudinal MR $R_{ab}(H_{ab})$, and the corresponding NIMR effect is indeed stronger for this geometry, as seen from Fig. \ref{fig:comparison}. At $H\geq
H_s$ when the AFM state converges to FM order, the second term in Eq. (\ref{RMzF}) vanishes to
zero, and MR becomes positive. 

\subsection{Generalizations}

Similar qualitative idea should work not only for bilayer crystals with
A-type AFM with wave vector across the layers, but also for any
antiferromagnetic conductor. Indeed, in a commensurate AFM state the unit
cell doubles as compared to nonmagnetic state. Hence, in the new doubled
unit cell there are two electron states per each quasi-momentum, denoted by
indices $1$ and $2$. These two states are degenerate without AFM, which we
called the $\hat{Z}_2:1\leftrightarrow 2$ symmetry, but this degeneracy is
lifted by AFM. Without AFM the electron wave function is given by a
symmetric or antisymmetric superposition of these two states, as in Eq. (\ref%
{psipmS}). In the AFM state this symmetry is lifted, as in Eq. (\ref{psipmWF}%
). Repeating the arguments given above we obtain that the integral in Eq. (%
\ref{I}) is greater in AFM state, i.e. when the $1\leftrightarrow 2$
symmetry is lifted, so that the wave function $\Psi _{\boldsymbol{k}}\left( 
\boldsymbol{r}\right) $ is more localized. This gives to extra resistance
given by Eq. (\ref{drho}), where $t_{0}$ is now the transfer integral
between states $1$ and $2$, corresponding to electron hopping between two opposite AFM sublattices. However, in most AFM metals this transfer
integral $t_{0}$ is comparable to the bandwidth and much larger than $%
\Delta E_{Zex}$, so that the correction (\ref{drho}) is negligibly small.
However, if in a AFM conductor $\left( \Delta E_{Zex}/4t_{0}\right) ^{2}$
is not too small, a quadratic negative magnetoresistance similar to that 
in EuSn$_{2}$As$_{2}$ may appear.

The performed theoretical analysis is, usually, applicable to both single-band and multi-band metals. 
If there are several electronic bands on the Fermi level, the scattering rate given by Eq. (\ref{TauGFR}) is also valid, but the subband index $\zeta$ now numerates the crystalline electronic bands each split into two subbands given by Eq. (\ref{Epm}). Then the integral (\ref{I}) also contains the sum over electronic bands on the Fermi level. 
If the values of parameters  $t_{0}$ and $\Delta E_{Zex}$ are close for all electronic bands, the final result given by Eq. (\ref{drho}) does not change.

Often, the impurity distribution within elementary cell is not uniform and
described by a normalized distribution function $D\left( \boldsymbol{r}%
_{i}\right) $, $\int D\left( \boldsymbol{r}_{i}\right) d\boldsymbol{r}_{i}=V$%
. If the in-plane and interlayer coordinates in the electron wave function $%
u_{\boldsymbol{k}}\left( \boldsymbol{r}\right) $ do not disentangle, e.g.
for zigzag crystal structures, one should use Eq. (\ref{Tau1}) instead of (%
\ref{Tau2}). Then the formulas in the rest of this section are not as
simple, but the qualitative result is still valid.

\newpage

\section{Supplemental Materials}
\label{SM}
Supplemental data to this article can be found online at ... . 
It includes detailed information on the sample synthesis, XRD structural analysis, 
  computational details of electron magnetotransport, electronic band structure (ARPES data and calculations). 

\subsection{SM I: Samples}
\label{SM:samples}

The EuSn$_2$As$_2$ single crystals were synthesized from homogeneous
SnAs (99.99\% Sn + 99.9999\% As) precursor and elemental Eu (99.95\%)
in stoichiometric molar ratio (2:1) using the growth method, similar to
our previous works \cite{golov_JMMM_2022, eltsev_UFN_2014, vlasenko_SST_2020}. The initial high purity binary SnAs compound
was obtained by the solid state reaction technique in quartz ampoule with residual argon atmosphere. Next,
SnAs precursor and Eu were mixed with a 2:1 molar ratio; the mixture
was placed in an alumina crucible and sealed in a quartz tube with
residual argon pressure. The quartz tube was heated in a furnace up to
850 C$^\circ$, held at this temperature for 12\,h to homogenize melting, cooled
down to 550C$^\circ$ at a rate of 2$^\circ$/h, and annealed at this temperature
for 36\,h. After that, the furnace was turned off, and the quartz tube
was cooled down to room temperature inside the furnace.
The synthesis protocol was used in our previous works \cite{golov_JMMM_2022, vlasenko_SST_2020}; it is also similar to that discribed in Ref.~\cite{pakhira_PRB_2021}. The grown crystals had a shiny surface and could be easily exfoliated.
More details on the flakes exfoliated from the buk crystals are given in Ref.~\cite{maltsev_tbp}.

\subsection{SM II: XRD examination of the crystal structure.}
\label{SM:XRD}
In search for potential lattice defects that could cause  scattering and NIMR we performed precise XRD examination of the bulk sample - its crystal structure and lattice defects evaluation.

Measurements were made with PanAlytical Ex'pert Pro X-ray diffactometer equipped with multiple crystal analyzers.
\begin{figure}[h]
	\includegraphics[width=200pt]{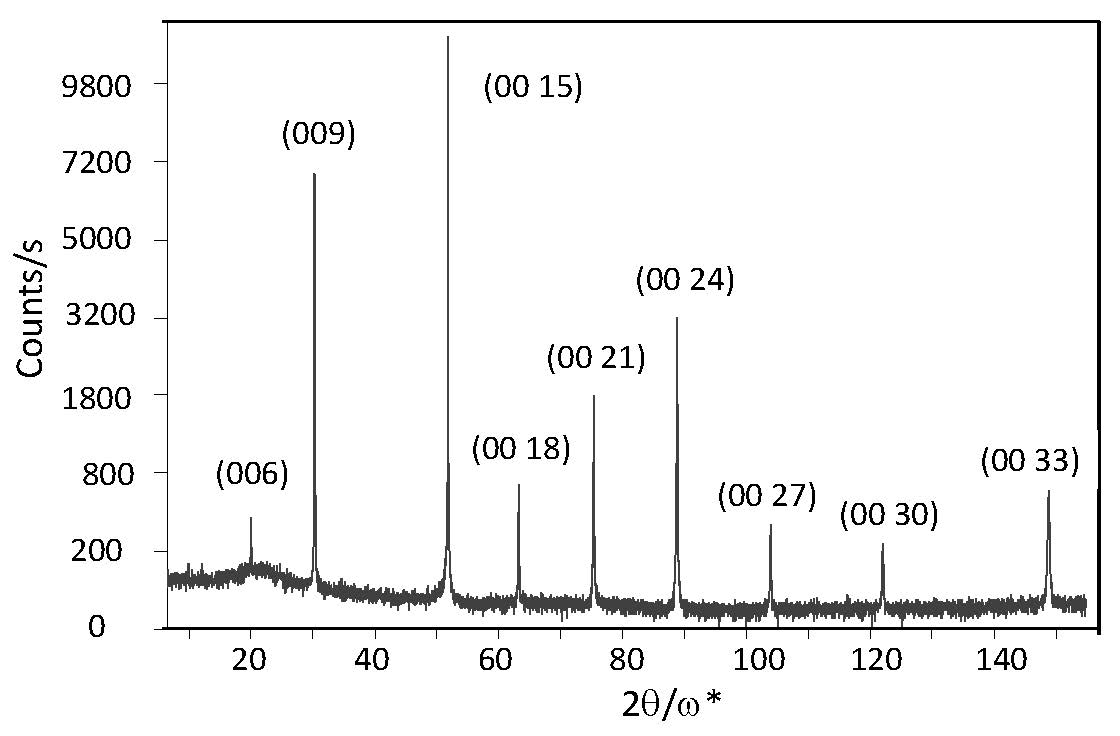}
	\caption{X-ray diffraction pattern on a square scale for the studied EuSn$_2$As$_2$ bulk crystal 
		from the side of the shiny surface. 
		Correction for hybrid monochromator $\Delta(2\theta)=0.0135^\circ$.}
	\label{fig:XRD1}
\end{figure}

\begin{figure}[ht]
	\includegraphics[width=240pt]{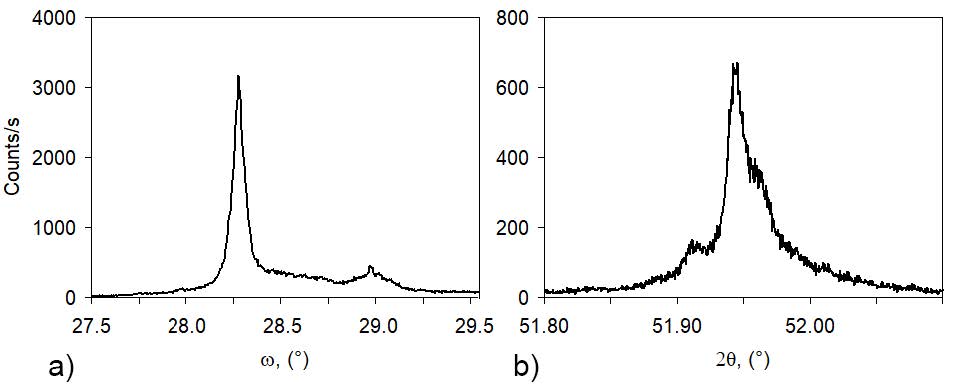}
	\caption{(a) Rocking curve on the strongest reflection (00\,15) with half-width $\Delta\omega=0.067^\circ$. (b) $2\theta -\omega$ scanning curve measured with the third analyzer crystal. }
	\label{fig:XRD2}
\end{figure}
Figure \ref{fig:XRD1} shows the presence of all possible peaks of the $(00 l)$ series for a 
rhombohedral EuSn$_2$As$_2$ lattice ($l=3n$, where $n$ is an integer), as well as the absence of a noticeable broadening of the farthest reflections; these results are the evidence of a high structural perfection of the sample in the direction perpendicular to the basal plane. 		

Rocking curve  (Fig.~\ref{fig:XRD2}a) measured with the strongest reflection (00 15)   demonstrates the absence of blocks in the bulk  crystal, but indicates slight ($\sim 1^{-3}$rad) bending of the $ab$-plane towards one of the sample edges.
Figure \ref{fig:XRD2}b measured with the 3rd analyzer crystal  reveals a small spread in the lattice parameter for individual layers along the $c$ axis. For the main maximum we find $2\theta=51.9436^\circ$, and taking into account the correction for the hybrid monochromator, eventually, obtain $c=26.3780\AA$.

\begin{figure}[h]
	\includegraphics[width=240pt]{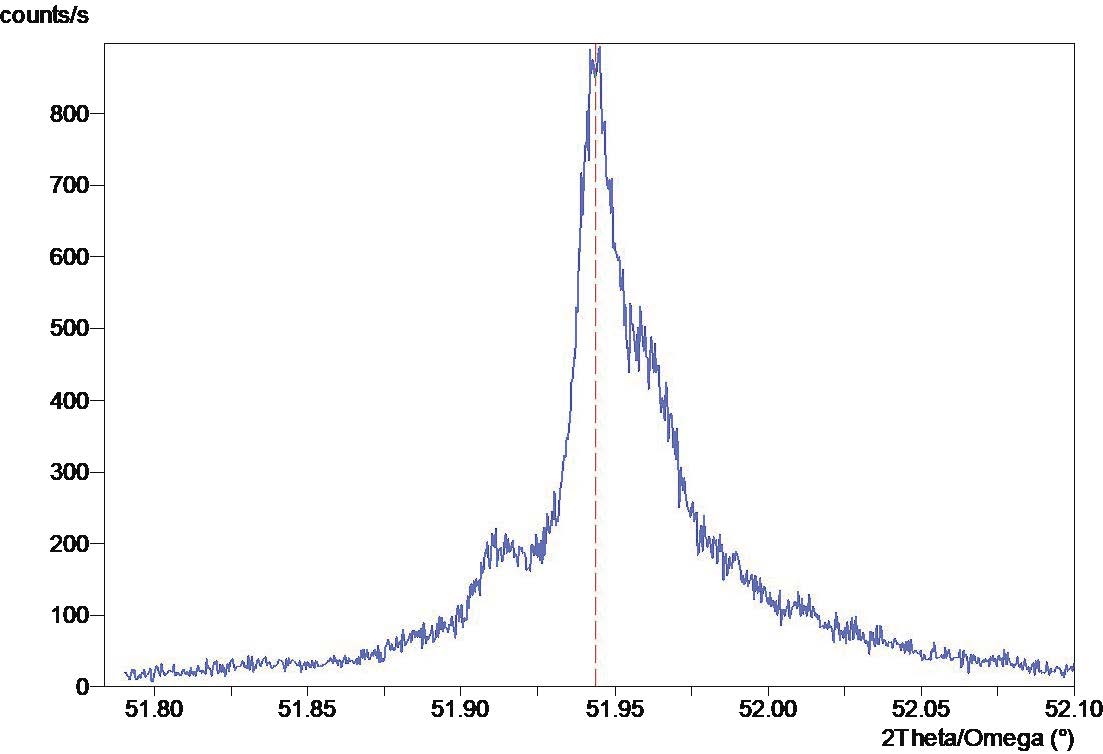}
	\caption{$2\theta\omega$ scanning curve measured with the third analyzer crystal. }
	\label{fig:XRD3}
\end{figure}

\begin{figure}[h]
	\includegraphics[width=240pt]{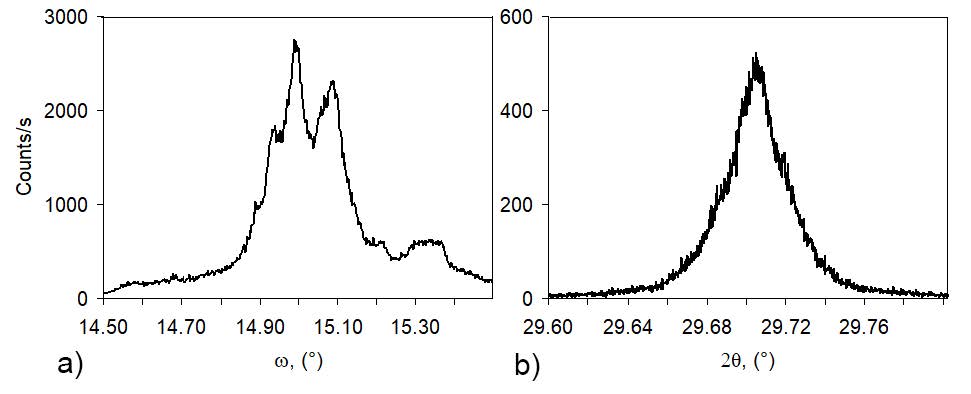}
	\caption{
		(a) Rocking curve on asymmetric reflection (10\,5) in the azimuthal position $\varphi=12^o$. 
		(b) The $(2\theta - \omega)$-scanning curve with the third crystal-analyzer on the asymmetric reflection (10\,5) in azimuthal position $\varphi=12^\circ$. }
	\label{fig:XRD4-5}
\end{figure}

Figure \ref{fig:XRD4-5}a reveals a block structure, which is hidden in the symmetrical reflection. The lattice parameter in the basal plane was determined  for the strongest peak. Note that all the curves without exception were measured under full illumination of the sample with X-ray beam.  Based on the angular position of $2\theta_{(10 5)}=29.7054^\circ$ in Fig.~\ref{fig:XRD4-5},  taking into account the correction for the hybrid monochromator, and the known value $c=26.3780$\AA, we find $a_{(10\,5)}=4.2190$\AA. Comparing with the literature data ($a=4.2071\AA$, $c=26.463 \AA$), we conclude that in the studied crystal the lattice parameter along the $c$ axis is lowered, and along the $a$ axis  is increased. 

\begin{figure}
	\includegraphics[width=240pt]{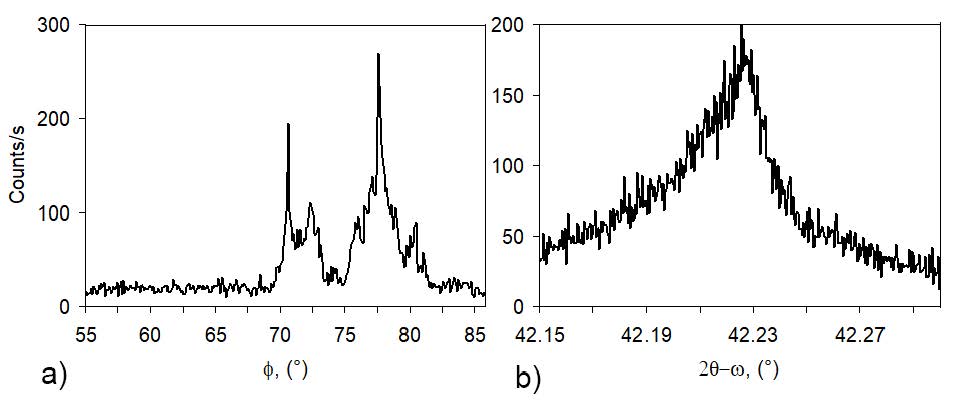}
	\caption{(a) The $\phi$-scanning curve about the normal to the basal plane on an asymmetric reflection (01\,10). 
		(b) $(2\theta -\omega)$ - scanning curve with the third crystal-analyzer on asymmetric reflection (01\, 10) in azimuthal position $\phi =70.4^{\circ}$. 	}
	\label{fig:XRD6-7}
\end{figure}
On the left panel of Fig.~\ref{fig:XRD6-7}a, instead of one peak anticipated at $\phi=72^\circ$ for the trigonal lattice, 
several peaks are seen, misoriented by $\Delta\phi=7^\circ$. This means that the main defect in the crystal is twisting of weakly interconnected layers about the normal to the basal plane.
From  the angular position of the $2\theta_{(01\,10)}=42.2254^\circ$ curve on the right panel of Fig.~\ref{fig:XRD6-7}, and taking into account the correction for the hybrid monochromator and the known value  $c=26.3780$\AA, 
we find $a_{(01\,10)}=4.2142$\AA. The asymmetry of the reflection is clearly seen, indicating the existence of layers with 
a large value of the lattice parameter in the basal plane.

\subsubsection{Conclusion from the XRD examination}
The studied crystal has one mirror shiny surface.
Our XRD measurements  evidence for the high quality of its crystal structure,
both the diffractogram and the rocking curve testify  the high structural perfection of the crystal.
In more detail:

\begin{enumerate}	
	\item 
	The lattice parameters of the studied crystal ($a_{(10\,5)}=4.2190$\AA, $c=26.3780$\AA) are slightly larger in
	the basal plane and smaller in the perpendicular direction compared to the literature data ($a=4.2071$\AA, $c=26.463$\AA) \cite{arguilla_InChemFront_2017, chen_ChPhysLet_2020}. The 0.3\%  diminishing of the $c$-axis spacing might be related to the presence of 3\% planar defects in which the  Eu-Eu spacing is smaller by 0.16\,nm  (see section \ref{sec:planar defect} and \cite{degtyarenko_tbp}). However, the diminished defect layer thickness may cause only a factor of 10 smaller $c$-parameter squeezing (0.16nm/26.46nm, that is  $0.7\% \times 3\% = 0.021\%$) insignificant to account for the measured 0.3\% difference in the $c$-axis parameter with literature data.
	\item
	The main defect of the studied crystal is the twisting (within $\approx 7^\circ$) of weakly coupled layers in the $a-b$ plane, perpendicular to the $c$-axis. 
\end{enumerate}

To summarize, 	no structural domains or blocks of different lattice orientations are found in the crystal.
Thus, our XRD examination shows the absence of such macroscopic structural defects  that can cause magnetic field dependent scattering of electrons. We note, however, that XRD is insensitive to purely magnetic defects.

	\subsection{ SM III: Energy spectrum measurements and calculations}
\label{SM:band structure}
The aim of the energy spectrum measurements  was (i) to verify  that the studied samples have the same band structure as earlier studied \cite{li_PRX_2019}, and (ii)  to quantify the  level splitting at the Fermi energy - the parameter required for the NIMR calculations. 
ARPES measurements were performed with UV He lamp source at 21.2eV photon energy. The samples were cleaved insitu 
in  vacuum of $10^{-10}$mBar. ARPES data were collected at sample temperature of 12\,K.

\begin{figure}[ht]
\vspace{0.2in}
	\includegraphics[width=180pt]{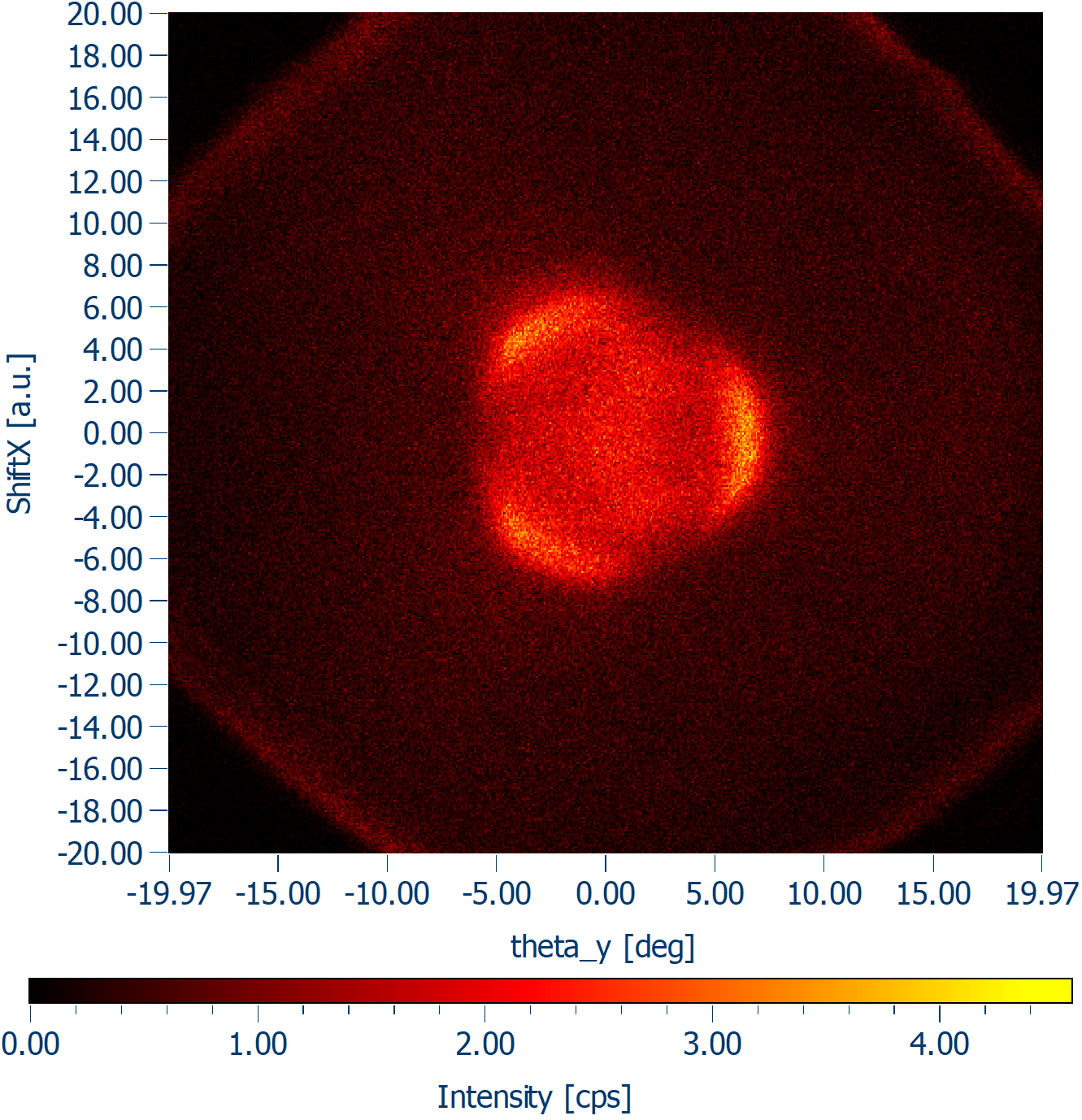}
	\includegraphics[width=180pt]{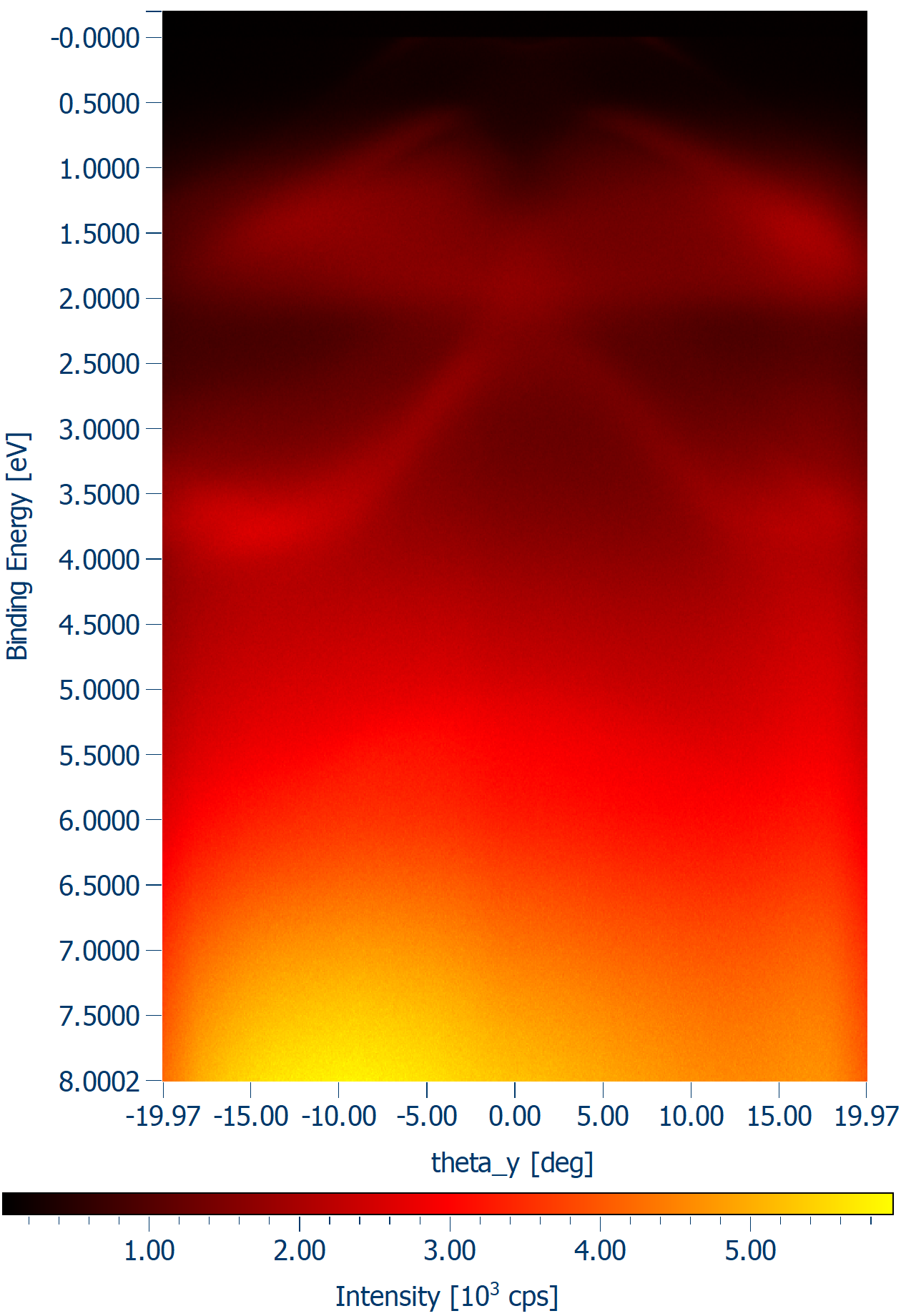}
	\caption{(a) Fermi surface map along $k_y-k_x$ across $k_z=0$ ($BE = 0$) at $h\nu=21.2$\,eV and  $T = 30$K.
		(b)	Electronic band dispersion along $M-\Gamma-M$; data are collected at $T=12$K with $h\nu=21.2$eV.}
	\label{fig:band structure}
\end{figure}
Figure~\ref{fig:band structure} shows the energy spectrum in the wide range of energies. From these data we conclude that the band structure
of the studied EuSn$_2$As$_2$ crystal is fully consistent with that in Ref.~\cite{li_PRX_2019}. 
The spectrum measured in the narrow  energy interval, in the vicinity of E$_F$, was shown in Fig.~\ref{fig:ARPES}.

\subsection{Planar nano-defects and weak ferromagnetism in the AFM ordered single crystal}
\label{sec:planar defect}
From TEM measurements on a thin (30\,nm) lamella cutout from the thick bulk crystal in the ac-plane  we detected  the presence of planar defects whose  density is sample-dependent, and roughly amounts to $\sim  3\%$. 
These investigations  are described in Ref.~\cite{degtyarenko_tbp}.
The planar defects have a local  composition approximated as EuSnAs$_ 2$. 
Such nanodefects are invisible by the conventional XRD technique.
The DFT calculations \cite{degtyarenko_tbp} show that the planar defect has a local ferromagnetic moment.
For the purpose of the current paper it is important only that	due to the small concentration of planar 
defects ($\sim 3-4$\%), and the small thickness  ($\approx 0.7$nm), much less than $\lambda_F$, 
they would rather unlikely to cause electron scattering and to affect the magnetotransport.
 However, their presence may induce a weak magnetization anisotropy  in the easy $ab$-plane. 
	
\end{document}